%% file: bodpin-schaefer-amsart.tex
\documentclass[11pt, a4paper]{amsart}
\usepackage{graphicx, a4wide}

\date{\today}
\newcommand{\citeyear}{\cite}

\title{Schaefer's theorem for graphs}

\author{Manuel Bodirsky}
\address{Institut f\"{u}r Algebra\\TU Dresden\\01062 Dresden\\Germany}
    \email{Manuel.Bodirsky@tu-dresden.de}
   \urladdr{http://www.math.tu-dresden.de/~bodirsky/}

\author{Michael Pinsker}
\address{Institut f\"{u}r Computersprachen\\Theory and Logic Group\\Technische Universit\"{a}t Wien\\Favoritenstrasse 9/E1852\\A-1040 Wien\\
Austria}
    \email{marula@gmx.at}
    \urladdr{http://dmg.tuwien.ac.at/pinsker/}
        
    \thanks{The research leading to these results has received funding from the European Research Council under the European Community's Seventh Framework Programme (FP7/2007-2013 Grant Agreement no. 257039). The second author has received funding through an APART fellowship of the Austrian Academy of Sciences as well as through project I836-N23 of the Austrian Science Fund (FWF). \\This is the journal version of an extended abstract that appeared at STOC 2011.}


\input macros.tex

\input theorems.tex

\newcommand{\cal}[1]{\mathcal #1}
\newcommand{\pf}{{Proof }}
\begin{document}

\maketitle
 \begin{abstract}
 \input abstract.tex
 \end{abstract}

\input motivationresult.tex

\input endos.tex
\input binary0.tex

\input binary1.tex

\input binary2.tex

\input binary3.tex

\input binary4.tex

\input algorithms.tex
\input classification.tex

\bibliographystyle{plain}
\bibliography{schaefer.bib}

\end{document}

%% file: macros.tex
\DeclareMathOperator{\Bool}{Boole}

\DeclareMathOperator{\NNE}{NNE}
\DeclareMathOperator{\NEN}{NEN}
\DeclareMathOperator{\ENN}{ENN}
\DeclareMathOperator{\NNN}{NNN}
\DeclareMathOperator{\EEE}{EEE}
\DeclareMathOperator{\EEN}{EEN}
\DeclareMathOperator{\ENE}{ENE}
\DeclareMathOperator{\NEE}{NEE}
\DeclareMathOperator{\EE}{EE}
\DeclareMathOperator{\NN}{NN}
\DeclareMathOperator{\EN}{EN}
\DeclareMathOperator{\NE}{NE}

\DeclareMathOperator{\EEQ}{E{=}}

\DeclareMathOperator{\ENEQ}{E{\neq}}
\DeclareMathOperator{\NEQ}{N{=}}
\DeclareMathOperator{\NEQEQ}{{\neq}{=}}
\DeclareMathOperator{\NEQNEQ}{\neq\neq}
\DeclareMathOperator{\NEQNEQNEQ}{\neq\neq\neq}

\DeclareMathOperator{\Aut}{Aut}

\newcommand{\ignore}[1]{}

\newcommand{\CSP}{\text{\rm CSP}}

\DeclareMathOperator{\sw}{sw}

\newcommand{\cproblem}[3]{
\vspace{.2cm}
\noindent {\bf #1} \\
INSTANCE: #2 \\
QUESTION: #3 \\}

\DeclareMathOperator{\inj}{inj}
\DeclareMathOperator{\Inv}{Inv}
\DeclareMathOperator{\Pol}{Pol}
\DeclareMathOperator{\Csp}{CSP}
\DeclareMathOperator{\Gsat}{Graph-SAT}
\DeclareMathOperator{\tp}{tp}
\DeclareMathOperator{\maxi}{max}
\DeclareMathOperator{\mini}{min}
\newcommand{\To}{\rightarrow}

\DeclareMathOperator{\xor}{xor}
\DeclareMathOperator{\xnor}{xnor}

%% file: theorems.tex
\theoremstyle{plain}

    \newtheorem{thm}{Theorem}
    \newtheorem{theorem}[thm]{Theorem}

    \newtheorem{lemma}[thm]{Lemma}

    \newtheorem{proposition}[thm]{Proposition}

    \newtheorem{corollary}[thm]{Corollary}

\theoremstyle{definition}

    \newtheorem{definition}[thm]{Definition}

%% file: abstract.tex
Schaefer's theorem is a complexity classification result for so-called
\emph{Boolean constraint satisfaction problems}: it states that every Boolean constraint satisfaction problem is either contained in one out of six classes and
can be solved in polynomial time,
or is NP-complete.

We present an analog of this dichotomy result for the \emph{propositional logic of graphs}
instead of Boolean logic. In this generalization of Schaefer's result,
the input consists of a set $W$ of variables and a conjunction $\Phi$ of statements (``constraints'') about these variables in the language of graphs, where each statement is taken from a fixed finite set $\Psi$ of allowed quantifier-free first-order formulas; the question is whether $\Phi$ is satisfiable in a graph.

We prove that either $\Psi$ is contained in one out of 17 classes of graph formulas and the corresponding problem can be solved in polynomial time, or the problem is NP-complete. This is achieved by a universal-algebraic approach, which in turn allows
us to use structural Ramsey theory. To apply the universal-algebraic
approach, we formulate the computational problems under consideration as constraint satisfaction problems (CSPs) whose templates
are first-order definable in the countably infinite random graph.
Our method for classifying the computational
complexity of those CSPs is based on a Ramsey-theoretic analysis of
functions acting on the random graph, and we develop general tools
suitable for such an analysis which are of independent mathematical
interest.

%% file: motivationresult.tex
\section{Motivation and the result}
In an influential paper in 1978, Schaefer~\citeyear{Schaefer} proved a
complexity classification for systematic restrictions of the Boolean
satisfiability problem. The way in which he restricts the Boolean satisfiability problem turned out to be very fruitful when restricting
other computational problems in theoretical computer science, and can be presented as follows.

Let $\Psi = \{\psi_1,\dots,\psi_n\}$ be
a finite set of propositional (Boolean) formulas.

\cproblem{Boolean-SAT$(\Psi)$}
{Given a finite set of variables $W$ and a propositional formula of the form
$\Phi = \phi_1 \wedge \dots \wedge \phi_l$ where each $\phi_i$ for $1 \leq i \leq l$ is obtained from one of the formulas $\psi$ in $\Psi$ by substituting the variables of $\psi$ by variables from $W$.}
{Is there a satisfying Boolean assignment to the variables of $W$ (equivalently, those of $\Phi$)?}

The computational complexity of this problem clearly depends on the
set $\Psi$, and is monotone in the sense that if $\Psi\subseteq \Psi'$, then solving Boolean-SAT($\Psi'$) is at least as hard as solving Boolean-SAT($\Psi$). Schaefer's theorem states that Boolean-SAT$(\Psi)$ can be solved in
polynomial time
if $\Psi$ is a subset of one of six sets of Boolean formulas (called \emph{0-valid, 1-valid, Horn, dual-Horn, affine}, and \emph{bijunctive}),
and is NP-complete otherwise. 

We prove a similar classification result, but for the propositional logic
of graphs instead of for propositional Boolean logic. More precisely, let $E$ be a relation symbol which denotes an antireflexive and symmetric binary relation and hence stands for the edge relation of a (simple, undirected) graph. We consider formulas that are constructed from atomic formulas of the form $E(x,y)$ and $x=y$ by the usual Boolean connectives (negation, conjunction, disjunction), and call formulas of this form \emph{graph formulas}. A graph formula $\Phi(x_1,\ldots,x_m)$ is \emph{satisfiable} if there exists a graph $H$ and an $m$-tuple $a$ of elements in $H$ such that $\Phi(a)$ holds in $H$.

The problem of deciding whether a given graph formula is satisfiable can be very difficult.
For example, the question whether or not
the Ramsey number $R(5,5)$ is larger than $43$ (which is an open problem, see e.g.~\cite{Exoo}) can be easily formulated 
in terms of satisfiability of a single graph formula.
Recall that $R(5,5)$ is the least number $k$ such
that every graph with at least $k$ vertices either contains
a clique of size $5$ or an independent set of size $5$. 
So the question whether or not $R(5,5)$ is greater than $43$
can be formulated as the question of satisfiability of a graph formula
using $43$ variables $x_1,\dots,x_{43}$ on which one imposes  
the following constraints: all variables denote different vertices in the graph,
and for every five-element subset of the variables we add a constraint that
forbids that the variables of this subset form a clique or an independent set; this can clearly be stated as a graph formula. 
If this graph formula is satisfiable, then this implies that $R(5,5) > 43$,
and otherwise $R(5,5) \leq 43$.

Let $\Psi = \{\psi_1,\dots,\psi_n\}$ be a finite set of graph formulas.
Then $\Psi$ gives
rise to the following computational problem.

\cproblem{Graph-SAT$(\Psi)$}
{Given a set of variables $W$ and a graph formula of the form
$\Phi = \phi_1 \wedge \dots \wedge \phi_l$ where each $\phi_i$ for $1 \leq i \leq l$ is obtained from one of the formulas $\psi$ in $\Psi$ by substituting the variables from $\psi$ by variables from $W$.}
{Is $\Phi$ satisfiable?}

As an example, let $\Psi$ be the set that just contains the formula
\begin{align}
 & (E(x,y) \wedge \neg E(y,z) \wedge \neg E(x,z))  \nonumber \\
\vee\; &
 (\neg E(x,y) \wedge E(y,z) \wedge \neg E(x,z))   \label{eq:rel} \\
 \vee \;  & (\neg E(x,y) \wedge \neg E(y,z) \wedge E(x,z)) \; . \nonumber
\end{align}
Then Graph-SAT$(\Psi)$ is the problem of deciding whether there
exists a graph such that certain prescribed subsets of its vertex set of cardinality at most three induce subgraphs with exactly one edge. This problem is NP-complete (the curious reader can check this by means of our classification in Theorem~\ref{thm:minimalTractableClones}).

Consider now the example where $\Psi$ consists of the formula 
\begin{align}
& (E(x,y) \wedge \neg E(y,z) \wedge \neg E(x,z)) \nonumber \\
 \vee \;  & (\neg E(x,y) \wedge E(y,z) \wedge \neg E(x,z)) \\
 \vee \; & (\neg E(x,y) \wedge \neg E(y,z) \wedge E(x,z)) \nonumber \\
 \vee \; & (E(x,y) \wedge E(y,z) \wedge E(x,z)) \; . \nonumber
\end{align}
In this example, Graph-SAT$(\Psi)$ is the problem of deciding whether there
exists a graph such that certain prescribed subsets of its vertex set of cardinality at most three induce either a subgraph with exactly one edge, or a complete triangle. This problem is tractable -- any instance is satisfiable in a clique. As we will see, the problem remains tractable if $\Psi$ additionally  contains the formula $\neg E(x,y)$.

The class of Graph-SAT problems generalizes 
the class of problems studied by Schaefer,
since to every set $\Psi$ of Boolean formulas
we can associate a set $\Psi'$ of graph formulas
such that Graph-SAT$(\Psi')$ and Boolean-SAT$(\Psi)$ are essentially the same problem. For every variable $x$ of $\Psi$ there are two variables $x_1,x_2$ in $\Psi'$. 
Then $\Psi'$ contains for every $\psi \in \Psi$ the graph formula obtained from
$\psi$ by replacing positive literals $x$ by $E(x_1,x_2)$,
and negative literals $\neg x$ by $N(x_1,x_2)$. 
An instance $\Phi$ of
Boolean-SAT$(\Psi)$ translates into an instance
$\Phi'$ of Graph-SAT$(\Psi')$ by modifying $\Phi$ in the same way; then $\Phi$ is satisfiable if and only if $\Phi'$ is satisfiable. 

It is obvious that the problem Graph-SAT$(\Psi)$ is
for all $\Psi$ contained in NP. 
The goal of this paper is to prove the following dichotomy result.
\begin{theorem}\label{thm:main}
For all $\Psi$, the problem $\Gsat(\Psi)$ is either NP-complete
or in P. Moreover, the problem
of deciding for given $\Psi$ whether $\Gsat(\Psi)$ is NP-complete or in P is decidable.  
\end{theorem}

One of the main contributions
of this paper is a novel general method combining concepts from
universal algebra and model theory with powerful tools of Ramsey theory.

\section{Discussion of our strategy}\label{sect:strategy}

We establish our result by translating Graph-SAT problems
into \emph{constraint satisfaction problems} (CSPs) over infinite domains.
More specifically, for every set of formulas $\Psi$ we present
an infinite relational structure $\Gamma_\Psi$ such that
Graph-SAT$(\Psi)$ is equivalent to $\Csp(\Gamma_\Psi)$; in a certain sense, Graph-SAT$(\Psi)$ and $\Csp(\Gamma_\Psi)$ are one and the same problem. The relational structure $\Gamma_\Psi$
has a first-order definition in the \emph{random graph $G$}, i.e., the (up to isomorphism) unique countably infinite universal homogeneous graph. 
This perspective allows us to use the so-called \emph{universal-algebraic approach}, and in particular \emph{polymorphisms} to classify the computational complexity of Graph-SAT problems.
In contrast to the universal-algebraic approach for
finite domain constraint satisfaction, our proof relies crucially on
strong results from structural Ramsey theory;
we use such results to find regular patterns in the behavior of polymorphisms of structures with a first-order definition in $G$, which in turn allows us to find analogies with polymorphisms of structures on a Boolean domain.

We call structures with a first-order definition in $G$ \emph{reducts} of $G$. While the classical  definition of a reduct of a relational structure $\Delta$  is a structure on the same domain obtained by forgetting some relations of $\Delta$, a reduct of $\Delta$ in our sense (following~\cite{RandomReducts}) is really a reduct of the expansion of $\Delta$ by all first-order definable relations.
It turns out that there is one class of reducts $\Gamma$ of $G$ for which $\Csp(\Gamma)$ is in P for trivial reasons; further, there are
16 classes of reducts $\Gamma$ for which $\Csp(\Gamma)$
(and the corresponding Graph-SAT problems)
can be solved by
non-trivial algorithms in polynomial time.

The presented algorithms are novel combinations of infinite domain constraint satisfaction techniques (such as used in~\cite{CJJK,ll,Maximal}) and reductions to
the tractable cases of Schaefer's theorem. Reductions of infinite domain CSPs in artificial intelligence
(e.g., in temporal and spatial reasoning~\cite{Duentsch}) to finite domain CSPs (where typically the domain consists of the elements of a so-called `relation algebra')
have been considered in the more applied
artificial intelligence literature~\cite{WestphalWoelfl}. Our results shed some light on
the question as to when such techniques can even lead to \emph{polynomial-time}
algorithms for CSPs.

The global classification strategy of the present paper is similar in spirit to the strategy presented in~\cite{tcsps-journal} for CSPs of reducts of $(\mathbb Q; <)$. But while in~\cite{tcsps-journal} the proof might
still have appeared to be very specific to constraint satisfaction over linear orders, with the present paper we demonstrate that in principle
such a strategy can be used for any class $\mathcal C$ of computational
problems that satisfies the following:
\begin{itemize}
\item All problems in $\mathcal C$ can be formulated as a CSP of a structure
which is first-order definable in a single 
structure $\Delta$;
\item $\Delta$ is homogeneous in a finite language and the class of finite substructures of $\Delta$ has the Ramsey property (as in~\cite{NesetrilSurvey}).
\end{itemize}
The subsequent survey article~\cite{BP-reductsRamsey}
is devoted to the application of the method of this paper
in this more general setting, providing further examples. 
We remark that in our case, the structure $\Delta$ above is the \emph{random ordered graph} (roughly the random graph equipped with the order of the rationals in 
a random way -- confer Section~\ref{sect:higher-arity}) rather than the random graph $G$ itself. The reducts of this structure have recently been classified~\cite{42}.

While in~\cite{tcsps-journal}, the classical theorem of Ramsey and its
product version were sufficient, the Ramsey theorems used in
the present paper are deeper and considerably more difficult
to prove~\cite{NesetrilRoedlOrderedStructures,AbramsonHarrington}.

\section{Tools from universal algebra and model theory}\label{sect:toolsua}

We now develop in detail the tools from universal algebra and model theory needed for our approach. We start by translating the problem Graph-SAT$(\Psi)$ into a constraint satisfaction problem for a reduct of the random graph $G$.

We write $G=(V;E)$ for the random graph. The graph $G$ is determined up to isomorphism by the two properties of being \emph{homogeneous} (i.e., any isomorphism between two finite induced subgraphs of $G$ can be extended to an automorphism of $G$), and \emph{universal} (i.e., $G$ contains all countable graphs as induced subgraphs). 
The random graph $G$ has the property of 
\emph{quantifier elimination}, that is, every first-order
formula is over $G$ equivalent to a quantifier-free first-order formula. 
Moreover, $G$ has the \emph{extension property}, which often is useful in combinatorial arguments: for all disjoint finite $U, U'\subseteq V$ there exists $v\in V$ such that $v$ is adjacent in $G$ to all members of $U$ and to none in $U'$. Up to isomorphism, there exists only one unique countably infinite graph which has this extension property, and hence the property can be used as an alternative definition of $G$.
The name of the random graph is due to the fact that if for a countably infinite vertex set, one chooses independently and with probability $\frac{1}{2}$ for each pair of vertices whether to connect the two vertices by an edge, then with probability $1$ the resulting graph is isomorphic to the random graph.
For
the many other remarkable properties of $G$ and its automorphism group $\Aut(G)$,
and various connections to many branches of mathematics, see e.g.~\cite{RandomCameron,RandomRevisitedCameron}.

Let $\Gamma$ be a structure with a finite relational signature $\tau$.
A first-order $\tau$-formula is called \emph{primitive positive} if
it is of the form $$\exists x_1,\dots,x_n.\; \psi_1 \wedge \dots \wedge \psi_m,$$ where the $\psi_i$ are \emph{atomic}, i.e., of the form $y_1=y_2$ or $R(y_1,\dots,y_k)$ for a $k$-ary relation symbol 
$R \in \tau$ and not necessarily distinct variables $y_i$. A $\tau$-formula is called a \emph{sentence} if it contains no free variables.

\begin{definition}
The \emph{constraint satisfaction problem for $\Gamma$}, denoted by $\Csp(\Gamma)$, is the computational problem of deciding for a given primitive positive $\tau$-sentence
$\Phi$ whether $\Phi$ is true in $\Gamma$.
\end{definition}

Let $\Psi=\{\psi_1,\ldots,\psi_n\}$ be a set of graph formulas. Then we define $\Gamma_\Psi$ to be the structure with the same domain $V$ as the random graph $G$ which has for each $\psi_i$ a relation $R_i$ consisting of those tuples in $G$ that satisfy $\psi_i$ (where the arity of $R_i$ is given by the number of variables that occur in $\psi_i$). Thus by definition, $\Gamma_\Psi$ is a reduct of $G$. Now given any instance $\Phi = \phi_1 \wedge \dots \wedge \phi_l$ with variable set $W$ of Graph-SAT$(\Psi)$, we construct a primitive positive sentence $\Phi'$ in the language of $\Gamma_\Psi$ as follows: In $\Phi$, we replace every $\phi_i$, which by definition is of the form $\psi_j(y_1,\ldots,y_m)$ for some $1\leq j\leq n$ and variables $y_k$ from $W$, by $R_j(y_1,\ldots,y_m)$; after that, we existentially quantify all variables that occur in $\Phi'$. It then follows immediately from the universality of $G$ that the problem Graph-SAT($\Psi$) has a positive answer for $\Phi$ if and only if the sentence $\Phi'$ holds in $\Gamma_\Psi$. Hence, every problem Graph-SAT$(\Psi)$ \emph{is} in fact of the form $\Csp(\Gamma)$, for a reduct $\Gamma$ of $G$ in a finite signature. We will thus henceforth focus on such constraint satisfaction problems in order to prove our dichotomy.

The following lemma has been first stated in~\cite{JeavonsClosure} for finite domain structures $\Gamma$ only, but the proof there also works for arbitrary infinite structures. It shows us how we can slightly enrich structures without changing the computational complexity of the constraint satisfaction problem they define too much.

\begin{lemma}\label{lem:pp-reduce}
Let $\Gamma = (D; R_1,\dots,R_l)$ be a relational structure,
and let $R$ be a
relation that has a primitive positive definition in $\Gamma$.
Then $\Csp(\Gamma)$ and
$\Csp(D; R,R_1,\dots,R_l)$ are polynomial-time equivalent.
\end{lemma}

The preceding lemma enables the so-called \emph{universal-algebraic approach} to constraint satisfaction, as exposed in the following.
We say that a $k$-ary function (also called \emph{operation})
$f \colon D^k \rightarrow D$ \emph{preserves} an $m$-ary relation
$R \subseteq D^m$ if for all $t_1,\dots,t_k \in R$ the tuple
$f(t_1,\dots,t_k)$ (calculated componentwise) is also contained in $R$. In that case, we also say that $R$ is \emph{invariant} under $R$.
If an operation $f$ does not
preserve a relation $R$, we say that $f$ \emph{violates} $R$.

If $f$ preserves all relations of a structure $\Gamma$, we say that $f$ is a \emph{polymorphism} of $\Gamma$
(it is also common to say that $\Gamma$ is \emph{closed
under $f$}, or that $f$ \emph{preserves} $\Gamma$). We write $\Pol(\Gamma)$ for the set of all polymorphisms of $\Gamma$. 
A unary polymorphism of $\Gamma$
is also called an \emph{endomorphism} of $\Gamma$.

Conversely, for a set $F$ of operations of finite arity defined on a set $D$ and a finitary relation $R$ on $D$, we say that $R$ is \emph{invariant} under $F$ if $R$ is invariant under all $f\in F$, and we write $\Inv(F)$ for the set of all finitary relations on $D$ that are invariant under $F$.

The set of all polymorphisms $\Pol(\Gamma)$ of a relational structure $\Gamma$ forms an algebraic object called a \emph{clone} (see~\cite{Szendrei}, \cite{GoldsternPinsker}), which is
a set of finitary operations defined on a set $D$ that is closed
under composition and that contains all projections.
Moreover, $\Pol(\Gamma)$ is closed under interpolation (see Proposition 1.6 in~\cite{Szendrei}):
we say that a $k$-ary operation $f$ on $D$ is \emph{interpolated}
by a set of operations $F$ on $D$
if for every finite subset $A$ of $D^k$
there is some $k$-ary operation $g \in F$ such that $g$ agrees with $f$ on $A$.
We say that $F$ \emph{locally generates} an operation $g$ if
$g$ is contained in the smallest clone that is closed under interpolation and
contains all operations in $F$.  Clones with the property that they contain all functions locally generated by their members are called \emph{locally closed}, \emph{local} or just \emph{closed}.

We can thus assign to every structure $\Gamma$ the closed clone $\Pol(\Gamma)$ of its polymorphisms. For certain $\Gamma$, this clone
captures the computational complexity of $\Csp(\Gamma)$:
a countable
structure $\Gamma$ is called \emph{$\omega$-categorical} if
every countable model of the first-order theory of $\Gamma$ is
isomorphic to $\Gamma$. It is well-known that the random graph $G$ is $\omega$-categorical, and that reducts of $\omega$-categorical structures are $\omega$-categorical as well (see for example~\cite{Hodges}).

\begin{theorem}[from~\cite{BodirskyNesetrilJLC}]
\label{conf:thm:inv-pol}
Let $\Gamma$ be an $\omega$-categorical structure.
Then the relations preserved by the polymorphisms of $\Gamma$, i.e., the relations in  $\Inv(\Pol(\Gamma))$, 
are precisely those having a primitive positive definition in $\Gamma$.
\end{theorem}

Clearly, this theorem together with Lemma~\ref{lem:pp-reduce} imply that if two $\omega$-categorical structures $\Gamma, \Delta$ with finite relational signatures have the same clone of polymorphisms, then their CSPs are polynomial-time
equivalent. Moreover, if $\Pol(\Gamma)$ is contained in $\Pol(\Delta)$, then $\CSP(\Gamma)$ is, up to polynomial time, at least as hard as $\CSP(\Delta)$.

Recall that we have only defined $\Csp(\Gamma)$
for structures $\Gamma$ with a finite relational signature. But we now see that it makes sense (and here we follow conventions from finite domain constraint satisfaction, see e.g.~\cite{JBK}) to say for arbitrary $\omega$-categorical structures $\Gamma$ that $\Csp(\Gamma)$ is \emph{(polynomial-time) tractable} if the CSP for 
every finite signature structure
$\Delta$ with $\Pol(\Delta)\supseteq \Pol(\Gamma)$ is in P,
and to say that $\Csp(\Gamma)$ is \emph{NP-hard} if $\Csp(\Delta)$ is NP-hard for some finite signature structure $\Delta$ with $\Pol(\Delta)\supseteq \Pol(\Gamma)$.

Note that the \emph{automorphisms} of a structure $\Delta$
are just the bijective unary polymorphisms of $\Delta$ which preserve all relations
and their complements; the set of all automorphisms of $\Delta$
is denoted by $\text{Aut}(\Delta)$. It follows from the theorem of Ryll-Nardzewski (cf.~\cite{Hodges}) that for $\omega$-categorical structures $\Delta$, the closed clones containing $\Aut(\Delta)$ are precisely the polymorphism clones of reducts $\Gamma$ of $\Delta$. Therefore, in order to determine the computational complexity of the CSP of all reducts $\Gamma$ of $G$, it suffices to determine for every closed clone ${\mathcal C}$ containing $\Aut(G)$ the complexity of $\Csp(\Gamma)$ for some reduct $\Gamma$ of $G$ with $\Pol(\Gamma)={\mathcal C}$; then the complexity for all reducts with the same polymorphism clone is polynomial-time equivalent to $\Csp(\Gamma)$.

The following proposition is the analog to Theorem~\ref{conf:thm:inv-pol} on the ``operational side'', and characterizes the local generating process of functions on a domain $D$ by the operators $\Inv$ and $\Pol$.

\begin{proposition}[Corollary 1.9 in~\cite{Szendrei}]\label{prop:locclos}
    Let $F$ be a set of functions on a domain $D$, and let $g$ be a function on $D$. Then $F$ locally generates $g$
    if and only if $g$ preserves all relations that are invariant under $F$, i.e., if and only if $g\in\Pol(\Inv(F))$.
\end{proposition}

For some reducts, we will find that their CSP is equivalent to a CSP of a structure that has already been studied, by means of the following basic observation.

\begin{proposition}\label{prop:homoEqui}
    Let $\Gamma, \Delta$ be \emph{homomorphically equivalent}, i.e., they have the same signature and there exist homomorphisms $f \colon \Gamma\To\Delta$ and $g \colon \Delta\To\Gamma$. Then $\Csp(\Gamma)=\Csp(\Delta)$.
\end{proposition}

We finish this section with a technical general lemma that we will refer to on numerous occasions; it allows to restrict the arity of functions violating a relation. For a structure $\Gamma$ with domain $D$ and a tuple $t \in D^k$, the \emph{orbit of $t$} in $\Gamma$ is the set
$\{ \alpha(t) \; | \; \alpha \in \Aut(\Gamma) \}$.

\begin{lemma}[from~\cite{tcsps-journal}]\label{lem:arity-reduction}
    Let $\Gamma$ be a relational structure with domain $D$, and
    suppose that $R\subseteq D^k$ intersects not more than $m$ orbits of $k$-tuples in $\Gamma$. Suppose that an operation $f$ on $D$ violates $R$. Then $\{f\}\cup\Aut(\Gamma)$ locally generates an at most $m$-ary operation that violates $R$.
\end{lemma}

\section{Overview of the proof}

The method for proving Theorem~\ref{thm:main} can be described as follows. We remark that in principle, a similar strategy could work for reducts of other structures than the random graph; confer the end of Section~\ref{sect:strategy} for a description of the conditions we require.

The first step is providing hardness proofs for certain relations with a first-order definition over $G$. More precisely, we define seven relations $E_6$, $N_6$, $I_6$, $H_1$, $H_1'$, $H_2$, and $H_2'$ which have first order-definitions in $G$, and show hardness for the CSP defined by each of these relations by reduction of known NP-hard problems. We then know from Lemma~\ref{lem:pp-reduce} that if the CSP for a reduct $\Gamma$ is not NP-hard, then there is no primitive positive definition of any of these relations in $\Gamma$. This implies that there are polymorphisms of $\Gamma$ which violate the relations, by Theorem~\ref{conf:thm:inv-pol}.

We then analyze the polymorphisms of $\Gamma$ which violate the relations $E_6$, $N_6$, $I_6$, $H_1$, $H_1'$, $H_2$, and $H_2'$. The first, rather basic tool here is Lemma~\ref{lem:arity-reduction}, which we use in order to get bounds on the arity of such polymorphisms. The deeper part of our analysis is the simplification of the polymorphisms by means of Ramsey theory. It turns out that the polymorphisms can be assumed to behave regularly in a certain sense with respect to the base structure $G$ (the technical term for functions showing such regular behavior will be \emph{canonical}), making them accessible to case-by-case analysis. In order to be able to use results from Ramsey theory, we have to expand the structure $G$ generically by a linear order $\prec$ on $V$ which is isomorphic to the order of the rational numbers.

Finally, the presence of canonical polymorphisms is used in two ways: in the case of canonical unary polymorphisms, the image under such a polymorphism sometimes is a structure $\Delta$ for which the CSP has already been classified, and then one can refer to Proposition~\ref{prop:homoEqui} to argue that the $\CSP(\Gamma)$ is polynomial-time equivalent to the CSP of this structure $\Delta$. The second, and in our case considerably more important way of employing canonical polymorphisms,  is to prove tractability of $\CSP(\Gamma)$ by using the polymorphisms to design algorithms. Here, we adapt known algorithms showing that certain polymorphisms on a Boolean domain imply tractability of Boolean CSPs in order to prove that the same holds for their canonical counterparts on the random graph.

For reasons of efficiency, we present our proof in a slightly different fashion, albeit the above strategy describes our intuition behind it. We first cite known results on automorphism groups and endomorphism monoids of reducts of $G$, in particular from~\cite{Thomas96} and~\cite{RandomMinOps}.  These older results have been obtained using Ramsey theory, and thus by building on them we outsource  the  Ramsey-theoretic analysis of unary polymorphisms of reducts. Putting them together, we obtain a statement saying that for any reduct $\Gamma$ of $G$, either 
$\Gamma$ has a constant endomorphism, and its CSP is tractable, or
$\Gamma$ is homomorphically equivalent to a structure with a first-order definition in $(V;=)$, in which case the complexity of its CSP is known, or its endomorphisms are locally generated by $\Aut(\Gamma)$ (Section~\ref{sect:endos}). The latter case splits into four subcases, corresponding to the precisely four proper subgroups of the full symmetric group on $V$ which are automorphism groups of reducts of $G$.

In Section~\ref{sect:higher-arity}, we consider each of those four possibilities for $\Aut(\Gamma)$. Working under the assumption that the endomorphisms of $\Gamma$ are locally generated by $\Aut(\Gamma)$, we analyze the higher arity polymorphisms of $\Gamma$ to a level of detail not present in the literature (although we do also draw on earlier results on such higher arity polymorphisms from~\cite{RandomMinOps}). It is here where we apply Ramsey theory directly in our paper. We show that  in all four cases, either one of the hard relations $H_1$, $H_1'$, $H_2$, or $H_2'$ has a primitive positive definition in $\Gamma$, or $\Gamma$ has binary or ternary canonical polymorphisms with particular properties. Each of the four hard relations $H_1$, $H_1'$, $H_2$, and $H_2'$ corresponds to one of the possible cases for $\Aut(\Gamma)$.

Finally, Section~\ref{sect:algorithms} presents polynomial-time algorithms for reducts having these particular canonical polymorphisms.

The proof of the dichotomy claimed in Theorem~\ref{thm:main} is followed by Section~\ref{sect:classification} in which the classification is stated in more detail and the decidability part of the theorem is derived.

\section{Additional conventions}
When working with relational structures $\Gamma$,
we often use the same symbol for a relation of $\Gamma$ and its relation symbol. In particular, we use the symbol $E$ to denote both the edge relation of $G$ and the corresponding symbol in graph formulas. 

Since all our polymorphism clones contain the automorphism group $\Aut(G)$ of the random graph, we will abuse the notion of \emph{generates} from Section~\ref{sect:toolsua}, and use it as follows: for a set of functions $F$ and a function $g$ on the domain $V$, we say that $F$ \emph{generates} $g$
when $F\cup \Aut(G)$ locally generates
$g$; also, we say that a function $f$ \emph{generates} $g$ if $\{f\}$ generates $g$. That is, in this paper we consider the automorphisms of $G$ be present in all sets of functions when speaking about the local generating process.

The binary relation $N(x,y)$ on $V$ is defined by the formula $\neg E(x,y)\wedge x\neq y$. We use $\neq$ both in logical formulas to denote the negation of equality, and to denote the corresponding binary relation on $V$.

When $t$ is an $n$-tuple, we refer to its entries by $t_1,\ldots,t_n$. When $f \colon A \rightarrow B$ is a function and $C \subseteq A$, we write $f[C]$ for $\{f(a) \; | \; a \in C\}$.

%% file: endos.tex
\section{Endomorphisms} \label{sect:endos}
The goal of this section is the proof of Proposition~\ref{prop:endos}, which will in particular allow us to reduce the classification task
to the classification of those structures whose 
automorphism generate its endomorphisms. To state the proposition, we first define the following unary functions on $V$ that will play an important role throughout the paper. 

If we flip edges and non-edges of $G$, then the resulting graph is
isomorphic to $G$: it is straightforward to verify the extension
property. Let $-$ be such an isomorphism. 

For any finite subset $S$ of $V$, if we flip edges and non-edges
between $S$ and $V \setminus S$ in $G$, then the resulting graph is
isomorphic to $G$; again, this follows by verifying the extension
property. Let $\sw_S$ be such an isomorphism for each non-empty
finite $S$. Any two such functions
generate one another~\cite{RandomReducts}. We also write $\sw$ for $\sw_{\{0\}}$, where $0 \in V$ is any
fixed element of $V$.

There are automorphisms $\alpha,\beta$ of $G$ such that $x\mapsto \alpha(-(x))$ and $x\mapsto \beta(\sw(x))$ are the inverse functions of the functions $-$ and $\sw$, respectively; this follows readily from the definitions. 
Hence, if $-$ or $\sw$ preserve a relation $R$ with a first-order definition in $G$, they automatically preserve also the complement of $R$, and thus are automorphisms of the structure $(V;R)$.

The graph $G$ contains all countable graphs as
induced subgraphs. In particular, it contains an infinite complete
subgraph.
The homogeneity of $G$ implies 
that any two injective unary operations on $V$ 
whose images induce complete subgraphs in $G$  generate one another (see, e.g.,~\cite{RandomMinOps}); 
let $e_E$ be one such operation.
Similarly,
$G$ contains an infinite independent set.
Let $e_N$ be an injective unary operation on $V$ whose image induces an infinite independent set in $G$.

\begin{proposition}\label{prop:endos}
Let $\Gamma$ be a reduct of $G$. Then at least one of the following holds.
\begin{enumerate}
\item[(a)] $\Gamma$ has a constant endomorphism, and $\Csp(\Gamma)$ is in P. 
\item[(b)] $\Gamma$ has $e_E$ or $e_N$ among its endomorphisms, and $\Gamma$ 
is homomorphically equivalent to a countably infinite structure that is preserved by all permutations of its domain. In this case the complexity of $\Csp(\Gamma)$ has been classified in~\cite{ecsps}, and is either in P or NP-complete.
\item[(c)] The endomorphisms of $\Gamma$
are precisely the functions generated by $\{-\}$. 
\item[(d)] The endomorphisms of $\Gamma$
are precisely the functions generated by $\{\sw\}$. 
\item[(e)] The endomorphisms of $\Gamma$
are precisely the functions generated by $\{-,\sw\}$. 
\item[(f)] The endomorphisms of $\Gamma$ are precisely the functions generated by $\Aut(G)$, i.e., all endomorphisms of $\Gamma$ preserve $E$ and $N$. 
\end{enumerate}
\end{proposition}

Proposition~\ref{prop:endos} follows from two results about unary functions on $G$. The first result is from~\cite{Thomas96}; its reformulation from~\cite{RandomMinOps} reads as follows.

\begin{theorem}\label{thm:endos}
Let $\Gamma$ be a reduct of $G$. Then one
of the following cases applies.
\begin{enumerate}
\item $\Gamma$ has a constant endomorphism.
\item $\Gamma$ has the endomorphism $e_E$.
\item $\Gamma$ has the endomorphism $e_N$.
\item The endomorphisms of $\Gamma$ are generated by $\Aut(\Gamma)$.
\end{enumerate}
\end{theorem}

The second result we use, from~\cite{RandomReducts}, states that there exist precisely five
permutation groups on $V$ that contain $\Aut(G)$ and which are closed in the sense that they contain all permutations which they interpolate. By the theorem of Ryll-Nardzewski (confer also the discussion in Section~\ref{sect:toolsua}), these groups correspond precisely to the automorphism groups of reducts of $G$. Thus, the last case of Theorem~\ref{thm:endos} splits into five subcases, one for each group of the form $\Aut(\Gamma)$. We will next cite the theorem that lists them.

\begin{definition}
For $k\geq 1$, let $R^{(k)}$ be the $k$-ary relation that contains a tuple $(x_1,\dots,x_k) \in V^k$ if $x_1,\dots,x_k$ are pairwise distinct, and the number of edges
between these $k$ vertices is odd.
\end{definition}

\begin{definition}
    We say that two structures $\Gamma, \Delta$ on the same domain are \emph{first-order interdefinable} if all relations of $\Gamma$ have a first-order definition in $\Delta$ (without parameters) and vice-versa.
\end{definition}

\begin{theorem}[from~\cite{RandomReducts}]\label{thm:reducts}
Let $\Gamma$ be a reduct of $G$. Then exactly one of the following is true.
\begin{enumerate}
\item $\Gamma$ is first-order interdefinable with $(V;E)$; \\
equivalently, $\Aut(\Gamma)=\Aut(G)$.
\item $\Gamma$ is first-order interdefinable with $(V;R^{(4)})$; \\
equivalently, $\Aut(\Gamma)$ contains $\{-\}$, but not $\{\sw\}$.
\item $\Gamma$ is first-order interdefinable with $(V;R^{(3)})$; \\
equivalently, $\Aut(\Gamma)$ contains $\{\sw\}$, but not $\{-\}$.
\item $\Gamma$ is first-order interdefinable with
$(V;R^{(5)})$; \\
equivalently, $\Aut(\Gamma)$ contains $\{-,\sw\}$, but not all permutations of $V$.
\item $\Gamma$ is first-order interdefinable with $(V;=)$;  \\
equivalently, $\Aut(\Gamma)$ contains all permutations of $V$.
\end{enumerate}
\end{theorem}
\begin{proof}[of Proposition~\ref{prop:endos}]
If $\Gamma$ has a constant endomorphism, 
then $\Csp(\Gamma)$ is trivial, and in P. 
Otherwise, by Theorem~\ref{thm:endos},
$\Gamma$ is preserved by $e_N$, $e_E$, or 
the endomorphisms of $\Gamma$ are generated by $\Aut(\Gamma)$.  

We claim that if $\Gamma$ has the endomorphisms $e_E$ or $e_N$, then $\Gamma$ is homomorphically equivalent to an infinite structure that is preserved by all permutations of its domain. But this is clear since $e_E[V]$ and $e_N[V]$ induce structures in $G$ which are invariant under all permutations of their domain.

If the endomorphisms of $\Gamma$ are generated by $\Aut(\Gamma)$, then the statement follows
from Theorem~\ref{thm:reducts}: this is clear for the first four cases of the theorem; in the last case, $\Gamma$ has all unary injections among its endomorphisms, and in particular the functions $e_E$ and $e_N$.
\end{proof}

%% file: binary0.tex
\section{Higher arity polymorphisms}\label{sect:higher-arity}
In this section we will be concerned with reducts $\Gamma$ of $G$ where the endomorphisms
of $\Gamma$ are either the endomorphisms of $(V;E,N)$, or precisely the functions generated by $\{-\}$, by $\{\sw\}$, or by $\{-,\sw\}$, since for all
other reducts $\Gamma$ of $G$ the complexity 
of $\Csp(\Gamma)$ has already been determined in Proposition~\ref{prop:endos}. 
We first introduce the general concepts which allow us to analyze polymorphisms of reducts of $G$ using Ramsey theory (Section~\ref{sect:canonical}). These concepts will be crucial in all four cases which we shall then approach in Sections~\ref{sect:c1} to~\ref{sect:c4}.

\subsection{Canonical Behavior}\label{sect:canonical}
It will turn out that the relevant polymorphisms have, in a certain sense, regular behavior with respect to the structure of $G$; combinatorially, this is due to the fact that the set of finite ordered graphs is a \emph{Ramsey class}, and that one can find regular patterns in any arbitrary function on the random graph. We make this idea more precise.

\begin{definition}
    Let $\Delta$ be a structure. The \emph{type} $\tp(a)$ of an $n$-tuple $a$ of elements in $\Delta$ is the set of first-order formulas with
     free variables $x_1,\dots,x_n$ that hold for $a$ in $\Delta$. For structures $\Delta_1,\ldots,\Delta_k$ and tuples $a^1,\ldots,a^n\in\Delta_1\times\cdots\times\Delta_k$, the type of $(a^1,\ldots,a^n)$ in $\Delta_1\times\cdots\times\Delta_k$, denoted by $\tp(a^1,\ldots,a^n)$, is the $k$-tuple containing the types of $(a^1_i,\ldots,a^n_i)$ in $\Delta_i$ for each $1\leq i\leq k$.
\end{definition}

We bring to the reader's attention the well-known fact that in homogeneous structures, in particular in the random graph, two $n$-tuples have the same type if and only if their orbits coincide.

\begin{definition}
    Let $k\geq 1$ and let $\Delta_1,\ldots,\Delta_k, \Lambda$ be structures. A \emph{type condition} between $\Delta_1\times\cdots\times\Delta_k$ and $\Lambda$ is a pair $(t,s)$, where $t$ is a type of an $n$-tuple in $\Delta_1\times\cdots\times\Delta_k$, and $s$ is a type of an $n$-tuple in $\Lambda$, for some $n\geq 1$. 
      A function $f \colon\Delta_1\times\cdots\times\Delta_k\To \Lambda$ \emph{satisfies} a type condition $(t,s)$ between $\Delta_1\times\cdots\times\Delta_k$ and $\Lambda$ if for all tuples $a^1,\ldots, a^n\in \Delta_1\times\cdots\times\Delta_k$ with $\tp(a^1,\ldots,a^n)=t$ the $n$-tuple $(f(a^1),\ldots,f(a^n))=(f(a^1_1,\ldots,a^1_k),\ldots,f(a^n_1,\ldots,a^n_k))$ has type $s$ in $\Lambda$. A \emph{behavior} is a set of type conditions between a product of structures $\Delta_1\times\cdots\times\Delta_k$ and a structure $\Lambda$. A function from $\Delta_1\times\cdots\times\Delta_k$ to $\Lambda$ \emph{has behavior $B$} if it satisfies all the type conditions of $B$.
\end{definition}
\ignore{
\begin{definition}
    A \emph{type condition} between two structures $\Delta_1,\Delta_2$ is a pair $(t_1,t_2)$, where each $t_i$ is a type of an $n$-tuple in $\Delta_i$. A function $f \colon\Delta_1\To \Delta_2$ \emph{satisfies} a type condition $(t_1,t_2)$ if for all $n$-tuples $(a_1,\ldots,a_n)$ of type $t_1$, the $n$-tuple $(f(a_1),\ldots,f(a_n))$ is of type $t_2$. A \emph{behavior} is a set of type conditions between two structures. A function \emph{has behavior $B$} if it satisfies all the type conditions of the behavior $B$.
\end{definition}
}
\begin{definition}
    Let $\Delta_1,\ldots,\Delta_k, \Lambda$ be structures. An operation $f \colon \Delta_1\times\cdots\times\Delta_k \To \Lambda$ is \emph{canonical} if for all $n\geq 1$ and all types $t$ of $n$-tuples in $\Delta_1\times\cdots\times\Delta_k$ there exists a type $s$ of an $n$-tuple in $\Lambda$ such that $f$ satisfies the type condition $(t,s)$. In other words, $n$-tuples of equal type in $\Delta_1\times\cdots\times\Delta_k$ are sent to $n$-tuples of equal type in $\Lambda$ under $f$, for all $n\geq 1$.
\end{definition}

We remark that since $G$ is homogeneous and has only binary relations, the type of an $n$-tuple $a$ in $G$ is determined by its binary subtypes, i.e., the types of the pairs $(a_i, a_j)$, where $1\leq i,j\leq n$. In other words, the type of $a$ is determined by which of its components  are equal, and between which of its components there is an edge.
Therefore, a function $f \colon G^k\To G$ is canonical iff it satisfies the condition of the definition for types of 2-tuples. 

The polymorphisms proving tractability of reducts of $G$ will be canonical. We now define some behaviors that some of these canonical functions will have. For $m$-ary relations
$R_1,\dots,R_k$ over $V$,
we will in the following write $R_1\cdots R_k$ for the $m$-ary relation on $V^k$ that holds between $k$-tuples $x^1,\dots,x^m \in V^k$ iff $R_i(x^1_i,\dots,x^m_i)$ holds for all $1\leq i \leq k$. We start with behaviors of binary functions.

\begin{definition}
    We say that a binary injective operation $f \colon V^2\To V$ is
    \begin{itemize}
        \item \emph{balanced in the first argument} if for all $u,v\in V^2$ we have that $\EEQ(u,v)$ implies $E(f(u),f(v))$ and $\NEQ(u,v)$ implies $N(f(u),f(v))$;         
        \item \emph{balanced in the second argument} if $(x,y) \mapsto f(y,x)$ is balanced in the first argument;
        \item \emph{balanced} if $f$ is balanced in both arguments, and
        \emph{unbalanced} otherwise;
        \item \emph{$E$-dominated ($N$-dominated) in the first argument} if for all $u,v \in V^2$ with $\NEQEQ(u,v)$
        we have that $E(f(u),f(v))$ ($N(f(u),f(v))$);
        \item \emph{$E$-dominated ($N$-dominated) in the second argument} if
        $(x,y) \mapsto f(y,x)$ is $E$-dominated ($N$-dominated) in the first argument;
        \item \emph{$E$-dominated ($N$-dominated)} if it is $E$-dominated ($N$-dominated) in both arguments;
        \item \emph{of type $\mini$} if for all $u,v\in V^2$ with $\NEQNEQ(u,v)$ we have
        $E(f(u),f(v))$ if and only if $\EE(u,v)$;
        \item \emph{of type $\maxi$} if for all $u,v\in V^2$ with $\NEQNEQ(u,v)$ we have
        $N(f(u),f(v))$ if and only if $\NN(u,v)$;
        \item \emph{of type $p_1$} if for all $u,v \in V^2$ with $\NEQNEQ(u,v)$ we have
        $E(f(u),f(v))$ if and only if $E(u_1,v_1)$;
        \item \emph{of type $p_2$} if $(x,y) \mapsto f(y,x)$ is of type $p_1$;
        \item \emph{of type projection} if it is of type $p_1$ or $p_2$.
    \end{itemize}
\end{definition}

    It is easy to see that each of those properties describes the set of all functions of a certain behavior. We explain this for the first item defining functions which are balanced in the first argument, which can be expressed  by the following two type conditions. Let $t$ be the type of any $u,v\in V^2$ with $\EEQ(u,v)$, and let $s$ be the type of any $x,y\in V$ with $E(x,y)$. Then the first type condition is $(t,s)$. Now 
   let $t'$ be the type of any $u,v\in V^2$ with $\NEQ(u,v)$,
 and let $s'$ be the type of any $x,y \in V$ with $N(x,y)$.  The second type condition is $(t',s')$.

Note that a binary injection of type $\maxi$ is reminiscent of the
Boolean maximum function on $\{0,1\}$, 
where $E$ takes the role of $1$
and $N$ the role of $0$: for $u,v\in V^2$ with $\NEQNEQ(u,v)$, we have
$E(f(u),f(v))$ if $u,v$ are connected by an edge in at least one coordinate, and $N(f(u),f(v))$ otherwise. The names ``$\mini$'' and ``projection'' can be explained similarly.

Also note that, for example, being of type $\maxi$ is a behavior of binary functions that does not force a function to be canonical, since the condition only talks about certain types of pairs in $G^2$, but not all such types:
for example, it does not tell us whether or not
$E(f(u),f(v))$ for $u,v \in V^2$ with $u_1=v_1$. 
However, being both of type $\maxi$ (or of type $\mini$) and balanced does mean that a function is canonical. 

The next definition contains some important behaviors of ternary functions.

\begin{definition}
    An injective ternary function $f \colon V^3\To V$ is of type
     \begin{itemize}
        \item \emph{majority} if for all $u,v\in V^3$ with ${\neq}{\neq}{\neq}(u,v)$ we have that $E(f(u),f(v))$ if and only if $\EEE(u,v)$, $\EEN(u,v)$, $\ENE(u,v)$, or $\NEE(u,v)$;

        \item \emph{minority} if for all $u,v\in V^3$ with ${\neq}{\neq}{\neq}(u,v)$ we have $E(f(u),f(v))$ if and only if $\EEE(u,v)$, $\NNE(u,v)$, $\NEN(u,v)$, or $\ENN(u,v)$.
     \end{itemize}
\end{definition}

%% file: binary1.tex
\subsection{When the endomorphisms of a reduct are generated by $\Aut(G)$} \label{sect:c1}
\label{sect:higherArity}
We investigate Case~(f) of Proposition~\ref{prop:endos}. In this situation, the following lemma states that we may assume that the reduct contains the relations $E$ and $N$.

\begin{lemma}\label{lem:ENdef}
Let $\Gamma$ be a reduct of $G$. Then the endomorphisms of $\Gamma$ are generated by $\Aut(G)$ if and only if the relations $E$ and $N$ are primitive positive definable in $\Gamma$.
\end{lemma}
\begin{proof}
If these relations are primitive positive definable in $\Gamma$, then they are preserved by all endomorphisms of $\Gamma$ by Theorem~\ref{conf:thm:inv-pol}. Hence, the restriction of any endomorphism to a finite set is a partial isomorphism of $G$, and thus extends to an automorphism of $G$ by homogeneity. It follows that any endomorphism can be interpolated by an element of $\Aut(G)$ on any finite set, and hence it is generated by $\Aut(G)$.

If the endomorphisms of $\Gamma$ are generated by $\Aut(G)$, then $E$ and $N$ are primitive positive definable in $\Gamma$ by Theorem~\ref{conf:thm:inv-pol} and Lemma~\ref{lem:arity-reduction}.
\end{proof}

The following relation characterizes the NP-complete cases in the situation of this section.

\begin{definition}\label{defn:H}
We define a 6-ary relation 
$H_1(x_1,y_1,x_2,y_2,x_3,y_3)$ on $V$ by
\begin{align}
& \bigwedge_{i,j \in \{1,2,3\}, i \neq j, u \in \{x_i,y_i\}, v \in \{x_j,y_j\}} N(u,v) \nonumber \\
\wedge & \; \big((E(x_1,y_1) \wedge N(x_2,y_2) \wedge N(x_3,y_3))
\nonumber \\ 
& \vee \; (N(x_1,y_1) \wedge E(x_2,y_2) \wedge N(x_3,y_3)) \nonumber\\  
& \vee \; (N(x_1,y_1) \wedge N(x_2,y_2) \wedge E(x_3,y_3)) \big)\; . \nonumber
\end{align}
\end{definition}

Our goal for Section~\ref{sect:c1} is to prove the following proposition, which states that if $\Gamma = (V; E,N, \dots)$ is a reduct of $G$, then either $H_1$ has a primitive positive definition in $\Gamma$, and  $\Csp(\Gamma)$ is NP-complete, or $\Gamma$ has a canonical polymorphism with a certain behavior. Each of the listed canonical polymorphisms implies tractability for $\Csp(\Gamma)$, and we will present algorithms proving this in Section~\ref{sect:algorithms}.

\begin{theorem}\label{thm:higherArity}
    Let $\Gamma$ be a reduct of $G$ whose endomorphisms are generated by $\Aut(G)$. Then at least one of the following holds:
    \begin{enumerate}
\item[(a)] There is a primitive positive definition of $H_1$ in $\Gamma$.
\item[(b)] $\Pol(\Gamma)$ contains a canonical ternary injection of type minority, as well as a canonical binary injection which is of type $p_1$ and either $E$-dominated or $N$-dominated in the second argument.
\item[(c)] $\Pol(\Gamma)$ contains a canonical ternary injection of type majority, as well as a canonical binary injection which is of type $p_1$ and either $E$-dominated or $N$-dominated in the second argument.
\item[(d)] $\Pol(\Gamma)$ contains a canonical ternary injection of type minority, as well as a canonical binary injection which is balanced and of type projection.
\item[(e)] $\Pol(\Gamma)$ contains a canonical ternary injection of type majority, as well as a canonical binary injection which is balanced and of type projection.
\item[(f)] $\Pol(\Gamma)$ contains a canonical binary injection of type $\maxi$ or $\mini$.
\end{enumerate}
\end{theorem}

The remainder of this section contains the proof of Theorem~\ref{thm:higherArity}, and is organized as follows: we first show that the relation $H_1$ is hard. We then prove that if $H_1$ does not have a primitive positive definition in a reduct $\Gamma$ as in Theorem~\ref{thm:higherArity}, then $\Gamma$ has the polymorphisms of one of the Cases~(b) to~(f) of the theorem.

\subsubsection{Hardness of $H_1$}\label{subsect:hardnessOfH}

We present the hardness proof of the relation in Case~(a) of Theorem~\ref{thm:higherArity}.

\begin{proposition}\label{prop:h-hard}
    $\Csp(V; H_1)$ is NP-hard.
\end{proposition}
\begin{proof}
    The proof is a reduction from positive 1-in-3-3SAT (one of the hard
    problems in Schaefer's classification; also see~\cite{GareyJohnson}).
    Let $\Phi$ be an instance of positive 1-in-3-3SAT, that is, a
    set of clauses, each having three positive literals.
    We create from $\Phi$ an instance $\Psi$ of $\Csp(V;H_1)$ as follows.
    For each variable $x$ in $\Phi$ we have a pair $u_x,v_x$ of
    variables in $\Psi$. When $\{x,y,z\}$ is a clause in $\Phi$, then
    we add the conjunct $H(u_x,v_x,u_y,v_y,u_z,v_z)$ to $\Psi$. Finally, we existentially quantify all variables of the conjunction in order to obtain a sentence.
    Clearly, $\Psi$ can be computed from $\Phi$ in linear time.

    Suppose now that $\Phi$ is satisfiable, i.e., there exists a mapping $s$ from the variables of $\Phi$ to $\{0,1\}$ such that in each clause exactly one of the literals is set to $1$; we claim that $(V;H_1)$ satisfies $\Psi$. To show this, let $F$ be the graph
    whose vertices are the variables of $\Psi$, and that has an
    edge between $u_x$ and $v_x$ if $x$ is set to 1 under the mapping $s$, and that has no other edges. By universality of $G$ we may assume that $F$ is a subgraph of $G$. It is then enough to show that $F$ satisfies the conjunction of $\Psi$ in order to show that $(V;H_1)$ satisfies $\Psi$.
    Indeed, let $H(u_x,v_x,u_y,v_y,u_z,v_z)$ be a clause from $\Psi$. By definition of $F$, the conjunction in the first line of the definition of $H_1$ is clearly
    satisfied; moreover, from the disjunction in the remaining lines
    of the definition of $H_1$ exactly one disjunct will be true,
    since in the corresponding clause $\{x,y,z\}$ of $\Phi$ exactly
    one of the values $s(x),s(y),s(z)$ equals $1$.
    This argument can easily be inverted to see that every
    solution to $\Psi$ can be used to define a solution to $\Phi$ (in which for a variable $x$ of $\Phi$ one sets $s(x)$ to $1$ iff in the solution to $\Psi$ there is an edge between $u_x$ and $v_x$).
\end{proof}

\subsubsection{Producing canonical functions}\label{subsect:producingCanonical}

We now show that if $\Gamma=(V;E,N,\ldots)$ is a reduct of $G$ such that there is no primitive positive definition of $H_1$ in $\Gamma$, then
 one of the other cases of Theorem~\ref{thm:higherArity} applies.  By Theorem~\ref{conf:thm:inv-pol}, $\Gamma$ has a polymorphism that violates $H_1$. 
 
 \begin{definition}
    A function $f \colon V^n\To V$ is called \emph{essentially unary} if it depends on only one of its variables; otherwise, it is called \emph{essential}.
\end{definition}

Note that any essentially unary function preserving both
$E$ and $N$ preserves all relations with a first-order definition in
$G$, and in particular $H_1$; this is because by Lemma~\ref{lem:ENdef} any such operation is generated by the automorphisms of $G$, which have this property (cf.~\cite{Hodges}).  Therefore we have that if a polymorphism
$f$ of $\Gamma$ violates $H_1$, then it must be essential.
Thus the following theorem from~\cite{RandomMinOps} applies. Before stating it, it is convenient to define the dual of an operation $f$ on $G$, which can be imagined as the function obtained from $f$ by exchanging the roles of $E$ and $N$.

\begin{definition}\label{defn:dual}
    The \emph{dual} of a function $f(x_1,\ldots,x_n)$ on $G$ is the function $-f(-x_1,\ldots,-x_n)$.
\end{definition}

\begin{theorem}[from~\cite{RandomMinOps}]\label{thm:minimal-ops}
    Let $f$ be an essential operation on $G$ preserving $E$ and $N$. Then it generates one of the following binary functions.
    \begin{itemize}
    \item a canonical injection of type $p_1$ which is balanced;
    \item a canonical injection  of type $\maxi$ which is balanced;
    \item a canonical injection of type $p_1$ which is $E$-dominated;
    \item a canonical injection of type $\maxi$ which is $E$-dominated;
    \item a canonical injection of type $p_1$ which is balanced in the first and
    $E$-dominated in the second argument;
    \end{itemize}
    or the dual of a function of the last four classes (the first class is self-dual).
\end{theorem}

 It follows from Theorem~\ref{thm:minimal-ops} that indeed, if $H_1$ does not have a primitive positive definition in  a reduct $\Gamma=(V;E,N,\ldots)$, then $\Gamma$ has one of the binary canonical polymorphisms mentioned in Theorem~\ref{thm:higherArity}. In order to complete the proof of Theorem~\ref{thm:higherArity}, we have to additionally show that when $f$ does not generate a binary injection of type $\mini$ or $\maxi$, it generates a ternary canonical injection of type minority or majority. That is, we have to prove the following.
 
 \begin{proposition}\label{prop:maxMinMajorityMonority}
    Suppose that $f$ is an operation on $G$ that preserves the relations $E$ and $N$ and violates the relation $H_1$.
    Then $f$ generates a binary canonical injection of type $\mini$ or $\maxi$, or a ternary canonical injection of type minority or majority.
\end{proposition}

The remainder of Section~\ref{sect:c1} will be devoted to the proof of this proposition. This will be achieved by refining the Ramsey-theoretic methods developed in~\cite{RandomMinOps} which are suitable for
investigating functions on $G$ in several variables.

In our proof of 
Proposition~\ref{prop:maxMinMajorityMonority}, we
really would like to take one of the ``nice'' functions $g$ which we know is generated by $f$ of Theorem~\ref{thm:minimal-ops}, and then
show that $g$ generates one of the functions of
Proposition~\ref{prop:maxMinMajorityMonority}. However, the problem
with this are the canonical binary injections of type $p_1$, since
functions of type $p_1$ do not violate $H_1$ anymore. 
Hence, when simply
passing to a function of the theorem, 
we lose the information that our
$f$ violates $H_1$, 
which we must use at some point, since $H_1$ is a hard
relation.
We are thus
obliged to improve Theorem~\ref{thm:minimal-ops} for functions violating $H_1$.
Before that, let us observe that
Theorem~\ref{thm:minimal-ops} implies that we can restrict our attention to binary and ternary injections.

\begin{lemma}\label{lem:Ternary}
 Let $f$ be an operation on $G$ which preserves $E$ and $N$ and violates $H_1$. Then $f$ generates a ternary injection which shares the same properties.
\end{lemma}
\begin{proof}
Since the relation $H_1$ consists of three orbits of 6-tuples with respect to $G$, Lemma~\ref{lem:arity-reduction} implies that $f$ generates an at most ternary function that violates $H_1$,
    and hence we can assume that $f$ itself is at most ternary; by adding a dummy variable if necessary, we may assume that $f$ is actually ternary. Moreover, 
    $f$ must certainly be essential, since essentially unary operations that
    preserve $E$ and $N$ also preserve $H_1$. Applying Theorem~\ref{thm:minimal-ops}, we get that $f$ generates a binary canonical injection $g$ of type $\mini$, $\maxi$, or $p_1$. In the first two cases the function $h(x,y,z):=g(x,g(y,z))$ is a ternary injection which violates $H_1$. So consider the last case where $g$ is of type $p_1$, and set $$h(x,y,z):= g(g(g(f(x,y,z),x),y),z)\; .$$ 
Then $h$ is clearly injective, and still violates $H_1$ -- the latter can easily be verified combining the facts that $f$ violates $H_1$, $g$ is of type $p_1$, and all tuples in $H_1$ have pairwise distinct entries.
\end{proof}

It will turn out that just as in the proof of Lemma~\ref{lem:Ternary}, there are two cases for $f$ in the proof of Proposition~\ref{prop:maxMinMajorityMonority}: either all binary canonical injections generated by $f$ are of type projection, and $f$ generates a ternary canonical injection of type majority or minority, or $f$ generates a binary canonical injection which is not of type projection, in which case it even generates a binary canonical injection of type $\mini$ or $\maxi$. We start by considering the first case, which is combinatorially less involved.

\subsubsection{Producing majorities and minorities}

\begin{definition}\label{defn:arbitrarilyLarge}
 Let $\Delta_1,\ldots,\Delta_k$ and $\Lambda$ be structures, $f \colon\Delta_1\times\cdots\times\Delta_k\To\Lambda$ be a function, and let $(t,s)$ be a type condition for such functions. If $S$ is a subset of $\Delta_1\times\cdots\times\Delta_k$, then we say that $f$ \emph{satisfies the type condition $(t,s)$ on $S$} if for all tuples $a^1,\ldots, a^n\in S$ with $\tp(a^1,\ldots,a^n)=t$ in $\Delta_1\times\cdots\times\Delta_k$ the $n$-tuple $(f(a^1_1,\ldots,a^1_k),\ldots,f(a^n_1,\ldots,a^n_k))$ has type $s$ in $\Lambda$. We say that $f$ \emph{satisfies a behavior $B$ on $S$} if it satisfies all type conditions of $B$ on $S$.
 
Finally, we say that $f$ satisfies $B$ \emph{on arbitrarily large (finite) substructures} of $\Delta_1\times\cdots\times\Delta_k$ if for all finite substructures $F_i$ of $\Delta_i$, where $1\leq i\leq k$, there exist isomorphic copies $F_i'$ of $F_i$ in $\Delta_i$ such that $f$ satisfies $B$ on the product $F_1'\times\cdots\times F_k'$ of these copies.
 \end{definition}

In the following general proposition we exceptionally use the notion ``locally generates'' in its original sense (see Section~\ref{sect:toolsua}). The proof is a standard compactness argument, which we include nonetheless for the convenience of the reader. Similar proofs can be found, for example, in~\cite{BodPinTsa} for arbitrary homogeneous structures in a finite language, or for the random graph in~\cite{RandomMinOps}.

\begin{proposition}\label{prop:RamseyCore}
    Let $\Delta_1,\ldots,\Delta_k$ and $\Lambda$ be homogeneous structures on the same countably infinite domain $D$, and assume that 
     $\Lambda$ has a finite language.
     Let moreover    
    $B$ be a behavior for functions from $\Delta_1\times\cdots\times\Delta_k$ to $\Lambda$, and let $f \colon D^k\To D$ be a function which satisfies $B$ on arbitrarily large substructures of $\Delta_1\times\cdots\times\Delta_k$. Then $\{f\}\cup\Aut(\Lambda)\cup\Aut(\Delta_1)\cup\cdots \cup\Aut(\Delta_k)$  locally generates a function from $D^k$ to $D$ which satisfies $B$ everywhere.
\end{proposition}
\begin{proof}
 Write $D=\{d_0,d_1,\ldots\}$. We construct a sequence $(g_i)_{i\in\omega}$ such that for all $i\in\omega$
 \begin{itemize}
 \item[(i)] $g_i$ is a function from $D^k$ to $D$ locally generated by $\{f\}\cup\Aut(\Lambda)\cup\Aut(\Delta_1)\cup\cdots \cup\Aut(\Delta_k)$; 
 \item[(ii)] $g_i$ satisfies $B$ on $\{d_0,\ldots,d_i\}^k$;
 \item[(iii)] $g_{i+1}$ agrees with $g_i$ on $\{d_0,\ldots,d_i\}^k$.
 \end{itemize}
 
 The sequence then defines a function $g:D^k\To D$ by setting $g(d_{i_1},\ldots,d_{i_k}):=g_m(d_{i_1},\ldots,d_{i_k})$, for any $m\geq i_1,\ldots,i_k$. This function $g$ is clearly locally generated by $\{g_i: i\in\omega\}$ by local closure, and satisfies $B$ everywhere.
 
 To construct the sequence, we first construct a sequence $(h_i)_{i\in\omega}$ which only satisfies~(i) and~(ii) of the requirements for the sequence $(g_i)_{i\in\omega}$.  Let $i\in\omega$ be given. There exist subsets $F_1,\ldots, F_k$ of $D$ such that $F_j$ is isomorphic with $\{d_0,\ldots,d_i\}$ as substructures of $\Delta_j$ for all $1\leq j\leq k$ and such that $f$ satisfies $B$ on $F_1\times\cdots\times F_k$. Let $\alpha_j$ be an automorphism of $\Delta_j$ sending $\{d_0,\ldots,d_i\}$ onto $F_j$, for all $1\leq j\leq k$; these automorphisms exist by the homogeneity of the $\Delta_j$. Then we can set $h_i(x_1,\ldots,x_k):=f(\alpha_1(x_1),\ldots,\alpha_k(x_k))$.
 
Now to obtain the sequence $(g_i)_{i\in\omega}$ from the sequence $(h_i)_{i\in\omega}$, let $a=(a_0,a_1,\ldots)$ be an enumeration of $D^k$ such that the elements of $\{d_0,\ldots,d_i\}^k$ are an initial segment of this enumeration for each $i\in\omega$ (that is, they constitute the first $(i+1)^k$ entries). Denote for all $i,j\in\omega$ by $b^{i,j}$ the $(i+1)^k$-tuple which is obtained by applying $h_j$ to each of the first $(i+1)^k$ entries of the enumeration $a$. Set $t^{i,j}$ to be the type of $b^{i,j}$ in $\Lambda$. For $i,j,r,s\in\omega$ set $t^{i,j}\leq t^{r,s}$ if $i\leq r$ and $t^{i,j}, t^{r,s}$ agree on the variables they have in common, i.e., the restriction of $b^{r,s}$ to its initial segment of length $(i+1)^k$ has the same type as $b^{i,j}$ in $\Lambda$. This relation defines a tree on the types $t^{i,j}$. Since $\Lambda$ is homogeneous in a finite language, for every $i\in\omega$ there are only finitely many different types of $(i+1)^k$-tuples in $\Lambda$. Hence, for every $i\in\omega$, there are only finitely many distinct types $t^{i,j}$, 
and so this tree is finitely branching.
Moreover, there exists a $q\in\omega$ 
such that $t^{i,s}=t^{i,q}$ for infinitely many $s\in\omega$. Deleting all elements of the tree which do not enjoy this latter property, we are thus still left with an infinite tree. Hence by K\H{o}nig's lemma it has an infinite branch $(t^{0,j_0},t^{1,j_1},\ldots)$. Since we have reduced the tree to its ``infinite'' nodes, we may assume that the $j_i$ are strictly increasing, and in particular that $j_i\geq i$ for all $i\in\omega$. Since $\Lambda$ is homogeneous and by definition of the tree, we can pick for all $i\in\omega$ an automorphism $\alpha_i$ of $\Lambda$ which sends the initial segment of length $(i+1)^k$ of $b^{i+1,j_{i+1}}$ to $b^{i,j_i}$. Then setting $g_i:=h_{j_i}\circ \alpha_{i-1}\circ\cdots\circ \alpha_0$ for all $i\in\omega$ yields the desired sequence: (i) is obvious. (ii) holds since $h_i$ satisfies (ii), $h_{j_i}$ still satisfies (ii) since $j_i\geq i$, and $g_i$ satisfies (ii) since the property is preserved under applications of automorphisms of $\Lambda$. (iii) is by construction.
 \end{proof}

\begin{proposition}\label{prop:generatesMajority}
	Let $f$ be an operation on $G$ that preserves $E$ and $N$ and violates $H_1$. Suppose moreover that all binary injections  generated by $f$ are of type projection. Then $f$ generates a ternary canonical injection of type majority or minority.
\end{proposition}

\begin{proof}
By Lemma~\ref{lem:Ternary}, we can assume that $f$ is  a ternary injection. 
Because $f$ violates $H_1$, there are $x^1,x^2,x^3 \in H_1$ such that $f(x^1,x^2,x^3) \notin H_1$.
In the following, we will write $x_i := (x^1_i,x^2_i,x^3_i)$ for $1\leq i\leq 6$. So $(f(x_1),\dots,f(x_6)) \notin H_1$.

If there were an automorphism $\alpha$ of $G$ such that $\alpha(x^i) = x^j$ for $1\leq i \neq j \leq 3$,
then $f$ would generate a binary injection that still violates $H_1$, which contradicts the 
assumption that all binary injections generated by $f$ are of type projection. 
By permuting arguments of $f$ if necessary, we can therefore 
assume without loss of generality that 
\begin{align*}
	\ENN(x_1,x_2),\, \NEN(x_3,x_4),\,\text{and } \NNE(x_5,x_6).
\end{align*}

We set $$S := \{ y \in V^3 \; | \; \NNN(x_i,y) \text{ for all } 1\leq i \leq 6 \} \; .$$
Consider the binary relations $Q_1Q_2Q_3$ on $V^3$, where $Q_i\in\{E,N\}$ for $1\leq i\leq 3$.
We claim that for each such relation $Q_1Q_2Q_3$, whether $E(f(u),f(v))$ or $N(f(u),f(v))$ holds for $u,v \in S$ with $Q_1Q_2Q_3(u,v)$ does not depend on $u,v$; that is, whenever $u,v,u',v'\in S$ satisfy $Q_1Q_2Q_3(u,v)$ and $Q_1Q_2Q_3(u',v')$, then $E(f(u),f(v))$ if and only if $E(f(u'),f(v'))$. We go through all possibilities of $Q_1Q_2Q_3$.
\begin{enumerate}
\item $Q_1Q_2Q_3=\ENN$. Let $\alpha \in \Aut(G)$ be such that $(x^2_1,x^2_2,u_2,v_2)$ is mapped
to $(x^3_1,x^3_2,u_3,v_3)$; such an automorphism exists since $\NNN(x_1, u), \NNN(x_1, v),
\NNN(x_2, u), \NNN(x_2, v)$, and since $(x^2_1,x^2_2)$ has the same type as
$(x^3_1,x^3_2)$, and $(u_2,v_2)$ has the same type as $(u_3,v_3)$.
By assumption, the operation $g$ defined by $g(x,y):=f(x,y,\alpha(y))$ must
be of type projection. Hence, $E(g(u_1,u_2),g(v_1,v_2))$ iff $E(g(x_1^1,x_1^2),g(x_2^1,x_2^2))$. Combining this with the equations $(f(u),f(v))=(g(u_1,u_2),g(v_1,v_2))$ and 
$(g(x_1^1,x_1^2),g(x_2^1,x_2^2))=(f(x_1),f(x_2))$, we get that $E(f(u),f(v))$ iff $E(f(x_1),f(x_2))$, and so we are done.
\item $Q_1Q_2Q_3=\NEN$ or $Q_1Q_2Q_3=\NNE$. These cases are analogous to the previous case.
\item $Q_1Q_2Q_3=\NEE$. Let $\alpha$ be defined as in the first case. 
By assumption, the operation defined by $f(x,y,\alpha(y))$ must
be of type projection. Reasoning as above, one gets that $E(f(u),f(v))$ iff $N(f(x_1),f(x_2))$.
\item $Q_1Q_2Q_3=\ENE$ or $Q_1Q_2Q_3=\EEN$. These cases are analogous to the previous case.
\item $Q_1Q_2Q_3= \EEE$ or $Q_1Q_2Q_3=\NNN$. These cases are trivial since $f$ preserves $E$ and $N$.
\end{enumerate}

To show that $f$ generates an operation of type majority or minority, by Proposition~\ref{prop:RamseyCore} 
it suffices to prove that $f$ generates a function of type majority or minority on $S$, since
$S$ contains copies of products of arbitrary finite substructures of $G$. We show this by another case distinction, based on the fact that $(f(x_1),\dots,f(x_6)) \notin H_1$. 
\begin{enumerate}
\item Suppose that $E(f(x_1),f(x_2)), E(f(x_3),f(x_4)),E(f(x_5),f(x_6))$. 
Then by the above, $f$ itself is of type minority on $S$.
\item Suppose that $N(f(x_1),f(x_2)),N(f(x_3),f(x_4)),N(f(x_5),f(x_6))$. 
Then $f$ behaves like a majority on $S$.
\item Suppose that $E(f(x_1),f(x_2)),E(f(x_3),f(x_4)),N(f(x_5),f(x_6))$. 
Let $e$ be a self-embedding of $G$ such that for all $w \in V$, all $1\leq j\leq 3$, and all $1\leq i \leq 6$ 
we have that $N(x_i^j,e(w))$. 
Then $(u_1,u_2,e(f(u_1,u_2,u_3))) \in S$ for all 
$(u_1,u_2,u_3) \in S$. 
Hence, by the above, 
the ternary operation defined by $f(x,y,e(f(x,y,z)))$ is of type majority on $S$.
\item Suppose that $E(f(x_1),f(x_2)),N(f(x_3),f(x_4)),E(f(x_5),f(x_6))$, or 
$N(f(x_1),f(x_2))$, $E(f(x_3),f(x_4))$, $E(f(x_5),f(x_6))$. 
These cases are analogous to the previous case.
\end{enumerate}

Let $h(x,y,z)$ be a ternary function of type majority or minority generated by $f$;  it remains to make $h$ canonical and injective. By Theorem~\ref{thm:minimal-ops}, $f$ generates a binary canonical injection $g(x,y)$, which is of type projection by our assumption on $f$. Set $t(x,y,z):=g(x,g(y,z))$. Then the function $h(t(x,y,z),t(y,z,x),t(z,x,y))$ is still of type majority or minority, and canonical and injective; we leave the straightforward verification to the reader.
\end{proof}

\subsubsection{Producing $\maxi$
and $\mini$}
Having proven Proposition~\ref{prop:generatesMajority}, it is enough to show the following proposition in order to obtain a full proof of Proposition~\ref{prop:maxMinMajorityMonority}.

\begin{proposition}\label{prop:nonProjGeneratesMin}
    Let $f \colon V^2 \rightarrow V$ be a binary injection preserving $E$ and $N$ that is not of type projection. Then $f$ generates a binary canonical injection of type $\mini$ or of type $\maxi$.
\end{proposition}

We will now prove this proposition by Ramsey theoretic analysis of $f$, which requires the following definitions and facts from~\cite{RandomMinOps}.

Equip $V$ with a total order $\prec$ in such a way that $(V;E,\prec)$ is the random ordered graph, i.e., the unique countably infinite homogeneous totally ordered graph containing all finite totally ordered graphs (for existence and uniqueness of this structure, see e.g. \cite{Hodges}). The order $(V;\prec)$ is then isomorphic to the order $(\mathbb{Q};<)$ of the rationals. The ordered random graph has the advantage of being a so-called \emph{Ramsey structure}, i.e., it enjoys a certain combinatorial property (which the random graph without the order does not) -- see for example~\cite{BP-reductsRamsey}. Using this Ramsey property, starting from a function on $(V;E,\prec)$ one can generate a canonical function whilst keeping such information as violation of a relation. Our combinatorial tool will be the following proposition, which has first been used in~\cite{RandomMinOps} in a slightly simpler form, and which has been stated in full generality for ordered homogeneous Ramsey structures in~\cite{BodPinTsa}.

\begin{proposition}\label{prop:orderedRandomCanonical}
	Let $f \colon V^k\To V$ be a function, and let $c^1,\ldots,c^m\in V^k$. Then $f$ generates a function which is canonical as a function from $(V; E,\prec,c^1_1, \ldots, c^m_1)\times\cdots\times (V; E,\prec,c^1_k, \ldots, c^m_k)$ to {$(V;E,\prec)$}, and which is identical with $f$ on $\{c_1^1,\ldots,c_1^m\}\times\cdots\times\{c_k^1,\ldots,c_k^m\}$. Moreover, if $f$ is injective, then the generated canonical function can be chosen to be injective as well.
\end{proposition}

The global strategy behind what follows now is to take a binary injection $f$ and fix a finite number of constants $c^i\in V^2$ which witness that $f$ is not of type projection. Then, using Proposition~\ref{prop:orderedRandomCanonical}, we generate a binary canonical function which is identical with $f$ on all $c^i$; this canonical function then still is not of type projection, and can be handled more easily as it is canonical. However, we do not present the proof like that for the reason that there would be too many possibilities of canonical functions for primitive case-by-case analysis. What we do instead is rule out behaviors of canonical functions more systematically, for example before even adding constants to the language. As in~\cite{RandomMinOps}, let us define the following behaviors for functions from $(V;E,\prec)^2$ to $(V;E)$.
We write $\succ$ for the relation $\{(a,b) \; | \; b \prec a\}$. 

\begin{definition}
    Let $f \colon V^2 \rightarrow V$ be injective. If for all $u,v\in V^2$ with $u_1\prec v_1$ and $u_2\prec v_2$ we have
    \begin{itemize}
        \item $E(f(u),f(v))$ if and only if $\EE(u,v)$, then we say that \emph{$f$ behaves like $\mini$ on input $(\prec,\prec)$}.
        \item $N(f(u),f(v))$ if and only if $\NN(u,v)$, then we say that \emph{$f$ behaves like $\maxi$ on input $(\prec,\prec)$}.
        \item $E(f(u),f(v))$ if and only if $E(u_1,v_1)$, then we say that \emph{$f$ behaves like $p_1$ on input $(\prec,\prec)$}.
        \item $E(f(u),f(v))$ if and only if $E(u_2,v_2)$, then we say that \emph{$f$ behaves like $p_2$ on input $(\prec,\prec)$}.
    \end{itemize}
    Analogously, we define behavior on input $(\prec,\succ)$ using pairs $u,v\in V^2$ with $u_1 \prec v_1$ and $u_2\succ v_2$.
\end{definition}

Of course, we could also have defined ``behavior on input $(\succ ,\succ)$'' and ``behavior on input $(\succ ,\prec )$''; however, behavior on input $(\succ ,\succ )$ equals behavior on input $(\prec ,\prec )$, and behavior on input $(\succ ,\prec )$ equals behavior on input 
$(\prec ,\succ)$ since graphs are symmetric.
Thus, there are only two kinds of inputs to be considered, namely the ``straight input" $(\prec,\prec)$ and the ``twisted input"  $(\prec ,\succ )$.

\begin{proposition}\label{prop:binaryBehaviourOnInputBLaBla}
Let $f \colon V^2 \rightarrow V$ be injective and canonical as a function from $(V;E,\prec)^2$ to $(V;E,\prec)$, and suppose it preserves $E$ and $N$. Then it behaves like $\mini$, $\maxi$, $p_1$ or $p_2$ on input $(\prec ,\prec )$ (and similarly on input $(\prec ,\succ)$).
\end{proposition}
\begin{proof}
    By definition of the term canonical; one only needs to enumerate all possible types of pairs $(u,v)$, where $u,v \in V^2$.
\end{proof}

\begin{definition}
    If an injection $f \colon V^2 \rightarrow V$ behaves like $X$ on input $(\prec ,\prec )$ and like $Y$ on input $(\prec ,\succ )$, where $X,Y\in\{\maxi,\mini, p_1,p_2\}$, then we say that $f$ is of \emph{type $X / Y$}.
\end{definition}

We would like to emphasize that the term ``canonical'' depends on the structures under consideration; that is, a function $f \colon V^2 \To V$ might be canonical as a function from $(V;E,\prec)^2$ to $(V;E,\prec)$, but not as a function from $(V;E)^2$ to $(V;E)$, and vice-versa. In the following, we will for this reason carefully specify the structures we have in mind when using this term.

Observe that canonical functions from $(V;E,\prec)^2$ to $(V;E,\prec)$ also behave regularly with respect to the order $\prec$: this implies, for example, that any such function which is injective is either strictly increasing or decreasing with respect to the pointwise order. 

The structures $(V;E,\prec)$ and $(V;E,\succ)$ are isomorphic by
the theory of homogeneous structures (see, e.g.,~\cite{Hodges}), since they
are both homogeneous and embed the same finite
structures. Fix
an isomorphism $\alpha$. Then $\alpha$ is an automorphism of $G$ which
reverses the order $\prec$. By applying $\alpha$ to a canonical function if necessary, we may (in the presence of $\Aut(G)$) always assume that all canonical functions $f$ we use are strictly increasing. Having that, one easily checks that one of the implications
$$
u_1\prec v_1\wedge u_2\neq v_2\rightarrow f(u)\prec f(v)
$$
and
$$
u_1\neq v_1\wedge u_2\prec v_2 \rightarrow f(u)\prec f(v).
$$
hold. In the first case, we say that \emph{$f$ obeys $p_1$ for the order}, in the second case \emph{$f$ obeys $p_2$ for the order}. By switching the variables of $f$, we can always achieve that $f$ obeys $p_1$ for the order.

\subsubsection{Eliminating mixed behavior} 
In the following lemmas, we
show that when we have an injective canonical binary function which
behaves differently on input $(\prec ,\prec )$ 
and on input $(\prec,\succ)$, 
then it generates a function 
which behaves the same on both inputs.

\begin{lemma}\label{lem:mixtyp:maxp}
    Suppose that $f \colon V^2 \rightarrow V$ is injective and canonical as a function from $(V;E,\prec)^2$ to $(V;E,\prec)$, and suppose that it is of type $\maxi / p_i$ or of type $p_i / \maxi$, where $i\in\{1,2\}$. Then $f$ generates a binary injection of type $\maxi$.
\end{lemma}
\begin{proof}
    Assume without loss of generality that $f$ is of type $\maxi / p_i$, and note that we may assume that $f$ obeys $p_1$ for the order. Set $h(x,y):=f(x,\alpha(y))$. Then $h$ behaves like $p_i$ on input
    $(\prec ,\prec )$ and like $\maxi$ on input $(\prec ,\succ )$; moreover, $h(u)\prec h(v)$ iff $f(u)\prec f(v)$, for all $u,v\in V^2$ with 
     $u_1\neq v_1$ and $u_2\neq v_2$. We then have that $g(x,y):=f(f(x,y),h(x,y))$ is of type $\maxi / \maxi$, which means that it is of type $\maxi$ when viewed as a function from $G^2$ to $G$.
\end{proof}

\begin{lemma}\label{lem:mixtyp:minp}
    Suppose that $f \colon V^2 \rightarrow V$ is injective and canonical as a function from $(V;E,\prec)^2$ to $(V;E,\prec)$, and suppose that it is of type $\mini / p_i$ or of type $p_i / \mini$, where $i\in\{1,2\}$. Then $f$ generates a binary injection of type $\mini$.
\end{lemma}
\begin{proof}
    The dual proof works.
\end{proof}

\begin{lemma}\label{lem:mixtyp:maxmin}
    Suppose that $f \colon V^2 \rightarrow V $ is injective and canonical as a function from $(V;E,\prec)^2$ to $(V;E,\prec)$, and suppose that it is of type $\maxi / \mini$ or of type $\mini / \maxi$. Then $f$ generates a binary injection of type $\maxi$ (and by duality, a binary injection of type $\mini$).
\end{lemma}
\begin{proof}
    Assume without loss of generality that $f$ is of type $\maxi / \mini$, and remember that we may assume that $f$ obeys $p_1$ for the order. Then $g(x,y):=f(x,f(x,y))$ is of type $\maxi / p_1$ and generates a binary injection of type $\maxi$ by Lemma~\ref{lem:mixtyp:maxp}.
\end{proof}

We next deal with the last remaining mixed behavior, $p_1/p_2$, by combining operational with relational arguments.

\begin{lemma}\label{lem:min-relationally}
Let $\Gamma=(V;E,N,\ldots)$ be a reduct of $G$ which is preserved by a binary injection of type $p_1$. 
Then the following are equivalent.
\begin{enumerate}
\item $\Gamma$ has a binary injective polymorphism of behavior $\mini$.
\item For every primitive positive formula $\phi$ over $\Gamma$, if 
$\phi \wedge N(x_1,x_2) \wedge \bigwedge_{1\leq i<j\leq 4} x_i \neq x_j$ and $\phi \wedge N(x_3,x_4) \wedge \bigwedge_{1\leq i<j\leq 4} x_i \neq x_j$ 
are satisfiable over $\Gamma$, then $\phi \wedge N(x_1,x_2) \wedge N(x_3,x_4)$ is satisfiable over $\Gamma$ as well. 
\item For every finite $F \subseteq V^2$ there exists a binary injective polymorphism of $\Gamma$ which behaves like $\mini$ on $F$.
\end{enumerate}
\end{lemma}

\begin{proof}
The implication from (1) to (2) follows directly by applying a binary injective polymorphism of behavior $\mini$ to tuples $r, s$ satisfying $\phi \wedge N(x_1,x_2) \wedge \bigwedge_{1\leq i<j\leq 4} x_i \neq x_j$ and $\phi \wedge N(x_3,x_4) \wedge \bigwedge_{1\leq i<j\leq 4} x_i \neq x_j$ respectively.

To prove that (2) implies (3), assume (2) and let $F \subseteq V^2$
be finite. Without loss of generality we can
assume that $F$ is of the form
$\{e_1,\dots,e_n\}^2$, for sufficiently large $n$. Let $\Delta$ be the structure induced by
$F$ in $\Gamma^2$. We construct an injective homomorphism $h$ from 
$\Delta$ to $\Gamma$ with the property that for all $u,v\in F$ with $\EN(u,v)$ or $\NE(u,v)$ we have $N(h(u),h(v))$. Any homomorphism from $\Delta$ to $\Gamma$, in particular $h$, can clearly be extended to a binary polymorphism of $\Gamma$, for example inductively by using the universality of $G$. Such an extension of $h$ then behaves like $\mini$ on $F$.

To construct $h$, 
consider the formula $\phi_0$ 
with variables $x_{i,j}$ for $1\leq i,j\leq n$ which is the conjunction over all
literals $R(x_{i_1,j_1},\dots,x_{i_k,j_k})$ such that $R$ is a relation in $\Gamma$ and 
$R(e_{i_1},\dots,e_{i_k})$ and $R(e_{j_1},\dots,e_{j_k})$ hold in $\Gamma$. So $\phi_0$ states precisely which relations hold in $\Gamma^2$ on elements from $F$.
Since $\Gamma$ is preserved by a binary injection, we have that
$\phi_1 := \phi_0 \wedge \bigwedge_{1\leq i,j,k,l \leq n, (i,j) \neq (k,l)} x_{i,j} \neq x_{k,l}$ 
is satisfiable in $\Gamma$.

Let $P$ be the set of pairs of the form $((i_1,i_2),(j_1,j_2))$
with $i_1,i_2,j_1,j_2 \in \{1,\dots, n\}$, $i_1 \neq j_1$, $i_2 \neq j_2$,
and where $N(e_{i_1},e_{j_1})$ or $N(e_{i_2},e_{j_2})$.
We show by induction on the size of $I \subseteq P$
that the formula $\phi_1 \wedge \bigwedge_{((i_1,i_2),(j_1,j_2)) \in I} N(x_{i_1,i_2},x_{j_1,j_2})$ is satisfiable over $\Gamma$. Note that this statement
applied to the set $I = P$ gives us the a
homomorphism $h$ from $\Delta$ to $\Gamma$ such that 
for all $a,b \in F$ we have $N(h(a),h(b))$ whenever
$\EN(a,b)$ or $\NE(a,b)$ by setting $h(e_i,e_j):= s(x_{i,j})$, where $s$ is the satisfying assignment for $\phi_1 \wedge \bigwedge_{((i_1,i_2),(j_1,j_2)) \in P} N(x_{i_1,i_2},x_{j_1,j_2})$.

For the induction beginning, let $p = ((i_1,i_2),(j_1,j_2))$ be any
element of $P$. 
 Let $r,s$ be the $n^2$-tuples defined as follows.
\begin{align*}
r & := (e_1,\dots,e_1,e_2,\dots,e_2,\dots,e_n,\dots,e_n) \\
s & := (e_1,e_2,\dots,e_n,e_1,e_2,\dots,e_n,\dots,e_1,e_2,\dots,e_n)
\end{align*}
In the following we use double indices for the entries of $n^2$-tuples;
for example, $r=(r_{1,1},\dots,r_{1,n},r_{2,1},\dots,r_{n,n})$. 
The two tuples $r$ and $s$ satisfy $\phi_0$. To see this observe that by definition of $\phi_0$ the tuple 
$$((e_1,e_1),(e_1,e_2),\dots,(e_1,e_n),(e_2,e_1),\ldots,(e_n,e_n))$$ satisfies $\phi_0$ in $\Gamma^2$; since $r$ and $s$ are obtained by applying projections to that tuple onto the first and second coordinate, respectively, and projections are homomorphisms, $r$ and $s$ satisfy  $\phi_0$ as well. Let $g$ be a binary injective polymorphism of $\Gamma$ which is of type $p_1$, and set $r' := g(r,s)$ and $s' := g(s,r)$. 
Then $r'$ and $s'$ satisfy $\phi_1$ since $g$ is injective. 
Since $p \in P$, 
we have that $N(e_{i_1},e_{j_1})$ or $N(e_{i_2},e_{j_2})$. 
Assume that $N(e_{i_1},e_{j_1})$;
the other case is analogous. Since $r_{i_1,i_2} = e_{i_1}, r_{j_1,j_2}=e_{j_1}$,
$r' := g(r,s)$, and $g$ is of type $p_1$,
we have that $N(r'_{i_1,i_2},r'_{j_1,j_2})$,
proving that $\phi_1 \wedge N(x_{i_1,i_2},x_{j_1,j_2})$ is satisfiable in $\Gamma$.

In the induction step, 
let $I \subseteq P$ be a set of cardinality $n \geq 2$, and assume that the
statement has been shown for subsets of $P$ of cardinality $n-1$. Pick any distinct $q_1,q_2\in I$.
Set $$\psi := \phi_1 \wedge \bigwedge_{((i_1,i_2),(j_1,j_2)) \in I \setminus \{q_1,q_2\}} N(x_{i_1,i_2},x_{j_1,j_2})$$
and observe that $\psi$ is a primitive positive formula over $\Gamma$ since $\Gamma$ contains $E$ and $N$ and since the binary relation $x \neq y$ can be defined in $\Gamma$ by the primitive positive formula $\exists z.\; (E(x,z)\wedge N(y,z))$.
Write $q_1 = ((u_1,u_2),(v_1,v_2))$ and 
$q_2 = ((u'_1,u'_2),(v'_1,v'_2))$. 
Then the inductive assumption shows that each of
$\psi \wedge N(x_{u_1,u_2},x_{v_1,v_2})$
and
$\psi \wedge N(x_{u'_1,u'_2},x_{v'_1,v'_2})$
is satisfiable in $\Gamma$. Note that
$\psi$ contains in particular conjuncts that state that
the four variables $x_{u_1,u_2},x_{v_1,v_2},x_{u'_1,u'_2},x_{v'_1,v'_2}$ denote 
distinct elements. Hence, by (2), the formula $\psi \wedge N(x_{u_1,u_2},x_{v_1,v_2}) \wedge N(x_{u'_1,u'_2},x_{v'_1,v'_2})$ is satisfiable over $\Gamma$ as well, which is what we had to show.

The implication from (3) to (1) follows from Proposition~\ref{prop:RamseyCore}.
\end{proof}

\begin{lemma}\label{lem:generatingMin:1}
Let $f \colon V^2 \rightarrow V$ be a binary injection of type $p_1/p_2$ which preserves $E$ and $N$. 
Then $f$ generates a binary injection of type $\mini$ and a binary injection of type $\maxi$.
\end{lemma}
\begin{proof}
By Theorem~\ref{thm:minimal-ops}, $f$ generates a binary injection of type $\maxi$, $\mini$, or $p_1$. 

Suppose first that it does not generate a binary injection of type $\maxi$ or $\mini$; we will lead this to a contradiction. Let $\Gamma$ be the reduct of $G$ which has all relations
that are first-order definable in $G$ and preserved by $f$. Since $f$ generates a binary injection of type $p_1$, we may apply implication (2) $\rightarrow$ (1) from
Lemma~\ref{lem:min-relationally}. Let $\phi$ be a primitive positive formula with variable set $S$, $\{x_1,\dots,x_4\} \subseteq S$,
such that the formulas 
$\phi \wedge N(x_1,x_2) \wedge \bigwedge_{1\leq i<j\leq 4} x_i \neq x_j$ 
and $\phi \wedge N(x_3,x_4) \wedge \bigwedge_{1\leq i<j\leq 4} x_i \neq x_j$ 
have in $\Gamma$ the satisfying assignments $r$ and $s$ from $S \rightarrow V$, respectively.

We can assume without loss of generality that $r(x_1) \prec r(x_2)$ and $r(x_3) \prec r(x_4)$;
otherwise, since $r(x_1),\dots,r(x_4)$ must be pairwise distinct, 
we can apply an automorphism of $G$ to $r$ such that the resulting map has the required
property.
Similarly, by applying an automorphism of $G$ to $s$,
we can assume without loss of generality that $s(x_1) \prec s(x_2)$ and $s(x_3) \succ s(x_4)$.
Then the mapping $t \colon  S \rightarrow V$ defined by $t(x) = f(r(x),s(x))$ 
shows that $\phi \wedge N(x_1,x_2) \wedge N(x_3,x_4)$ is satisfiable in $\Gamma$:
\begin{itemize}
\item The assignment $t$ satisfies $\phi$ since $f$ is a polymorphism of $\Gamma$.
\item We have that $N(t(x_1),t(x_2))$ since $r(x_1)\prec r(x_2)$, $s(x_1) \prec s(x_2)$,
 $f$ is of type $p_1$ on input $(\prec,\prec)$, and $N(r(x_1),r(x_2))$. 
\item We have that $N(t(x_3),t(x_4))$ since $r(x_3)\prec r(x_4)$, $s(x_3)\succ s(x_4)$,
$f$ is of type $p_2$ on input $(\prec,\succ)$, and $N(s(x_3),s(x_4))$. 
\end{itemize}
By Lemma~\ref{lem:min-relationally}, we conclude that $\Gamma$ is preserved by a binary injection of type $\mini$, and consequently $f$ generates a binary injection of type $\mini$ -- a contradiction.

Therefore, $f$ generates a binary injection of type $\maxi$ or $\mini$. Since the assumptions of the lemma are symmetric in $E$ and $N$, we infer \emph{a posteriori} that $f$ generates both a binary injection of type $\maxi$ and a binary injection of type $\mini$.
\end{proof}

\subsubsection{Behaviors relative to vertices}
Having ruled out some behaviors without constants, we now examine behaviors when we add constants to the language. 
In the sequel, we will also say that a function $f \colon V^2\To V$ has behavior $B$ \emph{between two points $u,v\in V^2$} if it has behavior $B$ on the structure induced by $\{u,v\}$.

\begin{lemma}\label{lem:generatingMin:2}
   Let $u \in V^2$, and set 
   $U := (V\setminus\{u_1\})\times (V\setminus\{u_2\})$. 
   Let $f \colon V^2 \To V$ be a binary injection which preserves $E$ and $N$, 
behaves like $p_1$ on $U$,
and which behaves like
$p_2$ between $u$ and all points in $U$. Then $f$ generates a binary injection of type
$\mini$ as well as a binary injection of type $\maxi$.
\end{lemma}

\begin{proof}
Let $\Gamma$ be the reduct of $G$ having all relations
that are first-order definable in $G$ and preserved by $f$. Since $U$ contains copies of products of arbitrary finite graphs, $f$ behaves like $p_1$ on arbitrarily large finite substructures of $G^2$, and hence generates a binary injection of type $p_1$ by Proposition~\ref{prop:RamseyCore}. Hence $\Gamma$ is also preserved by such a function, and we may apply the implication from (2) to (1) in Lemma~\ref{lem:min-relationally} to $\Gamma$. 

Let $\phi$ be a primitive positive formula with variable
set $S$, $\{x_1,\dots,x_4\} \subseteq S$, such that
$\phi \wedge N(x_1,x_2) \wedge \bigwedge_{1\leq i<j\leq 4} x_i \neq x_j$ and $\phi \wedge N(x_3,x_4) \wedge \bigwedge_{1\leq i<j\leq 4} x_i \neq x_j$ 
are satisfiable over $\Gamma$, witnessed by satisfying assignments
$r,s \colon S \rightarrow V$, respectively.

Let $\alpha$ be an automorphism of $G$ that maps $r(x_3)$ to $u_1$,
and let $\beta$ be an automorphism of $G$ that maps $s(x_3)$ to $u_2$. Then $(\alpha(r(x_3)),\beta(s(x_3)))=u$, 
and $v:=(\alpha(r(x_4)),\beta(s(x_4))) \in U$ since $\alpha(r(x_4)) \neq \alpha(r(x_3)) = u_1$ and $\beta(s(x_4)) \neq \beta(s(x_3)) = u_2$. Thus, $f$ behaves
like $p_2$ between $u$ and $v$, and since $s$ satisfies $N(x_3,x_4)$,
we have that $t \colon  S \rightarrow V$ defined by 
$$t(x) = f(\alpha(x),\beta(x))$$
satisfies $N(x_3,x_4)$, too. Since $\alpha,\beta,f$ are polymorphisms of 
$\Gamma$, the assignment $t$ also satisfies $\phi$.
To see that $t$ also satisfies $N(x_1,x_2)$, observe that
$\alpha(r(x_1)) \neq \alpha(r(x_3))$ and $\beta(s(x_1)) \neq \beta(s(x_3))$, and hence $p:=(\alpha(r(x_1)),\beta(s(x_1))) \notin U$. 
Similarly, $q:=(\alpha(r(x_2)),\beta(s(x_2))) \notin U$. 
Hence, $f$ behaves as $p_1$ between $p$ and $q$, and
since $N(r(x_1),r(x_2))$, so does $t$. 

By Lemma~\ref{lem:min-relationally} we conclude that $\Gamma$ is
preserved by a binary injection of type $\mini$, and
consequently $f$ generates a binary injection of type $\mini$.

Since our assumptions on $f$ were symmetric in $E$ and $N$, it follows that $f$ also generates a binary injection of type $\maxi$.
\end{proof}

\begin{lemma}\label{lem:generatingMin:3}
   Let $u \in V^2$, and set $U:= (V\setminus\{u_1\})\times (V\setminus\{u_2\})$. Let $f \colon V^2 \To V$ be a binary injection which preserves $E$ and $N$, 
behaves like $p_1$ on $U$,
and which behaves like
$\mini$ between $u$ and all points in $U$. Then $f$ generates a binary injection of type
$\mini$.
\end{lemma}
\begin{proof}
	The proof is identical with the proof in the preceding lemma; note that our assumptions on $f$ here imply more deletions of edges than the assumptions in that lemma, so it can only be easier to generate a binary injection of type $\mini$.
\end{proof}

\begin{lemma}\label{lem:generatingMin:4}
   Let $u, v\in V^2$ such that ${\neq}{\neq}(u,v)$ and set $W:=(V\setminus\{u_1,v_1\})\times (V\setminus\{u_2, v_2\})$. Let $f \colon V^2\To V$ be a binary injection that 
   \begin{itemize}
   \item behaves like $p_1$ on $W$
   \item behaves like $p_1$ between any point in $\{u,v\}$ and any point in $W$
   \item  does not behave like $p_1$ between $u$ and $v$.
   \end{itemize}
    Then $f$ generates $e_E$, $e_N$, or a binary
injection of type $\mini$ as well as binary injection of type $\maxi$.
\end{lemma}
\begin{proof}
Consider first the case where $\EE(u,v)$ and $N(f(u),f(v))$. Let $\alpha\in\Aut(G)$ send $u_1$ to $u_2$ and $v_1$ to $v_2$, and consider the function $h(x):=f(x,\alpha(x))$. Then $N(h(u_1),h(v_1))$, and $h$ preserves $E$ and $N$ between any point in $\{u_1,v_1\}$ and all points in $V\setminus\{u_1,v_1\}$, and so it generates $e_N$ by a standard iterative argument. Similarly, if $\NN(u,v)$ and $E(f(u),f(v))$ then $f$ generates $e_E$.

It remains to consider the case where $\EN(u,v)$ and $N(f(u),f(v))$,
and the case where $\NE(u,v)$ and $E(f(u),f(v))$. In the first case we prove that
$f$ generates a binary injection of type $\mini$; it then follows by duality that in the second case, $f$ generates a binary injection of type $\maxi$.

As in Lemma~\ref{lem:generatingMin:2}, we apply the implication (2) $\rightarrow (1)$ from Lemma~\ref{lem:min-relationally}. Let $\Gamma$, $\phi$, $S$, $x_1,\dots,x_4$, $r$, and $s$ be as in the proof of Lemma~\ref{lem:generatingMin:2}; by the same argument as before, $\Gamma$ is preserved by a binary injection of type $p_1$.

 If $N(r(x_3),r(x_4))$, then the assignment $r$ shows that  $\phi \wedge N(x_1,x_2) \wedge N(x_3,x_4)$ is satisfiable and we are done. Otherwise, since $r(x_3) \neq r(x_4)$, we have 
$E(r(x_3),r(x_4))$. Therefore, there is an $\alpha \in \Aut(G)$
such that $(\alpha(r(x_3)),\alpha(r(x_4)))=(u_1,v_1)$. Similarly, 
since $N(s(x_3),s(x_4))$ and $N(u_2,v_2)$,
there is a $\beta \in \Aut(G)$ such that $(\beta(s(x_3)),\beta(s(x_4)))=(u_2,v_2)$. 
We claim that the map $t \colon  S \rightarrow V$ defined by 
$$t(x) = f(\alpha(x),\beta(x))$$
is a satisfying assignment for $\phi \wedge N(x_1,x_2) \wedge N(x_3,x_4)$.
The assignment $t$ satisfies $\phi$ since $\alpha,\beta$ and $f$ are polymorphisms of $\Gamma$.
Then $N(t(x_3),t(x_4))$ holds because $(\alpha(r(x_3)),\beta(s(x_3)))=u$
and $(\alpha(r(x_4)),\beta(s(x_4)))=v$, and $N(f(u),f(v))$. 
To prove that $N(t(x_1),t(x_2))$ holds, observe that $r(x_1) \neq r(x_3)$
and $r(x_1) \neq r(x_4)$, and hence
$\alpha(r(x_1)) \notin \{\alpha(r(x_3)),\alpha(r(x_4))\} = \{u_1,v_1\}$. 
Similarly, $\beta(s(x_1)) \notin \{\beta(s(x_3)),\beta(s(x_4))\} = \{u_2,v_2\}$. 
Hence, $(\alpha(r(x_1)),\beta(s(x_1)) \in W$. A similar argument for
$x_2$ in place of $x_1$ shows that $(\alpha(r(x_2)),\beta(s(x_2)) \in W$.
Since $f$ behaves like $p_1$ 
on $W$,
and since $r$ satisfies $N(x_1,x_2)$, we have proved the claim. 
This shows that $\Gamma$ is preserved by a binary injection of type $\mini$, and hence $f$ generates such a function.

By symmetry of our assumptions on $f$ in $E$ and $N$, it follows that $f$ generates a binary injection of type $\mini$ if and only if it generates a binary injection of type $\maxi$.
\end{proof}

We are now set up to prove Proposition~\ref{prop:nonProjGeneratesMin}. This completes the proof of Proposition~\ref{prop:maxMinMajorityMonority}, and in turn the proof of Theorem~\ref{thm:higherArity}.

\begin{proof}[\pf of Proposition~\ref{prop:nonProjGeneratesMin}]
    Let $f$ be given. By Theorem~\ref{thm:minimal-ops}, $f$ generates a binary canonical injection $g$ of type projection, $\mini$, or $\maxi$. In the last two cases we are done, so consider the first case. We claim that $f$ also generates a binary function $h$ of type $\mini$ or $\maxi$. Then $h(g(x,y),g(y,x))$ is still of type $\mini$ or $\maxi$ and in addition canonical and injective, and the proposition follows.
    
    To prove our claim, fix a finite set $C:=\{c_1,\ldots, c_m\}\subseteq V$ such that the fact that $f$ does not behave like a projection is witnessed on $C$. Invoking Proposition~\ref{prop:orderedRandomCanonical}, we may henceforth assume that $f$ is canonical as a function from $(V;E,\prec, c_1,\ldots,c_m)^2$ to $(V;E,\prec)$ (and hence also to $(V;E)$ since tuples of equal type in $(V;E,\prec)$ have equal type in $(V;E)$).
    
In the following we will consider orbits of elements in the structure ${(V;E,\prec, c_1,\ldots,c_m)}$. The infinite orbits are precisely the sets of the form $$\{v\in V\; |\; Q_i(v,c_i) \text{ and } R_i(v,c_i) \text{ for all } 1\leq i\leq m\},$$ for $Q_1,\ldots,Q_m\in\{E,N\}$, and $R_1,\ldots,R_m\in\{\prec,\succ\}$. The finite orbits are of the form $\{c_i\}$ for some $1\leq i\leq m$. 
It is well-known that each infinite orbit of $(V;E,\prec, c_1,\ldots,c_m)$ contains copies of arbitrary linearly ordered finite graphs, and in particular, forgetting about the order, of all finite graphs. 
Therefore, if $f$ behaves like $\mini$ or $\maxi$ on any set of the form $O_1\times O_2$, where $O_1, O_2$ are infinite orbits of $(V;E,\prec, c_1,\ldots,c_m)$, then by Proposition~\ref{prop:RamseyCore} it generates a function which behaves like $\mini$ or $\maxi$ everywhere, and we are done. 

Moreover, if $f$ is of mixed type on any set of the form $O_1\times O_2$ as above, then, by Proposition~\ref{prop:RamseyCore}, $f$ generates a canonical function which has the same mixed behavior everywhere. But then we are done by Lemmas~\ref{lem:mixtyp:maxp}, \ref{lem:mixtyp:minp}, \ref{lem:mixtyp:maxmin}, and \ref{lem:generatingMin:1}. Hence, we may assume that $f$ behaves like a projection on every set of this form. Fix in the following infinite orbits $O_1, O_2$ and assume without loss of generality that $f$ behaves like $p_1$ on $O_1\times O_2$.

Let $W_1, W_2$ be arbitrary infinite orbits. 
Then since $f$ is canonical, 
it behaves like $p_1$, $p_2$, $\mini$, or $\maxi$
between all $u,v$ with $u \in O_1\times O_2$, $v  \in W_1\times W_2$
and $u_1 \prec v_1$ and $u_2 \prec v_2$. 
Consider the case where there exist infinite orbits $W_1, W_2$ such that $f$ behaves like $p_2$ between all points $u\in O_1\times O_2$ and $v\in W_1\times W_2$ for which $u_1\prec v_1$ and $u_2\prec v_2$. Then fix any $v\in W_1\times W_2$, and set $O_1':=\{o\in O_1 \; |\;  o\prec v_1 \}$ and $O_2':=\{o\in O_2\; |\; o\prec v_2 \}$. Set $U_1:=O_1'\cup\{v_1\}$ and $U_2:=O_2'\cup\{v_2\}$. We then have that $f$ behaves like $p_2$ between $v$ and any point $u$ of $(U_1\setminus\{v_1\})\times (U_2\setminus\{v_2\})$, and like $p_1$ between any two points of $(U_1\setminus\{v_1\})\times (U_2\setminus\{v_2\})$. Since $(U_i;E,v_i)$ contains copies of all finite substructures of $(V;E,v_i)$, for $i\in\{1,2\}$, by Proposition~\ref{prop:RamseyCore} we get that $f$ generates a function which behaves like $p_2$ between $v$ and any point $u$ of $(V\setminus\{v_1\})\times (V\setminus\{v_2\})$, and which behaves like $p_1$ between any two points of $(V\setminus\{v_1\})\times (V\setminus\{v_2\})$. Then Lemma~\ref{lem:generatingMin:2} implies that $f$ generates a binary injection of type $\mini$ and we are done.

This argument is easily adapted to any situation where there exist infinite orbits $W_1, W_2$ such that $f$ behaves like $p_2$ between all points $u\in O_1\times O_2$ and $v\in W_1\times W_2$ with $R_1(u_1, v_1)$ and $R_2(u_2, v_2)$, for $R_1, R_2\in\{\prec,\succ\}$.

When there exist infinite orbits $W_1, W_2$ such that $f$ behaves like $\mini$ between all points $u\in O_1\times O_2$ and $v\in W_1\times W_2$ with $R_1(u_1, v_1)$ and $R_2(u_2, v_2)$, then we can argue similarly, invoking Lemma~\ref{lem:generatingMin:3} at the end. Replacing $\mini$ by $\maxi$ we can use the dual argument, with the difference that $f$ generates a binary injection of type $\maxi$ rather than $\mini$.

Since $f$ is canonical, one of the situations described so far must occur. Putting this together, we conclude that we may assume that for all infinite orbits $W_1, W_2$ and all points $u\in O_1\times O_2$ and $v\in W_1\times W_2$, $f$ behaves like $p_1$ between $u$ and $v$. Having that, suppose that for some infinite orbits $W_1, W_2$, $f$ behaves like $p_2$ on $W_1\times W_2$. Then exchanging the roles of $O_1\times O_2$ and $W_1\times W_2$ and of $p_1$ and $p_2$ above, we can again conclude that $f$ generates a binary injection of type $\mini$. We may thus henceforth assume that $f$ behaves like $p_1$ on $(V\setminus C)^2$. 

Pick any $u\in C^2$. Suppose that there exists $v\in (V\setminus C)^2$ such that $f$ does not behave like $p_1$ between $u$ and $v$; say without loss of generality that $\EN(u,v)$ and $N(f(u),f(v))$. Let $O_i$ be the (infinite) orbit of $v_i$, for $i\in\{1,2\}$. Then for all $v\in O_1\times O_2$ we have $\EN(u,v)$ and $N(f(u),f(v))$ since $f$ is canonical. Now let $w\in O_2\times O_1$. We distinguish the two cases $E(f(u),f(w))$ and $N(f(u),f(w))$. In the first case, $f$ behaves like $p_2$ between $u$ and all $v\in (O_1\cup O_2)^2$. We can then argue as above and are done. In the second case, $f$ behaves like $\mini$ between $u$ and all $v\in (O_1\cup O_2)^2$, and we are again done by the corresponding argument above. We conclude that we may assume that for all $u\in C^2$ and all $v\in (V\setminus C)^2$, $f$ behaves like $p_1$ between $u$ and $v$ as well.

Now pick $u,v\in C^2$ such that $f$ does not behave like $p_1$ between $u$ and $v$, say without loss of generality $\EN(u,v)$ and $N(f(u),f(v))$; this is possible since the fact that $f$ does not behave like $p_1$ everywhere is witnessed on $C$. Set $W_i:=(V\setminus C)\cup\{u_i,v_i\}$ for $i=\{1,2\}$. Since $W_1$ and $W_2$ induce a structure isomorphic to the random graph in $G$, and $f$  behaves like $p_2$ between $u$ and $v$, and like $p_1$ between all points in  $\{u,v\}$ and all points $(W_1\setminus \{u_1,v_1\})\times (W_2\setminus \{u_2,v_2\})$, we are done by Lemma~\ref{lem:generatingMin:4}.
\end{proof}

%% file: binary2.tex
\subsection{When the endomorphisms of a reduct are generated by $\{-\}$} \label{sect:c2}
\label{sect:minus}
We next consider Case~({c}) of Proposition~\ref{prop:endos}. That is, we will assume that the endomorphisms of $\Gamma$ are exactly the functions generated by $\{-\}$. In particular, $\Aut(\Gamma)$ contains $-$ but not $\sw$, and
the automorphisms of $\Gamma$ generate
its endomorphisms.

\begin{definition}\label{def:H1p}
Let $H_1'$ be the smallest $6$-ary relation that
is preserved by $\{-\}$ and contains $H_1$.
\end{definition}

The following is an analog of Theorem~\ref{thm:higherArity} for the situation of this section.

\begin{theorem}\label{thm:minus}
Let $\Gamma$ be a reduct of $G$ whose endomorphisms are precisely the unary functions generated by $\{-\}$. Then either $H_1'$ is primitive positive definable in $\Gamma$, or one of the cases (b)-(e)
of Theorem~\ref{thm:higherArity} applies.
\end{theorem}
\begin{proof}
Note that $H_1'$ consists of three orbits
of 6-tuples in $\Aut(\Gamma)$,
and hence, if $H_1'$ is not primitive positive
definable in $\Gamma$, then
there exists by
Theorem~\ref{conf:thm:inv-pol}
and Lemma~\ref{lem:arity-reduction} a ternary polymorphism $f$ of
$\Gamma$ that violates $H_1'$.
That is, there are $t^1,t^2,t^3 \in H_1'$ such
that $f(t^1,t^2,t^3) \notin H_1'$. Note that
for each $t^j$, either $t^j$ or $-t^j \in H_1$.
In the first case we set $g_j$ to be the identity function on $V$, in the second
case we let $g_j$ be the operation $-$.
Now consider the function $f'$ defined by
$f'(x_1,x_2,x_3) := f(g_1(x_1),g_2(x_2),g_3(x_3))$.
We have that $s^{j} := g_j^{-1}(t^j) \in H_1$, but $f'(s^1,s^2,s^3) =
f(t^1,t^2,t^3)$ is not in $H_1'$. Consider the function $h(x):=
f'(x,x,x)$; since the endomorphisms of $\Gamma$ are generated by
$\{-\}$, $h$ either preserves $E$ and $N$, or it flips them. By
replacing $f'$ by $-(f')$ in the latter case we may assume that $h$ preserves
$E$ and $N$. Note that we still have that $f'(s^1,s^2,s^3)$ is not in
$H_1'$, and therefore not in $H_1$ either.  Hence,
$f'$ violates $H_1$.

Now suppose that $f'$ violates $E$ or $N$; we will derive a
contradiction. Say without loss of generality that there are $u,v\in
V^3$ with $\EEE(u,v)$ such that $E(f'(u),f'(v))$ does not hold. Pick distinct 
$a,b,c,d\in V$ such that $\{a,b,c,d\}$ induces a clique in $G$, and such that each element is connected to all entries on $u,v$ by an edge. Pick then
$\alpha_1,\alpha_2,\alpha_3\in\Aut(G)$ such that $\alpha_i(a)=u_i$ and
$\alpha_i(b)=v_i$ for all $i\in\{1,2,3\}$, and such that
$\alpha_1({c})=\alpha_2({c})=\alpha_3({c})=c$ and
$\alpha_1({d})=\alpha_2({d})=\alpha_3({d})=d$. We then have that the
function $x \mapsto f'(\alpha_1(x),\alpha_2(x),\alpha_3(x))$ maps
$(c,d)$ to an edge since $h(x)$ preserves $E$, but it does not map
$(a,b)$ to an edge, by our assumption on $u$ and $v$. This is,
however, impossible, since  the function must be generated by $\{-\}$.

Therefore, $f'$ preserves $E$ and $N$. Then
Theorem~\ref{thm:higherArity} implies that $f'$
generates functions with the desired properties, or a binary canonical injection 
of type $\maxi$ or $\mini$.
A binary canonical injection of type $\maxi$
together with $\{-\}$ generates a binary canonical injection of type $\mini$, and vice
versa. Then $\maxi(\mini(x,y),\mini(y,z),\mini(x,z))$ is a ternary canonical injection of type majority with the desired properties, and we are also
done in this case (since identifying two of its variables moreover yields a binary canonical injection of type projection).
\end{proof}

\begin{proposition}\label{prop:h1prime}
$\Csp(V;H_1')$ is NP-hard. 
\end{proposition}
\begin{proof}
One can show NP-hardness similarly as in the proof of Proposition~\ref{prop:h-hard}, by reduction from positive Not-all-three-equal-3SAT instead of positive 1-in-3-3SAT. 
\end{proof}

%% file: binary3.tex
\subsection{When the endomorphisms of a reduct are generated by $\{\sw\}$}\label{sect:c3}
In this section we will prove Theorem~\ref{thm:sw} below, which treats Case~(d) in Proposition~\ref{prop:endos}. 

\begin{definition}
For $k\geq 1$, let $S^{(k)}$ be the $k$-ary relation that holds
on $x_1,\dots,x_k \in V$ if $x_1,\dots,x_k$ are 
pairwise distinct, and the number of edges between these $k$ vertices is even. 
\end{definition}

Recall also the definition of $R^{(k)}$ from Section~\ref{sect:endos}. The structure of this section will be similar
to the one of Section~\ref{sect:higherArity}, but $R^{(3)}$ will take the role of $E$, and $S^{(3)}$ will take the role of $N$. The relation $H_1$ will be replaced by the following relation.

\begin{definition}\label{defn:H2}
Let $H_2$ be the smallest $9$-ary relation that
is preserved by $\{\sw\}$ and contains all tuples
$(x_1,y_1,z_1,x_2,y_2,z_2,x_3,y_3,z_3) \in V^9$ such that
\begin{align}
& \bigwedge_{i,j \in \{1,2,3\}, i \neq j, u \in \{x_i,y_i,z_i\}, v \in \{x_j,y_j,z_j\}} N(u,v) \nonumber \\
\wedge & \; \big((R^{(3)}(x_1,y_1,z_1) \wedge S^{(3)}(x_2,y_2,z_2) \wedge S^{(3)}(x_3,y_3,z_3))
\nonumber \\ 
& \vee \; (S^{(3)}(x_1,y_1,z_1) \wedge R^{(3)}(x_2,y_2,z_2) \wedge S^{(3)}(x_3,y_3,z_3)) \nonumber\\  
& \vee \; (S^{(3)}(x_1,y_1,z_1) \wedge S^{(3)}(x_2,y_2,z_2) \wedge R^{(3)}(x_3,y_3,z_3)) \big)\; . \nonumber
\end{align}
\end{definition}

\begin{theorem}\label{thm:sw}
Let $\Gamma$ be a reduct of $G$ whose endomorphisms are precisely the unary functions generated by $\{\sw\}$. 
Then either $H_2$ is primitive positive definable
in $\Gamma$, or $\Gamma$ satisfies item (b) or (d)
of Theorem~\ref{thm:higherArity}. 
\end{theorem}

\begin{proposition}\label{prop:h2}
$\Csp(V;H_2)$ is NP-hard.
\end{proposition}
\begin{proof}
This can be shown analogously to Proposition~\ref{prop:h-hard} by reduction from 1-in-3-3SAT,
but this time we represent 
1 by triples from $R^{(3)}$ instead of pairs that satisfy $E$, and 0 by triples from $S^{(3)}$, and then use
$H_2$ analogously as we have used $H_1$ in the proof
of Proposition~\ref{prop:h-hard}. 
\end{proof} 

\subsubsection{Producing canonical functions of type projection} 

We use a combination of Lemma 5.3 in~\cite{HornOrFull} with Lemma 42 in~\cite{RandomMinOps}. Those lemmas are stated below for the convenience of the reader.

\begin{lemma}[Lemma 5.3 in~\cite{HornOrFull}]
\label{lem:indep}
Let $\Gamma$ be a relational structure over an infinite domain $D$ 
such that the set of primitive positive definable binary relations in $\Gamma$ is
exactly $\{D^2,\neq,=,\emptyset\}$.
Suppose that $\Gamma$ contains an $n$-ary relation $Q$ such that there
are pairwise distinct $1 \leq i,j,k,l \leq n$ for which the following conditions hold:
\begin{enumerate}
\item $Q(x_1,\dots,x_n) \wedge x_i \neq x_j$ is satisfiable;
\item $Q(x_1,\dots,x_n) \wedge x_k \neq x_l$ is satisfiable;
\item $Q(x_1,\dots,x_n) \wedge x_i \neq x_j \wedge x_k \neq x_l$ is unsatisfiable.
\end{enumerate}
Then the relation $S_D := \{(x,y,z) \in D^3 \; | \; y \neq z \wedge (x=y \vee x=z)\}$
has a primitive positive definition in $\Gamma$.
\end{lemma}

\begin{lemma}[Lemma 42 in~\cite{RandomMinOps}]\label{lem:bin-inj}
Let $\Gamma$ be a countable $\omega$-categorical structure
in which $\neq$ is primitive positive definable.
Then the following are equivalent.
\begin{enumerate}
\item If $\phi$ is a primitive positive formula such that
both $\phi \wedge x \neq y$ and $\phi \wedge u \neq v$ are
satisfiable over $\Gamma$, then $\phi \wedge x \neq y \wedge u \neq
v$ is satisfiable over $\Gamma$ as well.
\item $\Gamma$ is preserved by a binary injective operation.
\end{enumerate}
\end{lemma}

We use the following combination of these two lemmata. 

\begin{proposition}\label{prop:2trans}
Let $\Gamma$ be an $\omega$-categorical structure with a 2-transitive
automorphism group (i.e., for which the relation $\neq$ equals one orbit of pairs). Then one of the following
applies.
\begin{enumerate}
\item All polymorphisms of $\Gamma$ are essentially unary. 
\item $\Gamma$ has a constant endomorphism.
\item $\Gamma$ has a binary injective endomorphism.
\end{enumerate}
\end{proposition}
\begin{proof}
Write $D$ for the domain of $\Gamma$. If $\Gamma$ has a non-injective endomorphism, 
then a straightforward iterative argument using the 2-transitivity of $\Aut(\Gamma)$ and local closure  shows that $\Gamma$ also has a constant endomorphism and there is nothing to show. Otherwise, since $\neq$ only consists of one orbit of pairs, it is preserved by all polymorphisms and hence primitive positive definable by Theorem~\ref{conf:thm:inv-pol}. By the 2-transitivity of $\Aut(\Gamma)$ it is now clear that the set of primitive positive definable binary relations in $\Gamma$ is
exactly $\{D^2,\neq,=,\emptyset\}$. Hence, by Lemma~\ref{lem:indep} one of the following holds:
\begin{itemize}
\item the first item of Lemma~\ref{lem:bin-inj} applies,
and hence by Lemma~\ref{lem:bin-inj} 
the structure $\Gamma$ has a binary injective polymorphism;
\item there is a formula which is a counterexample to first item of Lemma~\ref{lem:bin-inj}. In that case, the expansion of $\Gamma$ by the relation defined by this formula satisfies the hypotheses of Lemma~\ref{lem:indep}, and hence the relation $S_D$ is primitive positive definable in $\Gamma$. It then follows that all polymorphisms of $\Gamma$ are essentially unary (this can be shown as in Proposition~5.3.2 in~\cite{Bodirsky-HDR}).
\end{itemize}
\end{proof}

\begin{proposition}\label{prop:binary-without-EN}
Let $\Gamma$ be a reduct of $G$ with an essential polymorphism. Then $\Gamma$ is preserved by a constant function, 
$e_E$, $e_N$, or by a canonical binary injection of type $\mini$, 
$\maxi$, or $p_1$. 
\end{proposition}
\begin{proof}
If there is a primitive positive definition of $E$ and $N$, then the statement follows from Theorem~\ref{thm:minimal-ops}. 
So suppose that this is not that case; also suppose that
$\Gamma$ is not preserved by $e_E$, $e_N$, or a constant function. Then the automorphisms of $\Gamma$ generate its endomorphisms by Theorem~\ref{thm:endos}, and so they must violate $E$ and $N$ as otherwise these relations would have a primitive positive definition. By Theorem~\ref{thm:reducts}, we then see that $\Aut(\Gamma)$ is 2-transitive. 
By Proposition~\ref{prop:2trans}, 
    $\Gamma$ has a binary injective polymorphism $g$. By 
    Proposition~\ref{prop:orderedRandomCanonical},
      $g$ generates a binary function $h$ which is canonical as a function from $(V;E,\prec)^2$ to $(V;E,\prec)$; this function is again 
      injective. The function $x\mapsto h(x,x)$ either preserves $E$ and $N$, or behaves like $-$, $e_E$ or $e_N$. We can assume that it does not behave like $e_E$ or $e_N$, and if it behaves like $-$, we can replace $h$ by $-h$ and assume that $x\mapsto h(x,x)$ preserves $E$ and $N$. Now consider the function  $x\mapsto h(x,\alpha(x))$, where $\alpha\in\Aut(G)$ reverses $\prec$. Again, we may exclude the possibility that it behaves like $e_E$ or $e_N$. But then the function $(x,y)\mapsto h(h(x,y),h(y,x))$ preserves $E$ and $N$ and we can apply Theorem~\ref{thm:minimal-ops} to conclude that it generates a binary  injection which is canonical as a function from $G^2$ to $G$ and of type $\mini$, $\maxi$, or $p_1$. 
\end{proof}


\begin{corollary}\label{cor:r3-p1}
Let $\Gamma=(V;R^{(3)},S^{(3)},\ldots)$ be a reduct of $G$ with an essential polymorphism. 
Then $\Gamma$ is preserved by a binary canonical injection of type $p_1$.
\end{corollary}
\begin{proof}
Since $e_N$ and functions of type $\mini$ 
do not preserve $R^{(3)}$
and $e_E$ and functions of type $\maxi$ do not preserve $S^{(3)}$, 
Proposition~\ref{prop:binary-without-EN}
implies that $\Gamma$ is preserved by a binary canonical injection of type $p_1$.
\end{proof}


\subsubsection{Eliminating mixed behavior} 

\begin{lemma}\label{lem:r3-nomixed}
Let $f \colon V^2 \to V$ be a binary injection that preserves $R^{(3)}$ and $S^{(3)}$. Then $f$ is not of type 
$p_1/p_2$. 
\end{lemma}
\begin{proof}
Suppose for contradiction that $f$ does have the behavior $p_1/p_2$. 
Let $u_1,u_2,u_3 \in V$ with $u_1 \prec u_2 \prec u_3$,
$E(u_1,u_2)$, $N(u_2,u_3)$, and $N(u_1,u_3)$.
Let $v_1,v_2,v_3 \in V$ with $v_1 \prec v_2 \prec v_3$
and $N(v_1,v_2),E(v_2,v_3),N(v_1,v_3)$. 
Then $E(f(u_1,v_1),f(u_2,v_3))$ and $N(f(u_1,v_1),f(u_3,v_2))$ since $f$ behaves
like $p_1$ on input $(\prec,\prec)$. Moreover,
$E(f(u_2,v_3),f(u_3,v_2))$ since $f$ behaves 
like $p_2$ on input $(\prec,\succ)$. 
Then $(u_1,u_2,u_3) \in R^{(3)}$ 
and $(u_1,u_2,u_3)\in R^{(3)}$,
but $(f(u_1,v_2),f(u_2,v_3),f(u_3,v_2)) \notin R^{(3)}$, in contradiction to our assumptions. 
\end{proof}

\subsubsection{Behaviors relative to vertices}

\begin{lemma}\label{lem:r3-constants}
   Let $u \in V^2$, and set 
   $U := (V\setminus\{u_1\})\times (V\setminus\{u_2\})$. 
   Let $f \colon V^2 \To V$ be a binary injection which 
behaves like $p_1$ on $U$,
 and which behaves like
$p_2$ or $\maxi$ between $u$ and all points in $U$. Then $f$ does not preserve $R^{(3)}$. 
\end{lemma}
\begin{proof}
Let $v,w \in U$ be such that $\NE(u,v)$, $\EN(v,w)$, and $\NN(u,w)$. Then we have $E(f(u),f(v))$, 
$E(f(v),f(w))$, and $N(f(u),f(w))$. Hence,
$R^{(3)}(u_i,v_i,w)$ for $i\in\{1,2\}$, but $S^{(3)}(f(u),f(v),f(w))$. 
\end{proof}

\begin{definition}
We say that a binary injective function $f \colon V^2 \to V$ is
\begin{itemize}
\item of type $R^{(3)}$-$p_i$, for $i \in \{1,2\}$, 
iff for all $u,v,w \in V^2$ with $\NEQNEQ(u,v)$,
$\NEQNEQ(v,w)$, and $\NEQNEQ(u,w)$ we have $R^{(3)}(f(u),f(v),f(w))$ if and only if $R^{(3)}(u_i,v_i,w_i)$.
\item of type $R^{(3)}$-projection iff it is
of type $R^{(3)}$-$p_1$ or of type $R^{(3)}$-$p_2$. 
\end{itemize}
\end{definition}


\begin{proposition}\label{prop:r3-projection}
    Let $f \colon V^2 \rightarrow V$ be a binary injective polymorphism of $(V;R^{(3)},S^{(3)})$.
    Then $f$ is of type $R^{(3)}$-projection.
\end{proposition}
\begin{proof}
The proof is similar to the proof of Proposition~\ref{prop:nonProjGeneratesMin}. 
 Fix a finite set $C:=\{c_1,\ldots, c_m\}\subseteq V$ such that the fact that $f$ is not of type $R^{(3)}$-projection is witnessed on $C$. Invoking Proposition~\ref{prop:orderedRandomCanonical}, we may henceforth assume that $f$ is canonical as a function from $(V;E,\prec, c_1,\ldots,c_m)^2$ to $(V;E,\prec)$.
    
In the following we will consider orbits of elements in the structure ${(V;E,\prec, c_1,\ldots,c_m)}$. The infinite orbits are precisely the sets of the form $$\{v\in V\; |\; Q_i(v,c_i) \text{ and } R_i(v,c_i) \text{ for all } 1\leq i\leq m\},$$ for $Q_1,\ldots,Q_m\in\{E,N\}$, and $R_1,\ldots,R_m\in\{\prec,\succ\}$. The finite orbits are of the form $\{c_i\}$ for some $1\leq i\leq m$. 
Each infinite orbit of $(V;E,\prec, c_1,\ldots,c_m)$ is isomorphic to $(V;E,\prec)$. Therefore Proposition~\ref{prop:RamseyCore} implies that if $f$ has a certain behavior on such an infinite orbit, then it generates a canonical function which has the same behavior everywhere. Therefore we have for all infinite orbits $O$ that $f$
\begin{itemize}
\item cannot be of type $\mini$ or $\maxi$ on $O$ since it preserves $R^{(3)}$ and $S^{(3)}$;
\item cannot have behavior $\maxi/p_i$ or $p_i/\maxi$ for $i \in \{1,2\}$ on $O$, by Lemma~\ref{lem:mixtyp:maxp};
\item cannot have behavior $\mini/p_i$ or $p_i/\mini$ for $i \in \{1,2\}$ on $O$, by \ref{lem:mixtyp:minp};
\item it cannot have behavior $\maxi/\mini$ or $\mini/\maxi$ on $O$, by Lemma~\ref{lem:mixtyp:maxmin};
\item it cannot have behavior $p_1/p_2$ or $p_2/p_1$ on $O$, by Lemma~\ref{lem:r3-nomixed}. 
\end{itemize}
Hence, we may assume that $f$ behaves like a projection on every infinite orbit. Fix in the following an infinite orbit $O$ and assume without loss of generality that $f$ behaves like $p_1$ on $O$.

Let $W$ be any infinite orbit. 
Then since $f$ is canonical, 
it behaves like $p_1$, $p_2$, $\mini$, or $\maxi$
between all $u,v$ with $u \in O^2$, $v  \in W^2$
and $u_1 \prec v_1$ and $u_2 \prec v_2$. 
Consider the case where there exists an infinite orbit $W$ such that $f$ behaves like $p_2$ or $\maxi$ between all points $u\in O^2$ and $v\in W^2$ for which $u_1\prec v_1$ and $u_2\prec v_2$. Then fix any $v\in W^2$, and set $O_1:=\{o\in O\; |\;  o\prec v_1 \}$ and $O_2:=\{o\in O\; |\; o\prec v_2 \}$. Set $O_1':=O_1\cup\{v_1\}$ and $O_2':=O_2\cup\{v_2\}$. We then have that $f$ behaves like $p_2$ or $\maxi$ between $v$ and any point $u$ of $(O_1'\setminus\{v_1\})\times (O_2'\setminus\{v_2\})$, and like $p_1$ between any two points of $(O_1'\setminus\{v_1\})\times (O_2'\setminus\{v_2\})$. Since $(O_i';E,v_i)$ is isomorphic to $(V;E,v_i)$, for $i\in\{1,2\}$, by Proposition~\ref{prop:RamseyCore} we get that $f$ generates a function which behaves like $p_2$ or $\maxi$ between $v$ and any point $u$ of $(V\setminus\{v_1\})\times (V\setminus\{v_2\})$, and which behaves like $p_1$ between any two points of $(V\setminus\{v_1\})\times (V\setminus\{v_2\})$. This is impossible by Lemma~\ref{lem:r3-constants}.
This argument is easily adapted to any situation where there exists an infinite orbit $W$ such that $f$ behaves like $p_2$ between all points $u\in O^2$ and $v\in W^2$ with $R_1(u_1, v_1)$ and $R_2(u_2, v_2)$, for $R_1, R_2\in\{\prec,\succ\}$.
When there exists an infinite orbit $W$ such that $f$ behaves like $\mini$ between all points $u\in O^2$ and $v\in W^2$ with $R_1(u_1, v_1)$ and $R_2(u_2, v_2)$, then we can argue similarly.

Since $f$ is canonical, one of the situations described so far must occur. Putting this together, we conclude that for every infinite orbit $W$ and all points $u\in O^2$ and $v\in W^2$, $f$ behaves like $p_1$ between $u$ and $v$. Having that, suppose that for an infinite orbit $W$, $f$ behaves like $p_2$ on $W$. Then exchanging the roles of $O$ and $W$ and of $p_1$ and $p_2$ above, we again arrive at a contradiction. We may thus henceforth assume that $f$ behaves like $p_1$ on $(V \setminus C)^2$. 

Pick any $u\in C^2$. Suppose that there exists $v\in (V\setminus C)^2$ such that $f$ does not behave like $p_1$ between $u$ and $v$. Assume first that $\EN(u,v)$ and $N(f(u),f(v))$. Let $O_i$ be the (infinite) orbit of $v_i$, for $i\in\{1,2\}$. Then for all $v\in O_1\times O_2$ we have $\EN(u,v)$ and $N(f(u),f(v))$ since $f$ is canonical. Now let $w\in O_2\times O_1$. We distinguish the two cases $E(f(u),f(w))$ and $N(f(u),f(w))$. In the first case, $f$ behaves like $p_2$ between $u$ and all $v\in (O_1\cup O_2)^2$. We can then argue as above and are done. In the second case, $f$ behaves like $\mini$ between $u$ and all $v\in (O_1\cup O_2)^2$, and we are again done by the corresponding argument above. The dual argument works when $\NE(u,v)$ and $E(f(u),f(v))$. Now assume that $\EE(u,v)$ and $N(f(u),f(v))$. We claim that $\EE(u,v')$ implies $N(f(u),f(v'))$ and $\NN(u,v')$ implies $E(f(u),f(v'))$ for all $v'\in (V\setminus C)^2$. Suppose that $v'\in (V\setminus C)^2$ is a counterexample. We can find $v''\in (V\setminus C)^2$ such that $v_1',v_1''$ and $v_2',v_2''$ belong to the same orbit  and such that $R^{(3)}(u_i,v_i,v_i'')$ for $i\in\{1,2\}$. But then $S^{(3)}(f(u),f(v),f(v''))$, a contradiction. By applying a version of $\sw$ which switches edges and non-edges with respect to $f[C^2]$ to $f$ from the left, we may assume that $f$ behaves like $p_1$ between all $u\in C^2$ and all $v\in (V\setminus C)^2$

Since $f$ does not behave like $R^{(3)}$-$p_1$ on $C^2$, in particular it does not behave like $p_1$ on $C^2$. Pick $u,v\in C^2$ witnessing this. Then $f$ behaves like $p_1$ between any point in $\{u,v\}$ and any point in $(V\setminus C)^2$. Since $(V\setminus C)\cup\{u_i,v_i\}$ induces an isomorphic copy of the random graph for $i\in\{1,2\}$, we can refer to Lemma~\ref{lem:generatingMin:4} to arrive at a contradiction: $f$ generates $e_E$, $e_N$, or a binary injection of type $\mini$ or $\maxi$, all of which violate either $R^{(3)}$ or $S^{(3)}$.
\end{proof}

\begin{definition}
We say that a ternary injective function $f \colon V^3 \to V$ is
\begin{itemize}
\item \emph{of type $R^{(3)}$-majority}
iff for all $u,v,w \in V^3$ with $\NEQNEQNEQ(u,v)$, $\NEQNEQNEQ(u,w)$, $\NEQNEQNEQ(v,w)$ we have $R^{(3)}(f(u),f(v),f(w))$ if and only if $R^{(3)}R^{(3)}R^{(3)}(u,v,w)$, $R^{(3)}R^{(3)}S^{(3)}(u,v,w)$, $R^{(3)}S^{(3)}R^{(3)}(u,v,w)$, or $S^{(3)}R^{(3)}R^{(3)}(u,v,w)$. 
\item \emph{of type $R^{(3)}$-minority} iff 
for all $u,v,w \in V^3$ with $\NEQNEQNEQ(u,v)$, $\NEQNEQNEQ(u,w)$, $\NEQNEQNEQ(v,w)$ we have $R^{(3)}(f(u),f(v),f(w))$ if and only if $R^{(3)}R^{(3)}R^{(3)}(u,v,w)$, $R^{(3)}S^{(3)}S^{(3)}(u,v,w)$, $S^{(3)}R^{(3)}S^{(3)}(u,v,w)$, or $S^{(3)}S^{(3)}R^{(3)}(u,v,w)$. 
\end{itemize}
\end{definition}

\begin{lemma}\label{lem:r3-majority}
A function $f \colon V^3 \to V$ of type $R^{(3)}$-majority does not preserve $R^{(3)}$.
\end{lemma}
\begin{proof}
Let $u^1,u^2,u^3 \in V^4$ be such that 
\begin{itemize}
\item $E(u^1_1,u^1_2)$ and $N(u^1_i,u^1_j)$ for all pairs
$(i,j)$ of distinct elements from $\{1,\dots,4\}$ that
are distinct from $(1,2)$. 
\item $E(u^2_2,u^2_3)$ and $N(u^1_i,u^1_j)$ for all pairs $(i,j)$ of distinct elements from $\{1,\dots,4\}$ that
are distinct from $(2,3)$.  
\item $E(u^3_1,u^3_3)$ and $N(u^3_i,u^3_j)$ for all pairs $(i,j)$ of distinct elements from $\{1,\dots,4\}$ that are distinct from $(1,3)$.  
\end{itemize}
Since $f$ is of type $R^{(3)}$-majority,
we have $S^{(3)}(f(u_1),f(u_2),f(u_4))$,
$S^{(3)}(f(u_1),f(u_3),f(u_4))$, and $S^{(3)}(f(u_2),f(u_3),f(u_4))$. Since for all four-element subsets 
of $V$ there must always be an
even number of three-element subsets in 
$R^{(3)}$, we then have $S^{(3)}(f(x_1),f(x_2),f(x_3))$,
and hence $f$ does not preserve $R^{(3)}$. 
\end{proof}

\begin{lemma}\label{lem:r3-minority}
Let $f \colon V^3 \to V$ be of type $R^{(3)}$-minority. Then $\{f,\sw\}$ generates a function of
type minority.
\end{lemma}
\begin{proof}
Let $g$ be any ternary injection of type minority, and let $u,v,w \in V^3$ with $\NEQNEQNEQ(u,v),\NEQNEQNEQ(u,w),\NEQNEQNEQ(v,w)$ be given. We will show that $R^{(3)}(g(u),g(v),g(w))$
if and only if $R^{(3)}(f(u),f(v),f(w))$. 
Recall that $R^{(3)}(f(u),f(v),f(w))$ 
if and only if
$R^{(3)}S^{(3)}S^{(3)}(u,v,w)$,  
$S^{(3)}R^{(3)}S^{(3)}(u,v,w)$,
$S^{(3)}S^{(3)}R^{(3)}(u,v,w)$, or
$R^{(3)}R^{(3)}R^{(3)}(u,v,w)$.
This is in turn the case if and only if the cardinality of the set $$E \cap \bigcup_{i \in \{1,2,3\}}\{(u_i,v_i),(u_i,w_i),(v_i,w_i)\}$$ is odd. This in turn
is the case if and only if $E \cap \{(g(u),g(v)),(g(u),g(w)),(g(v),g(w))\}$ is odd, which is the case if and only if $R^{(3)}(g(u),g(v),g(w))$ holds. 

By Corollary~\ref{cor:r3-p1}, $f$ generates a binary canonical injection $s(x,y)$ of type $p_1$. Set $t(x,y,z):=s(x,s(y,z))$. As in the proof of 
 Proposition~\ref{prop:generatesMajority}
 the function $p(x,y,z):=f(t(x,y,z),t(y,z,x),t(z,x,y))$ is still of type $R^{(3)}$-minority, and the function $q(x,y,z):=g(t(x,y,z),t(y,z,x),t(z,x,y))$ is still of type minority. Moreover, by the above we have $R^{(3)}(p(u),p(v),p(w))$ if and only if $R^{(3)}(q(u),q(v),q(w))$ for all $u,v,w \in V^3$, since $t$ is injective. Therefore, the homogeneity of $(V;R^{(3)})$ implies that for all finite $S\subseteq V^3$ there exists a unary operation $a$
generated by $\{\sw\}$ such that the ternary function $a(p(x,y,z))$ agrees with $q(x,y,z)$ on $S$. By local closure, $q$ is thus generated by $\{f,\sw\}$.
\end{proof}

\begin{lemma}\label{lem:Ternary2}
Let $\Gamma=(V;R^{(3)},S^{(3)},\ldots)$ be a reduct of $G$ 
such that $H_2$ is not primitive positive definable.
Then $\Gamma$ has a ternary injective polymorphism which violates $H_2$. 
\end{lemma}
\begin{proof} 
Since the relation $H_2$ consists of three orbits of 9-tuples in $\Aut(V;R^{(3)})$, 
Lemma~\ref{lem:arity-reduction} implies that $f$ generates an at most ternary function that violates $H_2$, and hence we can assume that $f$ itself is at most ternary; by adding a dummy variable if necessary, we may assume that $f$ is actually ternary. Moreover, 
    $f$ must certainly be essential, since essentially unary operations that
    preserve $R^{(3)}$ and $S^{(3)}$ are generated by $\{\sw\}$ and hence
    also preserve $H_2$.  
    Corollary~\ref{cor:r3-p1} implies that 
       $\Gamma$ is preserved by a binary canonical injection $g$ of type $p_1$. Consider $$h(x,y,z):= g(g(g(f(x,y,z),x),y),z)\; .$$ 
Then $h$ is clearly injective, and still violates $H_2$ -- the latter can easily be verified combining the facts that $f$ violates $H_2$, $g$ is of type $p_1$, and all tuples in $H_2$ have pairwise distinct entries.
\end{proof}

\begin{proposition}\label{prop:violatesh2}
Let $f$ be an operation on $G$ that preserves
$R^{(3)}$ and $S^{(3)}$ and violates $H_2$.
Then $\{f,\sw\}$ generates a ternary canonical injection of type minority.
\end{proposition}
\begin{proof}
The proof is similar to the proof of Proposition~\ref{prop:generatesMajority}.
By Lemma~\ref{lem:Ternary2}, we can 
assume that $f$ is a ternary injection. 
Because $f$ violates $H_2$, there are $x^1,x^2,x^3 \in H_2$ such that $f(x^1,x^2,x^3) \notin H_2$. In the following, we will write 
$x_i := (x_i^1,x_i^2,x_i^3)$ for $1 \leq i \leq 9$. 
So $(f(x_1),\dots,f(x_9)) \notin H_2$. If there were a
map $a$ generated by $\sw$ such that 
 $a(x^i) = x^j$ for $1\leq i \neq j \leq 3$,
then $\{f,\sw\}$ would generate a binary injection that still violates $H_2$. 
Proposition~\ref{prop:r3-projection} asserts that all binary injections generated 
by $\{f,\sw\}$ are of type $R^{(3)}$-projection, so we have reached a contradiction since 
operations of type $R^{(3)}$-projection preserve $H_2$. By permuting arguments of $f$ if necessary, 
we can therefore 
assume without loss of generality that 

\begin{align*}
	R^{(3)}S^{(3)}S^{(3)}(x_1,x_2,x_3),\, S^{(3)}R^{(3)}S^{(3)}(x_4,x_5,x_6),\,\text{and } S^{(3)}S^{(3)}R^{(3)}(x_7,x_8,x_9).
\end{align*}

We set $$S := \{ y \in V^3 \; | \; \NNN(x_i,y) \text{ for all } 1\leq i \leq 9 \} \; .$$
Consider the ternary relations $Q_1Q_2Q_3$ on $V^3$, where $Q_i \in \{R^{(3)},S^{(3)}\}$ for $1\leq i\leq 3$; each of these relations defines a 3-type in $(V;R^{(3)})$. 
We claim that for fixed  $Q_1Q_2Q_3$, whether or not $R^{(3)}(f(u),f(v),f(w))$ holds for $u,v,w\in S$ with $Q_1Q_2Q_3(u,v,w)$ does not depend on $u,v,w$. We go through all possibilities of $Q_1Q_2Q_3$.
\begin{enumerate}
\item $Q_1Q_2Q_3=R^{(3)}S^{(3)}S^{(3)}$. Let $\alpha \in \Aut(V;R^{(3)})$ be such that $(x^2_1,x^2_2,x^2_3,u_2,v_2,w_2)$ is mapped
to $(x^3_1,x^3_2,x^3_3,u_3,v_3,w_3)$; such an automorphism exists since $\NNN(x_1, u), \NNN(x_1, v),\NNN(x_1,w),
\NNN(x_2, u), \NNN(x_2, v),\NNN(x_2, w)$, and since $(x^2_1,x^2_2,x^2_3)$ has the same type as
$(x^3_1,x^3_2,x^3_3)$, and $(u_2,v_2,w_2)$ has the same type as $(u_3,v_3,w_3)$ in $(V;R^{(3)})$.
By Proposition~\ref{prop:r3-projection}, the operation $g$ defined by $g(x,y):=f(x,y,\alpha(y))$ must
be of type $R^{(3)}$-projection. Hence, $R^{(3)}(g(u_1,u_2),g(v_1,v_2),g(w_1,w_2))$ iff $R^{(3)}(g(x_1^1,x_1^2),g(x_2^1,x_2^2),g(x_3^1,x_3^2))$. Combining this with the equations $(f(u),f(v),f(w))=(g(u_1,u_2),g(v_1,v_2),g(w_1,w_2))$ and 
$(g(x_1^1,x_1^2),g(x_2^1,x_2^2),g(x_3^1,x_3^2))=(f(x_1),f(x_2),f(x_3))$, we get that $R^{(3)}(f(u),f(v),f(w))$ iff $R^{(3)}(f(x_1),f(x_2),f(x_3))$, and so we are done.
\item $Q_1Q_2Q_3=S^{(3)}R^{(3)}S^{(3)}$ or $Q_1Q_2Q_3=S^{(3)}S^{(3)}R^{(3)}$. These cases are analogous to the previous case.
\item $Q_1Q_2Q_3=S^{(3)}R^{(3)}R^{(3)}$. Let $\alpha$ be defined as in the first case. 
By Proposition~\ref{prop:r3-projection}, the operation defined by $f(x,y,\alpha(y))$ must
be of type projection. Reasoning as above, one gets that $R^{(3)}(f(u),f(v),f(w))$ iff $S^{(3)}(f(x_1),f(x_2),f(x_3))$.
\item $Q_1Q_2Q_3=R^{(3)}S^{(3)}R^{(3)}$ or $Q_1Q_2Q_3=R^{(3)}R^{(3)}S^{(3)}$. These cases are analogous to the previous case.
\item $Q_1Q_2Q_3= R^{(3)}R^{(3)}R^{(3)}$ or $Q_1Q_2Q_3=S^{(3)}S^{(3)}S^{(3)}$. These cases are trivial since $f$ preserves $R^{(3)}$ 
and $S^{(3)}$.
\end{enumerate}

To show that $f$ generates an operation of type minority, by Proposition~\ref{prop:RamseyCore} 
it suffices to prove that $f$ generates a function 
of type minority on  
$S$, since $S$ is the product of isomorphic copies of $G$. We show this by another case distinction, based on the fact that 
$(f(x_1),\dots,f(x_9)) \notin H_2$. 
\begin{enumerate}
\item Suppose that $R^{(3)}(f(x_1),f(x_2),f(x_3))$, 
$R^{(3)}(f(x_4),f(x_5),f(x_6))$,
and $R^{(3)}(f(x_7),f(x_8),f(x_9))$. 
By the above, note that $R^{(3)}(f(u),f(v),f(w))$ for $u,v,w \in S$
if and only if
$R^{(3)}S^{(3)}S^{(3)}(u,v,w)$,  
$S^{(3)}R^{(3)}S^{(3)}(u,v,w)$,
$S^{(3)}S^{(3)}R^{(3)}(u,v,w)$, or
$R^{(3)}R^{(3)}R^{(3)}(u,v,w)$.
Hence, $f$ behaves like an $R^{(3)}$-minority on $S$,
and we are done by Lemma~\ref{lem:r3-minority}. 
\item Suppose that 
$S^{(3)}(f(x_1),f(x_2),f(x_3))$, 
$S^{(3)}(f(x_4),f(x_5),f(x_6))$,
and $S^{(3)}(f(x_7),f(x_8),f(x_9))$.
Then $f$ behaves like an $R^{(3)}$-majority on $S$,
which is impossible by Lemma~\ref{lem:r3-majority}. 
\item Suppose that 
$R^{(3)}(f(x_1),f(x_2),f(x_3))$,
$R^{(3)}(f(x_4),f(x_5),f(x_6))$,
and $S^{(3)}(f(x_7),f(x_8),f(x_9))$. 
Let $e$ be a self-embedding of $G$ such that for all $w \in V$, all $1\leq j\leq 3$, and all $1\leq i \leq 9$ 
we have that $N(x_i^j,e(w))$. 
Then $(u_1,u_2,e(f(u_1,u_2,u_3))) \in S$ for all 
$(u_1,u_2,u_3) \in S$. 
Hence, by the above, 
the ternary operation defined by $f(x,y,e(f(x,y,z)))$ is of type $R^{(3)}$-majority on $S$; but this is impossible by Lemma~\ref{lem:r3-majority}.
\item Suppose that 
$R^{(3)}(f(x_1),f(x_2),f(x_3))$, 
$S^{(3)}(f(x_4),f(x_5),f(x_6))$, and $R^{(3)}(f(x_7),f(x_8),f(x_9))$
or $S^{(3)}(f(x_1),f(x_2),f(x_3))$, $R^{(3)}(f(x_4),f(x_5),f(x_6))$, and $R^{(3)}(f(x_7),f(x_8),f(x_9))$. 
These cases are analogous to the previous case.
\end{enumerate}
Let $h(x,y,z)$ be a ternary injection of type minority generated by $f$; 
 it remains to make $h$ canonical. 
 By Corollary~\ref{cor:r3-p1}, $f$ generates a binary canonical injection $g(x,y)$ of type $p_1$. Set $t(x,y,z):=g(x,g(y,z))$. As in the proof of 
 Proposition~\ref{prop:generatesMajority}
 the function $h(t(x,y,z),t(y,z,x),t(z,x,y))$ is still of type minority and canonical.
\end{proof}

\begin{proof}[of Theorem~\ref{thm:sw}]
Assume that $H_2$ is not primitive positive definable; by Theorem~\ref{conf:thm:inv-pol} there exists a polymorphism $f$ of $\Gamma$ that violates $H_2$. 
Since $\Aut(\Gamma)$ contains $\sw$, the relations $R^{(3)}$ and $S^{(3)}$ consist of only one orbit of triples in $\Gamma$. Therefore, since they are preserved by all endomorphisms of $\Gamma$, it follows by Theorem~\ref{conf:thm:inv-pol} and Lemma~\ref{lem:arity-reduction} that
these relations are primitive positive definable in $\Gamma$. 

We can now apply Proposition~\ref{prop:violatesh2}
and obtain that $\{f,\sw\}$ generates a ternary injection of type minority which is canonical as a function from $(V;E)$ to $(V;E)$. 
Corollary~\ref{cor:r3-p1} implies that $\Gamma$
is preserved by a binary injection of type $p_1$ which is canonical as a function from $(V;E)$ to $(V;E)$, and the statement follows from Theorem~\ref{thm:minimal-ops}.
\end{proof}

%% file: binary4.tex
\subsection{When the endomorphisms of a reduct are generated by $\{-,\sw\}$} \label{sect:c4}
We next consider Case~({e}) of Proposition~\ref{prop:endos}. That is, we will assume that the endomorphisms of $\Gamma$ are precisely the unary functions generated by $\{-,\sw\}$. In particular, $\Aut(\Gamma)$ contains $-,\sw$, and
the automorphisms of $\Gamma$ generate
its endomorphisms. The proof for this case is similar to that for Case~({c}) of Proposition~\ref{prop:endos}, presented in Section~\ref{sect:minus}. 

\begin{definition}\label{def:H2p}
Let $H_2'$ be the smallest $9$-ary relation that
is preserved by $-$ and contains $H_2$.
\end{definition}

\begin{proposition}\label{prop:h2prime}
$\Csp(V;H_2')$ is NP-hard. 
\end{proposition}
\begin{proof}
If $H_2'$ is primitive positive definable in 
$\Gamma$, then 
one can show similarly 
as in the proof of Proposition~\ref{prop:h-hard} 
that $\Csp(\Gamma)$ is NP-hard, 
by reduction from positive Not-all-three-equal-3SAT instead of positive 1-in-3-3SAT, 
and by simulating $1$ with $R^{(3)}$ instead of $E$,
and $0$ with $S^{(3)}$ instead of $N$.
\end{proof}

The following is an analog of Theorem~\ref{thm:higherArity} for the situation of this section.

\begin{theorem}\label{thm:minus-sw}
Let $\Gamma$ be a reduct of $G$ whose endomorphisms are precisely the unary functions generated by $\{-,\sw\}$. 
Then $H_2'$ is primitive positive definable
in $\Gamma$, or (b) or (d) from Theorem~\ref{thm:higherArity} applies. 
\end{theorem}
\begin{proof}
Note that $H_2'$ consists of three orbits
of 9-tuples in $\Aut(\Gamma)$, 
and hence, if $H_2'$ is not primitive positive
definable in $\Gamma$, then
there exists by 
Theorem~\ref{conf:thm:inv-pol} 
and Lemma~\ref{lem:arity-reduction} a ternary polymorphism $f$ of
$\Gamma$ that violates $H_2'$.  
That is, there are $t^1,t^2,t^3 \in H_2'$ such
that $f(t^1,t^2,t^3) \notin H_2'$. Note that 
for each $t^j$, either $t^j$ or $-t^j \in H_2$. 
In the first case we set $g_j$ to be the identity function on $V$, in the second
case we let $g_j$ be the operation $-$. 
Now consider the function $f'$ defined by
$f'(x_1,x_2,x_3) := f(g_1(x_1),g_2(x_2),g_3(x_3))$.
We have that $s^{j} := g_j^{-1}(t^j) \in H_2$, but $f'(s^1,s^2,s^3) = f(t^1,t^2,t^3)$ is not in $H_2'$, and therefore not in $H_2$ either.  Hence,
$f'$ violates $H_2$. 
The function 
$h(x):=f'(x,x,x)$ is generated by $\{-,\sw\}$, and hence $h$ either preserves $R^{(3)}$ and $S^{(3)}$,
or it flips them. Since $f'(s^1,s^2,s^3)$ is not in $H_2'$, neither is $-f'(s^1,s^2,s^3)$, and in particular not in $H_2$, so also $-f'$ violates $H_2$.  
Hence, by replacing $f'$ with $-f'$ if necessary, 
we may assume 
that $h$ preserves $R^{(3)}$ and $S^{(3)}$. 

We claim that $f'$ preserves $R^{(3)}$ and $S^{(3)}$. 
Suppose for contradiction that there are $u,v,w \in V^3$ with $R^{(3)}(u_i,v_i,w_i)$ for all $i \in \{1,2,3\}$ such that
$R^{(3)}(f'(u),f'(v),f'(w))$ does not hold;
the case where $f'$ violates $S^{(3)}$ can be treated similarly. 
If $(u_1,v_1,w_1)$, $(u_2,v_2,w_2)$, and $(u_3,v_3,w_3)$ all lie in the same orbit of triples in $G$, then we choose 
$a,b,c \in V$ with $R^{(3)}(a,b,c)$
such that $N(x,y)$ for $x \in \{a,b,c\}$ 
and $y \in \{u_1,v_1,w_1,u_2,v_2,w_2,u_3,v_3,w_3\}$. Then by the homogeneity of $G$ there is for each $i \in \{2,3\}$ a unary operation 
$\alpha_i \in \Aut(G)$ such that
$\alpha_i(u_1,v_1,w_1,a,b,c) = (u_i,v_i,w_i,a,b,c)$. 
We then have that the unary function $g(x):= f'(x,\alpha_2(x),\alpha_3(x))$ maps $(u_1,v_1,w_1) \in R^{(3)}$ to $(f'(u),f'(v),f'(w)) \notin R^{(3)}$. But $g$ and the function $h$ above agree on $\{a,b,c\}$, and hence $g$ preserves $R^{(3)}$ on $\{a,b,c\}$, but violates it on $\{u_1,v_1,w_1\}$. 
This contradicts the assumption that $g$ is generated by $\{-,\sw\}$. 

So suppose in the following that $R^{(3)}(f'(u),f'(v),f'(w))$ for all $u,v,w \in V^3$ with $R^{(3)}(u_i,v_i,w_i)$ for all $i \in \{1,2,3\}$ such that $u,v,w$ belong to the same orbit of triples in $G$. We now show that $R^{(3)}(f'(u),f'(v),f'(w))$ for all $u,v,w \in V^3$ with $R^{(3)}(u_i,v_i,w_i)$ for all $i \in \{1,2,3\}$. 
To this end, note that for each $i \in \{2,3\}$ there is a subset $S_i$ of $\{u_i,v_i,w_i\}$
such that $(\sw_{S_i}(u_i),\sw_{S_i}(v_i),\sw_{S_i}(w_i))$ and $(u_1,v_1,w_1)$ belong to the same orbit in $G$.  
Hence, there is $\beta_i \in \Aut(G)$ such that
$\beta_i(\sw_{S_i}(u_1)) = u_i$,
$\beta_i(\sw_{S_i}(v_1)) = v_i$, and
$\beta_i(\sw_{S_i}(w_1)) = w_i$. 
Pick $a,b,c \in V \setminus \bigcup_{i \in \{1,2,3\}} 
\{u_i,v_i,w_i\}$. 
Note that for both $i \in \{2,3\}$ we have that the triples 
$(a,b,c)$ and $(\sw_{S_i}(a),\sw_{S_i}(b),\sw_{S_i}(c))$ lie in the same orbit. 
We then have that the function $x \mapsto f'(x,\beta_2(\sw_{S_2}(x)),\beta_3(\sw_{S_3}(x)))$ maps $(u_1,v_1,w_1) \in R^{(3)}$ to $(f'(u),f'(v),f'(w)) \notin R^{(3)}$. But the same unary function also maps $(a,b,c) \in R^{(3)}$ to a tuple in $R^{(3)}$ 
since $f'$ by assumption preserves $R^{(3)}$ on tuples 
$R^{(3)}$ that lie in the same orbit, and indeed we have
that for $i \in \{2,3\}$ the triples
$(a,b,c)$ and $(\beta_i(\sw_{S_i}(a)),\beta_i(\sw_{S_i}(b)),\beta_i(\sw_{S_i}(c)))$ lie in the same orbit.  
This again contradicts the assumption that the unary function is generated by
$\{-,\sw\}$. 

We therefore have that $f'$ preserves $R^{(3)}$ and $S^{(3)}$. Since it violates $H_2$, 
Proposition~\ref{prop:h2} implies that $\{f',\sw\}$ generates a ternary canonical injection of type minority,
and we are done. 
\end{proof}

%% file: algorithms.tex
\section{Algorithms}
\label{sect:algorithms}

We now prove that if one of the Cases~(b) to~(f) of Theorem~\ref{thm:higherArity} holds for a reduct $\Gamma$ of $G$ with a finite language, then $\Csp(\Gamma)$ is in P. Tractability of Cases~(b) and (c) is shown in Subsection~\ref{subsect:edgeMinMajUnbalanced}, tractability of Case~(d) in Subsection~\ref{subsect:edgeMinorityBalanced}, of Case~(e) in Subsection~\ref{subsect:edgeMajorityBalanced}, and finally tractability of Case~(f) in Subsection~\ref{subsect:maxMin}.


\subsection{Tractability of types minority / majority with unbalanced projections}\label{subsect:edgeMinMajUnbalanced}

We show tractability of the CSP for reducts $\Gamma$ as in Cases~(b) and~(c) of Theorem~\ref{thm:higherArity}.

\begin{proposition}\label{prop:tract:bc}
    Let $\Gamma$ be a finite language reduct of $G$, and assume that $\Pol(\Gamma)$ contains a ternary injection of type minority or majority, as well as a binary injection which is of type $p_1$ and either $E$-dominated or $N$-dominated in the second argument. Then $\Csp(\Gamma)$ is tractable.
\end{proposition}

It turns out that for such $\Gamma$, we can reduce $\Csp(\Gamma)$ to the CSP of the \emph{injectivization} of $\Gamma$. This implies in turn that the CSP can be reduced to a CSP over a Boolean domain.

\begin{definition}
	A tuple is called \emph{injective} if all its components
	are pairwise distinct.  
    A relation is called \emph{injective} if all its tuples are injective. A structure is called \emph{injective} if all its relations are injective. 
\end{definition}

 With the goal of reducing the CSP to injective structures, we define \emph{injectivizations} for relations, atomic formulas, and structures.

\begin{definition}\label{def:inj}
    \begin{itemize}

        \item Let $R$ be any relation. Then the \emph{injectivization of $R$}, denoted by $\inj(R)$, is the largest injective relation contained in $R$.

        \item Let $\phi(x_1,\ldots,x_n)$ be an atomic formula in the language of a reduct $\Gamma$, where $x_1,\ldots,x_n$ is a list of the variables that appear in $\phi$. Then
        the \emph{injectivization of $\phi(x_1,\dots,x_n)$} is the formula $R^{inj}_\phi(x_1,\ldots,x_n)$, where $R^{inj}_\phi$ is a relation symbol which denotes the injectivization of the relation defined by $\phi$.
        \item For any relational structure $\Gamma$
        with finite language
        we fix a relational structure $\inj(\Gamma)$
        with finite language and
        the same domain as $\Gamma$,
        and whose relations are those defined 
        by the injectivizations of the atomic formulas 
        over $\Gamma$ (note that there are finitely many relations that can be defined in this way). 
        We also call
        $\inj(\Gamma)$ the \emph{injectivization} 
        of $\Gamma$.
    \end{itemize}
\end{definition}

Note that $\inj(\Gamma)$ also contains the injectivizations of relations that are defined by atomic formulas in which one variable might appear several times. In particular, the injectivization of an atomic formula $\phi$ might have smaller arity than the relation symbol that appears in $\phi$.
For example, when $R$ is ternary, then
the atomic formula $R(x,x,y)$ defines a binary relation.

To state the reduction to the CSP of an injectivization, we also need the following operations on instances of $\CSP(\Gamma)$.

\begin{definition}
    Let $\Gamma$ be a structure in a finite language, $\Delta$ be the injectivization of $\Gamma$, and
    $\Phi$ be an instance of $\Csp(\Gamma)$. Then
    the \emph{injectivization of $\Phi$}, denoted by $\inj(\Phi)$, is the instance $\Psi$
    of $\Csp(\Delta)$ obtained from $\Phi$ by replacing each conjunct
    $\phi(x_1,\dots,x_n)$ of $\Phi$ by $R^{inj}_\phi(x_1,\ldots,x_n)$.
\end{definition}

We say that a constraint (=conjunct) in an instance of $\Csp(\Gamma)$ \emph{is false} if
it defines an empty relation in $\Gamma$.
Note that a constraint
 $R(x_1,\dots,x_n)$ might be false
even if the corresponding relation
 $R$ of $\Gamma$ is non-empty (simply because some of the variables from $x_1,\dots,x_n$
 might be equal).

\begin{figure*}[t]
\begin{center}
\small
\fbox{
\begin{tabular}{l}
{\rm // Input: An instance $\Phi$ of CSP$(\Gamma)$ with variables $U$} \\
While $\Phi$ contains a constraint $\phi$ that implies $x=y$ for $x,y \in U$ do \\
\hspace{.5cm} Replace each occurrence of $x$ by $y$ in $\Phi$.  \\
\hspace{.5cm} If $\Phi$ contains a false constraint then reject \\
Loop \\
Accept if and only if $\inj(\Phi)$ is satisfiable in $\Delta$.
\end{tabular}}
\end{center}
\caption{Polynomial-time reduction from the $\Csp(\Gamma)$
for $\Gamma$ closed under an unbalanced binary injection, to the CSP of its injectivization $\Delta$.}
\label{fig:alg-unbalanced}
\end{figure*}

\begin{lemma}\label{lem:inj-reduction}
    Let $\Gamma$ be a finite language reduct of $G$ which is preserved by a
    binary injection $f$ of type $p_1$ that is $E$-dominated or $N$-dominated in the second argument.
    Then the algorithm shown in Figure~\ref{fig:alg-unbalanced}
    is a polynomial-time reduction of $\Csp(\Gamma)$ to $\Csp(\Delta)$, where
    $\Delta$ is the injectivization of $\Gamma$.
\end{lemma}
\begin{proof}
	By duality, we may in the following assume that $f$ is $E$-dominated in the second argument.

    In the main loop, when the algorithm detects a constraint that is false and therefore rejects, then $\Phi$ cannot hold in $\Gamma$, because
     the algorithm only contracts variables $x$ and $y$
    when $x=y$ in all solutions to $\Phi$  -- and contractions are the
    only modifications performed on the input formula $\Phi$.
    So suppose that the algorithm does not reject, and let $\Psi$ be
    the instance of $\Csp(\Gamma)$ computed by the
    algorithm when it reaches the final line of the algorithm.

    By the observation we just made it suffices to show that
    $\Psi$ holds in $\Gamma$
    if and only if $\inj(\Psi)$ holds in $\Delta$.
    It is clear that when $\inj(\Psi)$ holds
    in $\Delta$ then $\Psi$ holds in $\Gamma$ (since the constraints in $\inj(\Psi)$ have been made stronger).
    We now prove that if $\Psi$ has a solution $s$ in $\Gamma$,
    then there is also a solution for $\inj(\Psi)$ in $\Delta$.

    Let
    $s'$ be any mapping from the variables of $\Psi$ to $G$ such that for all distinct variables $x,y$ of $\Psi$ we have that
    \begin{itemize}
    \item if $E(s(x),s(y))$ then $E(s'(x),s'(y))$;
    \item if $N(s(x),s(y))$ then $N(s'(x),s'(y))$;
    \item if $s(x)=s(y)$ then $E(s'(x),s'(y))$.
    \end{itemize}
    Clearly, such a mapping exists. We claim that $s'$ is a solution to $\Psi$
    in $\Gamma$. Since $s'$ must be injective, it is then clearly
    also a solution to $\inj(\Psi)$.

    To prove the claim, let $\phi = R(x_1,\dots,x_n)$ be a constraint in $\Psi$. Since we are at the final stage of the algorithm, we can conclude that
    $\phi$ does not imply equality of any of the variables $x_1,\dots,x_n$,
    and so there is for all $1\leq i < j \leq n$ an $n$-tuple $t^{(i,j)}$ such that $R(t^{(i,j)})$ and
    $t_i \neq t_j$ hold. Since $R(x_1,\ldots,x_n)$ is preserved by a binary injection, it is also preserved by injections of arbitrary arity (it is straightforward to build such terms from a binary injection). Application of an injection of arity $\binom{n}{2}$ to the tuples $t^{(i,j)}$ shows that $R(x_1,\ldots,x_n)$ contains an injective tuple $t=(t_1,\dots,t_n)$.

    Consider the mapping $r \colon \{x_1,\dots,x_n\} \rightarrow G$ given by
    $r(x_l) := f(s(x_l),t_l)$.
    This assignment has the property that  for all $1\leq i,j \leq n$, 
    if $E(s(x_i),s(x_j))$ then $E(r(x_i),r(x_j))$,
    and if $N(s(x_i),s(x_j))$ then $N(r(x_i),r(x_j))$, because $f$ is of type $p_1$ and because the entries of $t$ are distinct.
    Moreover, if $s(x_i)=s(x_j)$ then $E(r(x_i),r(x_j))$ because $f$ is $E$-dominated in the second argument.
    Therefore, $(s'(x_1),\dots,s'(x_n))$ and $(r(x_1),\dots,r(x_n))$ have the same type in $G$.
    Since $f$ is a polymorphism of $\Gamma$, we have that $(r(x_1),\dots,r(x_n))$ satisfies the constraint $R(x_1,\ldots,x_n)$. Hence, $s'$ satisfies
    $R(x_1,\ldots,x_n)$ as well. In this fashion we see that $s'$ satisfies all the constraints of $\Psi$, proving our claim.
\end{proof}

To reduce the CSP for injective structures to Boolean CSPs,
we make the following definition.

\begin{definition}
    Let $t$ be a $k$-tuple of distinct vertices of $G$, and let $q$ be ${k}\choose{2}$.
     Then $\Bool(t)$ is the $q$-tuple $(a_{1,2},a_{1,3},\dots,a_{1,k}$,
    $a_{2,3},\dots,a_{k-1,k}) \in \{0,1\}^q$
    such that $a_{i,j}=0$ if $N(t_i,t_j)$
     and $a_{i,j} = 1$ if $E(t_i,t_j)$.
    If $R$ is a $k$-ary injective relation, then $\Bool(R)$ is the $q$-ary Boolean relation $\{ \Bool(t) \; | \;  t \in R \}$.
    If $\phi$ is a formula that defines a relation
    $R$ over $G$, then we also write $\Bool(\phi)$ instead of $\Bool(\inj(R))$. Finally, for an injective reduct $\Gamma$, we write $\Bool(\Gamma)$ for the structure over a Boolean domain which has the relations of the form $\Bool(R)$, where $R$ is a relation of $\Gamma$.
\end{definition}

\begin{lemma}\label{lem:injectiveToBooleanReduction}
    Let $\Gamma$ be a finite language reduct of $G$ which is injective. Then $\Csp(\Gamma)$ can be reduced to $\Csp(\Bool(\Gamma))$ in polynomial time.
\end{lemma}
\begin{proof}
    Let $\Phi$ be an instance of $\Csp(\Gamma)$, with variable set $W$.
    We create an instance $\Psi$ of $\Csp(\Bool(\Gamma))$ as follows.
    The variable set of $\Psi$ is the set of unordered pairs of variables from $\Phi$.
    When $\phi = R(x_1,\dots,x_k)$ is a constraint in $\Phi$, then $\Psi$ contains the constraint $$\Bool(R)(x_{1,2},x_{1,3},\dots,x_{1,k},x_{2,3},\dots,x_{k-1,k}).$$
    It is straightforward to verify that $\Psi$ can be computed from $\Phi$
    in polynomial time, and that $\Phi$ is a satisfiable instance of $\Csp(\Gamma)$ if and only if $\Psi$ is a satisfiable instance of $\Csp(\Bool(\Gamma))$.
\end{proof}

The \emph{Boolean majority operation} is the unique ternary function $f$ on a Boolean domain satisfying $f(x,x,y)=f(x,y,x)=f(y,x,x)=x$. The \emph{Boolean minority operation} is the unique ternary function $f$ on a Boolean domain satisfying $f(x,x,y)=f(x,y,x)=f(y,x,x)=y$.

\begin{lemma}\label{lem:edgeMajorityImpliesBooleanMajority}
    Let $\Gamma$ be a finite language reduct of $G$ which is injective, and suppose it has an polymorphism of type minority (majority). Then $\Bool(\Gamma)$ has a minority (majority) polymorphism, and hence $\Csp(\Bool(\Gamma))$ can be solved in polynomial time.
\end{lemma}
\begin{proof}
    It is straightforward to show that $\Bool(\Gamma)$ has a minority (majority) polymorphism, and well-known (see~\cite{Schaefer}) that $\Csp(\Bool(\Gamma))$ can then be solved in polynomial time.
\end{proof}

Lemmas~\ref{lem:inj-reduction},~\ref{lem:injectiveToBooleanReduction}, and~\ref{lem:edgeMajorityImpliesBooleanMajority} together provide a proof of Proposition~\ref{prop:tract:bc}.


\subsection{Tractability of type minority with balanced projections}\label{subsect:edgeMinorityBalanced}

We move on to reducts as in Case~(d) of Theorem~\ref{thm:higherArity}.

\begin{proposition}\label{prop:tract:d}
    Let $\Gamma$ be a finite language reduct of $G$, and assume that $\Pol(\Gamma)$ contains a ternary injection of type minority, as well as a binary injection which is of type $p_1$ and balanced. Then $\Csp(\Gamma)$ is tractable.
\end{proposition}

We start by proving that the relations of the reducts under consideration can be defined in $G$ by first-order formulas of a certain restricted syntactic form; this normal form will later be essential for our algorithm.

A Boolean relation is called \emph{affine}
if it can be defined by a conjunction of linear equations modulo 2.
It is well-known that a Boolean relation is affine if and only if it
is preserved by the Boolean minority operation (for a neat proof, see e.g.~\cite{rendezvous}).

In the following, we denote the Boolean exclusive-or connective (xor) by $\oplus$.

\begin{definition}\label{defn:edgeAffine}
    A graph formula is called \emph{edge affine} if it is
    a conjunction of formulas of the form
    \begin{align*}
    x_1 \neq y_1 \; \vee \; & \dots \; \vee \; x_k \neq y_k \\
    \vee \; & \big (u_1 \neq v_1 \wedge \dots \wedge u_l \neq v_l \\
    & \quad \wedge E(u_1,v_1) \oplus \dots \oplus E(u_l,v_l) = p \big) \\
    \vee \; & (u_1=v_1 \wedge \dots \wedge u_l=v_l) \; ,
    \end{align*}
    where $p \in \{0,1\}$, variables need not be distinct, and each of $k$ and $l$ can be $0$.
\end{definition}

\begin{definition}\label{defn:balanced}
    A ternary operation $f \colon  V^3 \rightarrow V$
    is called \emph{balanced} if for every $c \in V$,
    the binary operations $(x,y) \mapsto f(x,y,c)$,
    $(x,z) \mapsto f(x,c,z)$, and $(y,z) \mapsto f(c,y,z)$ are balanced
    injections of type $p_1$.
\end{definition}

We remark that the existence of balanced operations and even balanced minority injections $f$ follows from the fact that $G$ contains all countable graphs as induced subgraphs. To see this, consider the graph defined on $V^3$ which has an edge between 
$(x,y,z)$ and $(x',y',z')$ if and only if 
$f(x,y,z)$ and $f(x',y',z')$ are supposed to have an
edge by the requirement that $f$ is a balanced minority.
Then this graph has an embedding into $G$ by
universality, and this embedding has the desired behavior. 

\begin{proposition}\label{prop:syntax-affine}
    Let $R$ be a relation with a first-order definition over $G$.
    Then the following are equivalent:
    \begin{enumerate}
        \item $R$ can be defined by an edge affine formula;
        \item $R$ is preserved by every ternary injection which is of type minority and balanced;
        \item $R$ is preserved by some ternary injection of type minority, and some balanced binary injection of type $p_1$.
    \end{enumerate}
\end{proposition}

\begin{proof}
    We first show the implication from $(1)$ to $(2)$, that $n$-ary relations $R$ defined by edge affine formulas $\Psi(x_1,\dots,x_n)$ are preserved by balanced injections $f$ of type minority.
    By injectivity of $f$, it is easy to see that we only have to show this for
    the case that $\Psi$ does not contain disequality disjuncts
    (i.e., $k=0$).
    Now let $\phi$ be a clause from $\Psi$, say
    \begin{align*}
    \phi :=  & \big (u_1 \neq v_1 \wedge \dots \wedge u_l \neq v_l \\
    & \quad \wedge \; (E(u_1,v_1) \oplus \dots \oplus E(u_l,v_l) = p) \big ) \\
    & \vee \; (u_1=v_1 \wedge \dots \wedge u_l=v_l) \; ,
    \end{align*}
	for $p \in \{0,1\}$ and $u_1,\dots,u_l,v_1,\dots,v_l \in \{x_1,\dots,x_n\}$. In the following, it will sometimes be notationally convenient to consider
    tuples in $G$ satisfying a formula as mappings from the variable set of the formula to $V$.
    Let $t_1,t_2,t_3 \colon \{x_1,\dots,x_n\} \rightarrow V$ be three mappings that satisfy $\phi$. We have to show that
     the mapping $t_0 \colon \{x_1,\dots,x_n\} \rightarrow V$ defined by  $t_0(x) = f(t_1(x),t_2(x),t_3(x))$ satisfies
    $\phi$.

    Suppose first that each of $t_1,t_2,t_3$ satisfies $u_1 \neq v_1 \wedge \dots \wedge u_l \neq v_l$. In this case,
    $t_0(u_1) \neq t_0(v_1) \wedge \dots \wedge t_0(u_l) \neq t_0(v_l)$, since $f$ preserves $\neq$.
     Note that $E(t_0(u_i),t_0(v_i))$, for $1 \leq i \leq l$,
     if and only if $E(t_1(u_i),t_1(v_i)) \oplus E(t_2(u_i),t_2(v_i)) \oplus E(t_3(u_i),t_3(v_i)) = 1$.
     Therefore, since each $t_1,t_2,t_3$ satisfies
     $E(u_1,v_1) \oplus \dots \oplus E(u_l,v_l) = p$,
     we find that $t_0$ also satisfies $E(u_1,v_1) \oplus \dots \oplus E(u_l,v_l) = p \oplus p \oplus p = p$.

    Next, suppose that one of $t_1,t_2,t_3$ satisfies $u_i = v_i$
    for some (and therefore for all) $1\leq i \leq l$.
    By permuting arguments of $f$, we can assume
    that $t_1(u_i)=t_1(v_i)$ for all $i \in \{1, \dots, l\}$.
    Since the function $f$ is balanced, the operation $g: (y,z) \mapsto f(t_1(u_i),y,z)$ is a balanced injection of type $p_1$.
    Suppose that $t_2(u_i)=t_2(v_i)$.
    Then
    $E(t_0(u_i),t_0(v_i))$ if and only if $E(t_3(u_i),t_3(v_i))$, since
    $g$ is balanced.
    Hence, $t_0$ satisfies $\phi$.
    Now suppose that $t_2(u_i) \neq t_2(v_i)$. If also $t_3(u_i) \neq t_3(v_i)$, then $E(t_0(u_i),t_0(v_i))$
    if and only if $E(t_2(u_i),t_2(v_i))$ since $g$ is of type $p_1$. If on the other hand $t_3(u_i) = t_3(v_i)$, then again $E(t_0(u_i),t_0(v_i))$
    if and only if $E(t_2(u_i),t_2(v_i))$ since $g$ is balanced. In either case, $t_0$ satisfies $\phi$.
    This shows that $f$ preserves $\phi$, and hence also $\Psi$.

    The implication from $(2)$ to $(3)$ is trivial, since every balanced injection of type minority generates
    a balanced binary injection of type $p_1$ by identification of two of its variables. It is also here that we have to check the existence of balanced  injections of type minority; as mentioned above, this follows easily from the universality of $G$.

    We show the implication from $(3)$ to $(1)$ by induction on the arity
    $n$ of the relation $R$.
    Let $g$ be the balanced binary injection of type $p_1$,
    and let $h$ be the operation of type minority.
    For $n=2$ the statement of the theorem holds, because all binary
    relations with a first-order definition in $G$ can be defined over $G$
    by expressions as in Definition~\ref{defn:edgeAffine}:
    \begin{itemize}
        \item For $x \neq y$ we set $k=1$ and $l=0$.
        \item For $\neg E(x,y)$ we can set $k=0, l=1,p=0$.
        \item For $\neg N(x,y)$ we can set $k=0, l=1,p=1$.
        \item Then, $E(x,y)$ can be expressed as $(x \neq y) \wedge \neg N(x,y)$.
        \item $N(x,y)$ can be expressed as $(x \neq y) \wedge \neg E(x,y)$.
        \item $x=y$ can be expressed as $\neg E(x,y) \wedge \neg N(x,y)$.
        \item The empty relation can be expressed as $E(x,y) \wedge N(x,y)$.
        \item Finally, $V^2$ can be defined by the empty conjunction.
    \end{itemize}

    For $n>2$, we construct the formula $\Psi$ that defines
    the relation $R(x_1,\dots,x_n)$ as follows.
    If there are distinct $i,j \in \{1,\dots,n\}$ such that for all  tuples
    $t$ in $R$ we have $t_i=t_j$, consider the relation
    defined by $\exists x_i. R(x_1,\dots,x_n)$. This relation is also
    preserved by $g$ and $h$, and by
    inductive assumption has a definition $\Phi$ as required.
    Then the formula $\Psi := (x_i = x_j \wedge \Phi)$ proves the claim.
    So let us assume that for all distinct $i,j$ there is a tuple $t \in R$
    where $t_i \neq t_j$. Note that since $R$ is preserved by the binary injective operation
    $g$, this implies that $R$ also contains an injective tuple.

    Since $R$ is preserved by an operation of type minority,
    the relation $\Bool(\text{inj}(R))$ is preserved by the Boolean minority
    operation, and hence has a definition by a conjunction of
    linear equations modulo 2.
    From this definition it is straightforward
    to obtain a definition $\Phi(x_1,\dots,x_n)$ of $\text{inj}(R)$ which is the conjunction
    of $\bigwedge_{i < j \leq n} x_i \neq x_j$ and of
    formulas of the form
    $$E(u_1,v_1) \oplus \dots \oplus E(u_l,v_l) = p\; ,$$
    for $u_1,\dots,u_l,v_1,\dots,v_l \in \{x_1,\dots,x_n\}$.
    It is clear that we can assume that none of the formulas of the form $E(u_1,v_1) \oplus \dots \oplus E(u_l,v_l) = p$ in $\Phi$
    can be equivalently replaced by a conjunction
    of shorter formulas of this form.

    For all $i,j \in \{1,\dots,n\}$ with $i<j$,
    let $R_{i,j}$ be the relation that holds for the tuple $(x_1,\ldots,x_{i-1},x_{i+1},\ldots,x_n)$ iff $R(x_1,\dots,x_{i-1},x_j,x_{i+1},\dots,x_n)$ holds. Because $R_{i,j}$ is preserved by $g$ and $h$, but has arity $n-1$, it has a definition
    $\Phi_{i,j}$ as in the statement by inductive assumption.
    We call the conjuncts of $\Phi_{i,j}$ also the \emph{clauses}
    of $\Phi_{i,j}$.
    We add to each clause of $\Phi_{i,j}$ a disjunct $x_i \neq x_j$.

    Let $\Psi$ be the conjunction composed of conjuncts from the following two groups:
    \begin{enumerate}
        \item all the modified clauses from all formulas $\Phi_{i,j}$;
        \item when $\phi = (E(u_1,v_1) \oplus \dots \oplus E(u_l,v_l) = p)$ is a conjunct
        of $\Phi$, then
        $\Psi$ contains the formula
        \begin{align*} & (u_1 \neq v_1 \wedge \dots \wedge u_l \neq v_l \;\wedge \; \phi) \\
        \vee & (u_1=v_1 \wedge \dots \wedge u_l=v_l) \; .
        \end{align*}
    \end{enumerate}
    Obviously, $\Psi$ is a formula in the required form.
    We have to verify that $\Psi$ defines $R$.

    Let $t$ be an $n$-tuple such that $t \notin R$.
    If $t$ is injective, then
    $t$ violates a formula of the form
    $$E(u_1,v_1) \oplus \dots \oplus E(u_l,v_l) = p$$
    from the formula $\Phi$ defining $\inj(R)$, and hence it violates a conjunct of $\Psi$ of the second group.
    If there are $i,j$ such that $t_i=t_j$
    then the tuple $t^i:=(t_1,\dots,t_{i-1},t_{i+1},\dots,t_n) \notin R_{i,j}$.
    Therefore some conjunct $\phi$ of
    $\Phi_{i,j}$ is not satisfied by $t^i$, and $\phi \vee x_i \neq x_j$
    is not satisfied by $t$. Thus, in this case $t$ does not satisfy $\Psi$ either.

    It remains to verify that all $t \in R$ satisfy $\Psi$.
    Let $\psi$ be a conjunct of $\Psi$ created from some clause in $\Phi_{i,j}$.
    If $t_i \neq t_j$, then $\psi$ is satisfied by $t$ because $\phi$
    contains $x_i \neq x_j$. If $t_i = t_j$, then $(t_1,\dots,t_{i-1},t_{i+1},\dots,t_n) \in R_{i,j}$ and thus this tuple satisfies $\Phi_{i,j}$. This also implies that $t$ satisfies $\psi$.
    Now, let $\psi$ be a conjunct of $\Psi$ from the second group.
    We distinguish three cases.
    \begin{enumerate}
    \item For all $1 \leq i \leq l$ we have that $t$ satisfies $u_i = v_i$. In this case we are clearly
    done since $t$ satisfies the second disjunct of $\psi$.

    \item For all $1 \leq i \leq l$ we have that $t$ satisfies $u_i \neq v_i$.
    Suppose for contradiction that $t$ does not satisfy $E(u_1,v_1) \oplus \dots \oplus E(u_l,v_l)=p$. Let $r\in R$ be injective, and consider the tuple $s:=g(t,r)$. Then $s\in R$, and $s$ is injective since the tuple $r$ and the function $g$ are injective. However, since $g$ is of type $p_1$, we have $E(s(u_i),s(v_i))$ if and only if $E(t(u_i),t(v_i))$, for all $1\leq i\leq l$. Hence, $s$ violates the conjunct $E(u_1,v_1) \oplus \dots \oplus E(u_l,v_l)=p$ from $\Phi$, a contradiction since $s\in\inj(R)$.

    \item The remaining case is that there is a proper non-empty subset $S$ of $\{1,\dots,l\}$ such that $t$ satisfies $u_i=v_i$ for all $i \in S$ and $t$ satisfies $u_i \neq v_i$ for all
    $i \in \{1,\dots,n\} \setminus S$. We claim that this case cannot occur.
    Suppose that all tuples $t'$ from $\inj(R)$ satisfy that
    $\bigoplus_{i \in S} E(u_i,v_i) = 1$. In this case we could have
    replaced $E(u_1,v_1) \oplus \dots \oplus E(u_l,v_l)=p$ by the two shorter formulas
    $\bigoplus_{i \in S} E(u_i,v_i) = 1$ and $\bigoplus_{i \in  [n]\setminus S}
    E(u_i,v_i) = p \oplus 1$, in contradiction to our assumption on $\Phi$.
     Therefore there is a tuple $s \in \inj(R)$ where $\bigoplus_{i \in S} E(u_i,v_i) = 1$.
    Now, for the tuple $g(t,s)$ we have
    \begin{align*}
    \bigoplus_{i \in [n]} E(u_i,v_i) = & \; \bigoplus_{i \in S} E(u_i,v_i) \oplus \bigoplus_{i \in [n] \setminus S} E(u_i,v_i) \\
    = & \; 1 \oplus p \\
    \neq & \; p 
    \end{align*}
    which is a contradiction since $g(t,s) \in \inj(R)$.
    \end{enumerate}
    Hence, all $t \in R$ satisfy all conjuncts $\psi$ of $\Psi$. We conclude
    that $\Psi$ defines $R$.
\end{proof}

We now present a polynomial-time algorithm for $\Csp(\Gamma)$
for the case that a reduct $\Gamma$ has finitely many edge affine relations.

\begin{definition}
    Let $\Gamma$ be a finite language reduct of $G$ which has only edge affine relations, and let $\Phi$ be an instance of $\Csp(\Gamma)$.
    Then the \emph{graph of $\Phi$} is the (undirected) graph
    whose vertices are unordered pairs of distinct variables of $\Phi$, and which
    has an edge between distinct sets $\{a,b\}$ and $\{c,d\}$ if $\Phi$ contains a
    constraint whose definition as in Definition~\ref{defn:edgeAffine}
    has a conjunct of the form
    \begin{align*}
    & \big (u_1 \neq v_1 \wedge \dots \wedge u_l \neq v_l
    \wedge (E(u_1,v_1) \oplus \dots \oplus E(u_l,v_l) = p) \big) \\
    \vee \; & (u_1=v_1 \wedge \dots \wedge u_l=v_l)
    \end{align*}
    such that $\{a,b\}=\{u_i,v_i\}$ and $\{c,d\} = \{u_j,v_j\}$ for
    some $i,j \in \{1,\dots,l\}$.
\end{definition}

It is clear that for $\Gamma$ with finite signature,
the graph of an instance $\Phi$ of $\Csp(\Gamma)$ can be computed
in linear time from $\Phi$.

\begin{definition}
     Let $\Gamma$ be a finite language reduct of $G$ which has only edge affine relations, and let $\Phi$ be an instance of $\Csp(\Gamma)$.
    For a set $C$ of 2-element subsets of variables of $\Phi$, we
    define $\inj(\Phi,C)$ to be the following affine Boolean
    formula.
    The set of variables of $\inj(\Phi,C)$ is
    $C$.
    The constraints of $\inj(\Phi,C)$ are obtained from
    the constraints $\phi$ of $\Phi$ as follows.
    If $\phi$ has a definition as in Definition~\ref{defn:edgeAffine} with a clause of the form
    \begin{align*}
    & \big (u_1 \neq v_1 \wedge \dots \wedge u_l \neq v_l
    \wedge (E(u_1,v_1) \oplus \dots \oplus E(u_l,v_l) = p) \big) \\
    \vee \; & (u_1=v_1 \wedge \dots \wedge u_l=v_l)
    \end{align*}
    where all pairs $\{u_i,v_i\}$ are in $C$,
    then $\inj(\Phi,C)$ contains the conjunct
    $\{u_1,v_1\} \oplus \dots \oplus \{u_l,v_l\} = p$.
\end{definition}

\begin{figure*}[t]
\begin{center}
\small
\fbox{
\begin{tabular}{l}
    {\rm // Input: An instance $\Phi$ of $\Csp(\Gamma)$ with variables $U$} \\
    Repeat \\
    \hspace{.3cm} For each connected component $C$ of the graph of $\Phi$ do \\
    \hspace{.6cm} Let $\Psi$ be the affine Boolean formula $\inj(\Phi,C)$. \\
    \hspace{.6cm} If $\Psi$ is unsatisfiable then \\
    \hspace{.9cm} For each $\{x,y\} \in C$ do \\
    \hspace{1.2cm} Replace each occurrence of $x$ by $y$ in $\Phi$. \\
    \hspace{.9cm} If $\Phi$ contains a false constraint then reject \\
    \hspace{.3cm} Loop \\
    Until $\inj(\Phi,C)$ is satisfiable for all components $C$ \\
    Accept
\end{tabular}}
\end{center}
\caption{A polynomial-time algorithm
for $\Csp(\Gamma)$ when $\Gamma$ is preserved by a balanced operation of type minority.}
\label{fig:balanced-minority}
\end{figure*}

Proposition~\ref{prop:tract:d} now follows from the following lemma and Proposition~\ref{prop:syntax-affine}.

\begin{lemma}\label{lem:alg-minority}
    Let $\Gamma$ be a finite language reduct of $G$ which has only edge affine relations. Then the algorithm shown in Figure~\ref{fig:balanced-minority} solves $\Csp(\Gamma)$ in polynomial time.
\end{lemma}

\begin{proof}
    We first show that when the algorithm detects a constraint that
    is false and therefore rejects in the innermost loop, then $\Phi$ must be unsatisfiable. Since variable contractions are the
    only modifications performed on the input formula $\Phi$,
    it suffices to show that the algorithm only
     equates variables $x$ and $y$
    when $x=y$ in all solutions.
    To see that this is true, assume that $\Psi := \inj(\Phi,C)$
    is an unsatisfiable Boolean formula for some connected component $C$. Hence, in any solution $s$ to $\Phi$ there must be a $\{x,y\}$ in
    $C$ such that $s(x)=s(y)$. It follows immediately from the definition
    of the graph of $\Phi$ that then $s(u)=s(v)$ for all
    $\{u,v\}$ adjacent to $\{x,y\}$ in the graph of $\Phi$.
    By connectivity of $C$, we have that $s(u)=s(v)$ for all $\{u,v\} \in C$.
    Since this holds for any solution to $\Phi$, the contractions in the
    innermost loop of the algorithm preserve satisfiability.

    So we only have to show that when the algorithm accepts,
    there is indeed a solution to $\Phi$. When the algorithm accepts,
    we must have that $\inj(\Phi,C)$ has a solution $s_C$ for all components
    $C$ of the graph of $\Phi$.
    Let $s$ be a mapping from the variables of $\Phi$ to $V$ such that $E(x_i,x_j)$ if $\{x_i,x_j\}$ is in component $C$ of the graph of $\Phi$ and $s_C(\{x_i,x_j\})=1$, and $N(x_i,x_j)$ otherwise.
    It is straightforward to verify that this assignment satisfies all conjuncts of $\Phi$. This solution also
    gives a solution to the original input instance
    by setting variables $x$ that have been replaced
    by $y$ during the course of the algorithm to
    the same value as $y$.
\end{proof}


\subsection{Tractability of type majority with balanced projections}\label{subsect:edgeMajorityBalanced}

We turn to reducts as in Case~(e) of Theorem~\ref{thm:higherArity}.

\begin{proposition}\label{prop:tract:e}
    Let $\Gamma$ be a finite language reduct of $G$, and assume that $\Pol(\Gamma)$ contains a ternary injection of type majority, as well as a binary injection which is of type $p_1$ and balanced. Then $\Csp(\Gamma)$ is tractable.
\end{proposition}

\begin{definition}
A formula is called \emph{graph bijunctive}
iff it is a conjunction of \emph{graph bijunctive clauses}, i.e., formulas of the form
\begin{align*}
x_1 \neq y_1 & \vee \cdots \vee x_k \neq y_k \vee \phi
\end{align*}
where $\phi$ is 
\begin{enumerate}
\item[(i)] of the form $u_1 = v_1$, 
\item[(ii)] of the form $L_1(u_1,v_1)$, 
\item[(iii)] of the form $L_1(u_1,v_1) \vee L_2(u_2,v_2)$, 
\item[(iv)] of the form $L_1(u_1,v_1) \vee u_1 = v_1$, or
\item[(v)] of the form $(L_1(u_1,v_1) \vee u_1=v_1 \vee L_2(u_2,v_2)) \wedge (u_1 \neq v_1 \vee L_2(u_2,v_2) \vee u_2 = v_2)$.
\end{enumerate}
for $L_1,L_2 \in \{E,N\}$, and $k \geq 0$. 
\end{definition}

\vspace{.3cm}
Note that when $M_1,M_2$ are such that $\{L_i,M_i\} = \{E,N\}$ for $i \in \{1,2\}$, then the graph bijunctive clause in item (v) can be equivalently written in the form 
$$(M_1(u_1,v_1) \Rightarrow L_2(u_2,v_2)) \wedge (u_1=v_1 \Rightarrow \neg M_2(u_2,v_2)) \; .$$ 

\begin{proposition}\label{prop:syntax}
Let $R$ be a relation with a first-order definition in $G$. Then the following are equivalent.
\begin{enumerate}
\item $R$ can be defined by a graph bijunctive formula;
\item $R$ is preserved by every ternary injection which is of type majority and balanced; 
\item $R$ is preserved by some 
ternary injection of type majority and some binary balanced injection of type $p_1$.
\end{enumerate}
\end{proposition}
\begin{proof}
We first show the equivalence of (1) and (2). 

For the implication (1) $\Rightarrow$ (2), 
let $\psi$ be a graph bijunctive clause.
It suffices to show that $\psi$ is preserved by every balanced injection $f$ of type majority. 
Let $t_1,t_2,t_3$ be three tuples that satisfy $\psi$.
If $\psi$ contains 
an inequality disjunct $x_i \neq y_i$,
and one of $t_1,t_2,t_3$ satisfies $x_i \neq y_i$,
then by injectivity of $f$ we have that $t_0 = f(t_1,t_2,t_3)$ satisfies $x_i \neq y_i$ and therefore also $\psi$. So we can focus on the case $k = 0$, i.e.,  
$\psi$ does not contain any inequality disjunct. 
If $\psi$ is of the form $u_1=v_1$, $\psi$ is clearly preserved. If $\psi$ is of the form $L_1(u_1,v_1)$ or 
of the form $\neg L_1(u_1,v_1)$, then $f$ preserves $\psi$ since it is of type majority and balanced. Suppose now that $\psi$ is of the form $L_1(u_1,v_1) \vee L_2(u_2,v_2)$ for $L_1,L_2 \in \{E,N\}$. 
Then at least two of $t_1,t_2,t_3$ satisfy $L_1(u_1,v_1)$, or at least two of $t_1,t_2,t_3$ satisfy $L_2(u_2,v_2)$. In the former case, $t_0$ satisfies $L_1(u_1,v_1)$, in the latter case $t_0$ satisfies $L_2(u_2,v_2)$, since $f$ is of type majority and balanced. 

Finally, suppose that $\psi$ is as in item (v) of the definition of graph bijunctive formulas.
If $t_0 = f(t_1,t_2,t_3)$ satisfies $\neg M_1(u_1,v_1) \wedge u_1 \neq v_1$, then $t_0$ satisfies
both conjuncts of $\psi$ and we are done. We thus may assume that $t_0$ satisfies either $u_1 = v_1$ or $M_1(u_1,v_1)$.

If $t_0$ satisfies $u_1 = v_1$,
then $t_0$ satisfies the first conjunct of $\psi$.
By injectivity of $f$ we must have that all of 
$t_1,t_2,t_3$ satisfy $u_1=v_1$, and therefore
all three tuples satisfy $L_2(u_2,v_2) \vee u_2=v_2$. Since $f$ is of type majority and balanced, also $t_0$ satisfies $L_2(u_2,v_2) \vee u_2=v_2$, which is the second conjunct of $\psi$, and we are done also in this case.

Suppose now that $t_0$ satisfies $M_1(u_1,v_1)$.
Since $f$ is of type majority and balanced, either 
\begin{enumerate}
\item[(a)] at least two out of $t_1,t_2,t_3$ satisfy $M_1(u_1,v_1)$, or 
\item[(b)] $t_1$ satisfies $M_1(u_1,v_1)$ and exactly one out
of $t_2,t_3$ satisfy $u_1=v_1$, or
\item[(c)] $t_1$ satisfies $u_1=v_1$ and $t_2$ satisfies $M_1(u_1,v_1)$.
\end{enumerate}

If at least two tuples out of $t_1,t_2,t_3$ satisfy
$M_1(u_1,v_1)$, then they also satisfy
$L_2(u_2,v_2)$, and
so does $t_0$ since $f$ is of type majority and balanced. We conclude that $t_0$ satisfies $\psi$. 

Now assume (b). Then $t_1$ satisfies $M_1(u_1,v_1)$, and therefore also satisfies $L_2(u_2,v_2)$. 
Moreover, one of $t_2,t_3$ satisfies $u_1 = v_1$, and therefore also
$L_2(u_2,v_2) \vee u_2 = v_2$. Since $f$ is balanced and of type majority we have that 
$t_0$ satisfies $L_2(u_2,v_2)$, and therefore also $\psi$.

Suppose finally that (c) holds, i.e., $t_1$ satisfies $u_1=v_1$ and $t_2$ satisfies $M_1(u_1,v_1)$.
In this case $t_1$ satisfies $L_2(u_2,v_2) \vee u_2 = v_2$ and $t_2$ satisfies
$L_2(u_2,v_2)$. Again, since $f$ is balanced and of type majority, we have that 
$t_0$ satisfies $L_2(u_2,v_2)$, and therefore also $\psi$.

We next show the implication (2) $\Rightarrow$ (1). 
Let $R$ be a relation preserved by a ternary injection $f$ which is of type majority and balanced.
Let $\Phi$ be a formula in CNF 
that defines $R$ over $(V;E,N)$ such that 
all literals of $\Phi$ are of the form $E(x,y)$, $N(x,y)$,
$x \neq y$, or $x=y$. This can be achieved by replacing literals of the form $\neg L(x,y)$ by $M(x,y) \vee x=y$, for $M$ such that $\{L,M\} = \{E,N\}$. 
Also suppose that 
$\Phi$ is \emph{minimal} in the sense that no clause $\phi$ of $\Phi$ can be replaced by a set of clauses such that
\begin{enumerate}
\item each replacing clause has fewer literals of the form $L(x,y)$ for $L \in \{E,N\}$ than $\phi$, or
\item each replacing clause has the same number of literals of the form $L(x,y)$, but fewer
literals of the form $x=y$ than $\phi$, or
\item each replacing clause has the same number of literals of the form $L(x,y)$ and of the form $x=y$, but fewer literals of the form $x \neq y$ than $\phi$. 
\end{enumerate}

Let $\Psi$ be the set of all graph bijunctive clauses that are implied by $\Phi$. To prove  {(2) $\Rightarrow$ (1)}, it suffices to show that  $\Psi$ implies
all clauses $\phi$ of $\Phi$. Let $\phi$ such a clause. 
In the entire proof we make the convention
that $L_1,\dots,L_n$ denote elements of 
$\{E,N\}$, and
$M_1,\dots,M_n$ are such that 
$\{L_i,M_i\} = \{E,N\}$, for all $i \leq n$.\\

{\it Observation 1: } The clause $\phi$ cannot contain two different literals of the form $x_1=y_1$ 
and $x_2=y_2$. 
Otherwise,
since $\Phi$ is minimal, the formula obtained by removing $x_1=y_1$ from $\phi$ is inequivalent to $\Phi$, and hence there exists a tuple $t_1$ that 
satisfies $\Phi$, and none of the literals in $\phi$ except for $x_1=y_1$. Similarly, there exists
a tuple $t_2$ that satisfies $\Phi$, and none of the literals in $\phi$ except for $x_2 = y_2$. 
By the injectivity of $f$, the tuple 
$t_0 = f(t_1,t_2,t_2)$ satisfies $x_1 \neq y_1$
and $x_2 \neq y_2$. 
Moreover, $t_0$ does not satisfy any other literal of $\phi$
because the fact that it is of type majority and balanced implies that $f$ preserves the negations of all literals of the form $x = y$, 
$E(x,y)$, $N(x,y)$, and $x \neq y$.
Therefore, $t_0$ satisfies
none of the literals in $\phi$, 
contradicting
the assumption that $f$ preserves $\Phi$. \\

{\it Observation 2: } The clause $\phi$ contains at most two literals of the form $L(x,y)$, where $L\in\{E,N\}$. Suppose to the contrary that $\phi$ contains three different literals of the form
$L_1(x_1,y_1)$, $L_2(x_2,y_2)$, 
and $L_3(x_3,y_3)$. Let $\theta$
be the clause obtained from $\phi$ by
removing those three literals from $\phi$.
Note that it is impossible that $\Phi$ has
satisfying assignments $t_1,t_2,t_3$ with
\begin{align*}
t_1 \models & M_2(x_2,y_2) \wedge M_3(x_3,y_3) \wedge \neg \theta \\
t_2 \models & M_1(x_1,y_1) \wedge M_3(x_3,y_3) \wedge \neg \theta \\
t_3 \models & M_1(x_1,y_1) \wedge M_2(x_2,y_2) \wedge \neg \theta \; .
\end{align*}
Otherwise, $t_0 = f(t_1,t_2,t_3)$ satisfies 
$M_1(x_1,y_1) \wedge M_2(x_2,y_2) \wedge M_3(x_3,y_3)$ since $f$ is of type majority and balanced. Moreover, $t_0$ satisfies $\neg \theta$, since $f$ preserves the negations of
literals of the form $x = y$, 
$E(x,y)$, $N(x,y)$, and $x \neq y$. 
Therefore, $t_0$ does not satisfy 
$\phi$, in contradiction to the assumption that $f$ preserves $\Phi$.

Suppose without loss of generality that there 
is no satisfying assignment $t_1$ as above. 
In other words,
$\Phi$ implies the clause 
\begin{align} \theta \vee L_2(x_2,y_2) \vee (x_2 = y_2) \vee L_3(x_3,y_3) \vee (x_3 = y_3) \; . \label{eq:no-good}
\end{align}
Note that $\Phi$ also implies the clauses
\begin{align} \theta \vee L_1(x_1,y_1) \vee L_2(x_2,y_2) \vee (x_3 \neq y_3)
\label{eq:eq1}
\end{align} 
\begin{align} \theta \vee L_1(x_1,y_1) \vee (x_2 \neq y_2) \vee L_3(x_3,y_3)
\label{eq:eq2}
\end{align} 
since they are obvious weakenings of $\phi$.
We claim that the clauses in (\ref{eq:no-good}), (\ref{eq:eq1}), and  (\ref{eq:eq2}) together imply $\phi$.
To see this, suppose they hold for a tuple $t$ which does not satisfy $\phi$. Then $t$ satisfies neither $\theta$ nor any of the $L_i$, and hence it satisfies both $(x_2 \neq y_2)$ and $(x_3 \neq y_3)$, by (\ref{eq:eq1}) and  (\ref{eq:eq2}). On the other hand, in this situation (\ref{eq:no-good}) implies $x_2=y_2\vee x_3=y_3$, a contradiction. Hence $\phi$ is equivalent to the conjunction of these three clauses. Now replacing $\phi$ by this conjunction in $\Phi$, we arrive at a contradiction to the minimality of $\Phi$.

Taking the two observations together, we conclude that $\phi$ contains at most one literal of the form $x=y$, and at most two literals of the 
form $L(x,y)$. If it has no literal of the form $x=y$ or no literal of the form $L(x,y)$ then it is itself graph bijunctive and hence an element of $\Psi$, and we are done. So assume henceforth that  
$\phi$ contains a literal $x_1=y_1$
and a literal of the form $L_2(x_2,y_2)$. It may or may not contain at most one more literal $L_3(x_3,y_3)$; all other literals of $\phi$ are of the form $x \neq y$.

Let us first consider the case where $\phi$ does not
contain the literal $L_3(x_3,y_3)$. 
Let $\theta$ be the clause obtained from $\phi$ by removing $x_1=y_1$ and $L_2(x_2,y_2)$; all
literals in $\theta$ are of the form $x \neq y$. 
We claim that $\Phi$ implies the following formula.
\begin{align}
\theta \vee (x_1 \neq y_1) \vee L_2(x_2,y_2) \vee (x_2=y_2)
\label{eq:important}
\end{align}
To show the claim, suppose for contradiction
that there is a tuple $t_1$ that
satisfies $\Phi \wedge \neg \theta \wedge (x_1 = y_1) \wedge M_2(x_2,y_2)$. By minimality of $\Phi$, there is also a
tuple $t_2$ that satisfies $\Phi \wedge \neg \theta \wedge (x_1 \neq y_1) \wedge L_2(x_2,y_2)$. Then 
$f(t_1,t_1,t_2)$ satisfies $\Phi \wedge \neg \theta \wedge x_1 \neq y_1 \wedge M_2(x_2,y_2)$ since $f$ is of type majority and balanced; but this is a contradiction
since such a tuple does not satisfy $\phi$. 
We next show that $\Phi$ implies the graph bijunctive formulas
\begin{align}
& \theta \vee (E(x_1,y_1) \vee x_1 = y_1 \vee L_2(x_2,y_2)) \wedge (x_1 \neq y_1 \vee L_2(x_2,y_2) \vee x_2 = y_2) 
\label{eq:case1-1} \\
& \theta \vee (N(x_1,y_1) \vee x_1 = y_1 \vee L_2(x_2,y_2)) \wedge (x_1 \neq y_1 \vee L_2(x_2,y_2) \vee x_2 = y_2) \; .
\label{eq:case1-2}
\end{align}
Since $\Phi$ implies~(\ref{eq:important}),
it suffices to show that $\Phi$ implies $\theta \vee E(x_1,y_1) \vee (x_1 = y_1) \vee L_2(x_2,y_2)$
and $\theta \vee N(x_1,y_1) \vee (x_1 = y_1) \vee L_2(x_2,y_2)$. But this is clear since those formulas are weakenings of $\phi$.
Hence, the formulas~(\ref{eq:case1-1}) and~(\ref{eq:case1-2}) are in $\Psi$.
As
$E(x_1,y_1) \vee (x_1 = y_1) \vee L_2(x_2,y_2)$
and 
$N(x_1,y_1) \vee (x_1 = y_1) \vee L_2(x_2,y_2)$
implies $(x_1 = y_1) \vee L_2(x_2,y_2)$,
the formulas~(\ref{eq:case1-1}) 
and (\ref{eq:case1-2}) 
imply $\phi$, and therefore $\Psi$ implies $\phi$. 

Finally, we consider the case where $\phi$ also contains a literal $L_3(x_3,y_3)$. 
Let $\theta$ be the clause obtained from $\phi$ by removing $x_1=y_1$, $L_2(x_2,y_2)$, and $L_3(x_3,y_3)$; all literals of $\theta$ are of the form $x \neq y$.  If $\Phi$ implies $\theta \vee \neg M_2(x_2,y_2)$, then we could have replaced
$\phi$ by the two clauses $\theta \vee L_2(x_2,y_2) \vee (x_2 = y_2)$ and 
$\theta \vee (x_1 = y_1) \vee (x_2 \neq y_2) \vee L_3(x_3,y_3)$ 
which together imply $\phi$, in contradiction to the minimality of $\Phi$. The same argument shows that $\Phi$ does not imply $\theta \vee \neg M_3(x_3,y_3)$.
Now observe that $\Phi$ implies the following.
\begin{align}
& \theta \vee x_1 = y_1 \vee x_2 \neq y_2 \vee x_3 \neq y_3 \label{eq:easy1} \\
& \theta \vee \neg M_2(x_2,y_2) \vee \neg M_3(x_3,y_3) \label{eq:easy2} \\
& \theta \vee x_2 \neq y_2 \vee L_3(x_3,y_3) \vee x_3 = y_3 \label{eq:guard1} \\
& \theta \vee x_3 \neq y_3 \vee L_2(x_2,y_2) \vee x_2 = y_2\; .
\label{eq:guard2}
\end{align}
This is obvious for~(\ref{eq:easy1}). For~(\ref{eq:easy2}), assume otherwise that there is an assignment $t$ satisfying $\Phi \wedge \neg \theta \wedge M_2(x_2,y_2) \wedge M_3(x_3,y_3)$. By minimality of
$\Phi$ there is also an assignment $t'$ satisfying 
$\Phi \wedge \neg \theta \wedge (x_1 \neq x_2)$. 
Then $f(t,t,t')$ satisfies none of the literals of $\phi$,
a contradiction. We now show that~(\ref{eq:guard1})
is implied; the proof for~(\ref{eq:guard2}) is symmetric.
Assume otherwise that $t$ satisfies 
$\Phi \wedge \neg \theta \wedge (x_2 = y_2) \wedge M_3(x_3,y_3)$.
There also exists a tuple $t'$
that satisfies $\Phi \wedge \neg \theta \wedge M_2(x_2,y_2)$ 
since $\Phi$ does not imply $\theta \vee \neg M_2(x_2,y_2)$ as we have observed above. Then $f(t,t,t')$ satisfies
$\neg \theta \wedge M_2(x_2,y_2) \wedge M_3(x_3,y_3)$,
which contradicts~(\ref{eq:easy2}).

We now claim that $\Phi$ also implies at least one of the following two formulas. 
\begin{align}
& \theta \vee L_2(x_2,y_2) \vee x_2=y_2 \vee L_3(x_3,y_3)
\label{eq:first} \\
& \theta \vee L_3(x_3,y_3) \vee x_3=y_3 \vee L_2(x_2,y_2)\; .
 \label{eq:second}
\end{align}
Otherwise, there would be a tuple $t$ satisfying
$\Phi \wedge \neg \theta \wedge M_2(x_2,y_2) \wedge \neg L_3(x_3,y_3)$ and a tuple $t'$ satisfying
$\Phi \wedge \neg \theta \wedge \neg L_2(x_2,y_2) \wedge M_3(x_3,y_3)$. Then $f(t,t',t')$ would satisfy
$\neg \theta \wedge M_2(x_2,y_2) \wedge M_3(x_3,y_3)$, which is impossible by~(\ref{eq:easy2}). Suppose without loss of generality that
$\Phi$ implies $\theta \vee L_2(x_2,y_2) \vee (x_2=y_2) \vee L_3(x_3,y_3)$. Since $\Phi$
also implies~(\ref{eq:guard1}),
we have that 
$\Psi$ contains the graph bijunctive formula
\begin{align}
\theta \vee \big ((L_2(x_2,y_2) \vee x_2=y_2 \vee L_3(x_3,y_3)) \wedge (x_2 \neq y_2 \vee L_3(x_3,y_3) \vee x_3 = y_3) \big) \; . \label{eq:last}
\end{align}
We finally show that $\Psi$ implies $\phi$.
Let $t$ be a tuple that satisfies $\Psi$.
If $t$ satisfies $\theta \vee (x_1 = y_1)$ there is nothing to show, so suppose otherwise. 
Then~(\ref{eq:easy1}), which is graph bijunctive and thereofore in $\Psi$, implies that either $x_2 \neq y_2$ or $x_3 \neq y_3$. If $x_2 \neq y_2$, then
by the first conjunct in~(\ref{eq:last}) we have
that $L_2(x_2,y_2)$ or $L_3(x_3,y_3)$, in which case $t$
satisfies $\phi$ and we are done. Otherwise, suppose
that $x_2 = y_2$. Then $x_3 \neq y_3$ as we have seen above. 
But then the second conjunct in~(\ref{eq:last}) implies 
that $L_3(x_3,y_3)$, and we are again done. 

The implication from (2) to (3) is trivial, since every balanced injection of type majority generates a balanced binary injection of type $p_1$ by identification of two of its variables. For the implication from (3) to (2), let $t$ be the ternary injection of type majority, and $p$ the binary balanced injection of type $p_1$. 
Set $s(x, y, z) := t(p(x, y), p(y, z), p(z, x))$ 
and $w(x, y, z) := s(p(x, y), p(y, z), p(z, x))$, 
and observe that $w$ is of type majority and balanced.  
\end{proof}

\begin{proposition}
Let $\Gamma$ be a reduct of $(V;E)$ with finite
relational signature, and suppose that $\Gamma$
has a balanced ternary polymorphism of type majority. 
Then $\CSP(\Gamma)$ can be solved in polynomial time.
\end{proposition}
\begin{proof}
Let $\Phi$ be an instance of $\CSP(\Gamma)$ with variables $S$, and let
$\Psi$ be the set of clauses obtained from $\Phi$ by replacing each constraint by its graph bijunctive definition over $(V;E,N)$ which exists by Proposition~\ref{prop:syntax}. 
Clearly, $\Phi$ is satisfiable in $\Gamma$ if and only if $\Psi$ is satisfiable in $(V;E,N)$. 


We associate to $\Psi$ a 2SAT instance $\psi = \psi(\Psi)$ as follows. 
For each unordered pair $\{u,v\}$ 
of distinct variables $u,v$ of $\Psi$ we have a
variable $x_{\{u,v\}}$ in $\psi(\Psi)$. Then
\begin{itemize}
\item if $\Psi$ contains the clause $E(u,v)$ or the clause $E(u,v) \vee u=v$ then $\psi(\Psi)$ contains the clause $\{x_{\{u,v\}}\}$;
\item if $\Psi$ contains the clause $N(u,v)$ or the clause $N(u,v) \vee u=v$ then $\psi(\Psi)$ contains the clause $\{\neg x_{\{u,v\}}\}$;
\item if $\Psi$ contains the clause $N(a,b) \vee E(c,d)$ then $\psi(\Psi)$ contains the clause $\{\neg x_{\{a,b\}},x_{\{c,d\}}\}$. Clauses of the form $L_1(a,b) \vee L_2(c,d)$ are translated correspondingly for all $L_1,L_2 \in \{E,N\}$;
\item if $\Psi$ contains the clause 
$(N(a,b) \vee a=b \vee E(c,d)) \wedge (a \neq b \vee E(c,d) \vee c = d)$
 then $\psi(\Psi)$ contains the clause $\{\neg x_{\{a,b\}}, x_{\{c,d\}}\}$. Clauses of the form $(L_1(u_1,v_1) \vee u_1=v_1 \vee L_2(u_2,v_2)) \wedge (u_1 \neq v_1 \vee L_2(u_2,v_2) \vee u_2 = v_2)$ are translated correspondingly for all $L_1,L_2 \in \{E,N\}$.
\end{itemize}
All other clauses of $\Psi$ are ignored for the definition of $\psi(\Psi)$. 

We recall an important and well-known 
concept to decide 
satisfiability of 2SAT instances $\psi$.
If $\psi$ contains clauses of size one,
we can reduce to the case where all clauses have size two by replacing the clause $\{x\}$ by $\{x,x\}$. 
 The
\emph{implication graph} $G_\psi$ of a conjunction $\psi$ of propositional clauses of size two is the directed
graph whose vertices $T$ are the variables $x,y,z,\dots$ of $\psi$, and the negations 
$\neg x, \neg y, \neg z$ of the variables. 
The edge set of $G_\psi$ contains $(x,x') \in V^2$ 
if $\psi$ contains the clause $\{\neg x, x'\}$ 
(here we identify $\neg ( \neg x)$ with $x$).
It is well-known that $\psi$ is unsatisfiable if and only if there exists $x \in T$ such that
$x$ and $\neg x$ belong to the same strongly connected component (SCC) of $G_\psi$. 

\begin{figure*}[h]
\begin{center}
\small
\fbox{
\begin{tabular}{l}
{\rm // Input: A set of graph bijunctive clauses $\Psi$} 
\\
Do \\
\hspace{0.5cm} While $\Psi$ contains a clause of the form $u=v$ do  \\
\hspace{1cm} Replace each occurrence of $v$ by $u$ in $\Psi$.  \\
\hspace{1cm} Remove literals of the form $E(u,u)$, $N(u,u)$, and $u \neq u$ from $\Psi$. \\
\hspace{1cm} If $\Psi$ contains an empty clause then reject. \\
\hspace{.5cm} Loop. \\
\hspace{.5cm} Compute the 2SAT instance $\psi = \psi(\Psi)$, and the graph $G_\psi$. \\
\hspace{.5cm} If $G_\psi$ contains $x_{\{u,v\}}$ such that $x_{\{u,v\}}$ and $\neg x_{\{u,v\}}$ are in the same SCC then \\
\hspace{1cm} Replace each occurrence of $v$ by $u$ in $\Psi$.  \\
\hspace{1cm} Remove literals of the form $E(u,u)$, $N(u,u)$, and $u \neq u$ from $\Psi$. \\
\hspace{1cm} If $\Psi$ contains an empty clause then reject. \\
Loop until $\Psi$ does not change any more. \\
Accept.
\end{tabular}}
\end{center}
\caption{Polynomial-time algorithm to test satisfiability of a given set of graph bijunctive clauses.}
\label{fig:graph-bijunctive-alg}
\end{figure*}

Now consider the algorithm displayed in Figure~\ref{fig:graph-bijunctive-alg}. We make the following claims. 
\begin{enumerate}
\item Whenever the algorithm replaces all occurrences of a variable $v$ in $\Psi$ by a variable $u$, then $u$ and $v$ must have the same value in all solutions of $\Psi$.
\item When the algorithm rejects an instance, then $\Psi$ is unsatisfiable. 
\item When the algorithm accepts, then
the input formula indeed is indeed satisfiable. 
\end{enumerate}
The first claim can be shown inductively over
the execution of the algorithm as follows.
When the algorithm replaces all occurrences of $v$ by $u$ in line 4 of the algorithm, the first claim is trivially true. The only other variable contraction can
be found in line 10 of the algorithm. 

So let $\Psi$ be the set of graph bijunctive clauses when we reach line 10, and 
suppose that $x_{\{u,v\}}$ and $\neg x_{\{u,v\}}$
lie in the same SCC
of $G_\psi$.
Since $x_{\{u,v\}}$ and $\neg x_{\{u,v\}}$ belong to the same SCC, there is a path $x_{\{u,v\}}=x_0,x_1,\dots,x_n=\neg x_{\{u,v\}}$ from $x_{\{u,v\}}$ 
to $\neg x_{\{u,v\}}$, and a path $\neg x_{\{u,v\}}=y_0,y_1,\dots,y_m=x$ from $\neg x_{\{u,v\}}$ to $x_{\{u,v\}}$.

Suppose that $\Psi$ has a solution $s \colon S \rightarrow V$. We have to show that $s(u)=s(v)$. 
Suppose otherwise that $s(u) \neq s(v)$;
without loss of generality, $E(s(u),s(v))$ holds. 
Let $\{u_i,v_i\}$ be the pair of variables of $\Phi$ that corresponds to $x_i$. We show by induction on $i$ that if $x_i$ is positive, then $E(s(u_i),s(v_i))$, and if $x_i$ is negative
then $N(s(u_i),s(v_i))$. 
Suppose without loss of generality that $x_i$
is positive, and suppose inductively that $E(s(u_i),s(v_i))$. 
There is a clause in
$\Psi$ that contributed the edge 
$(x_i,x_{i+1})$ to $G_\psi$. If $x_{i+1}$ is a positive literal, then this clause is
 either of the form $N(u_i,v_i) \vee E(u_{i+1},v_{i+1})$,
 or of the form
$$(N(u_i,v_i) \vee u_i=v_i \vee E(u_{i+1},v_{i+1})) \wedge (u_i \neq v_i \vee E(u_{i+1},v_{i+1}) \vee u_{i+1} = v_{i+1}) \; .\
 $$
 In both cases, the clause together with
 $E(s(u_i),s(v_i))$ implies that $E(s(u_{i+1}),s(v_{i+1}))$. The argument in the case
 that $x_{i+1}$ is a negative literals is similar. 
 For $i+1=n$ we obtain that
 $N(s(u),s(v))$, in contradiction to our assumption.
Therefore, 
we conclude that $s(u)=s(v)$, which concludes the proof of the first claim. 

Since the only modifications to $\Psi$ are variable contractions, the first claim implies that when at some stage during the execution of the algorithm the formula $\Psi$ contains an empty clause, then there is indeed no solution to the original input formula; this proves the second claim. 

To prove the third claim, suppose that 
the algorithm accepts. Let $\psi = \psi(\Psi)$ be
the 2SAT instance in the final round of the main loop of the algorithm, and let $T$ be the set of variables of $\psi$. 
The 2SAT formula $\psi$ must have a solution, since
otherwise the algorithm would have changed $\Phi$,
in contradiction to our assumptions.
From a solution 
$t \colon T \to \{0,1\}$ for $\psi$ we obtain a solution
$s \colon S \to V$ for the clause set $\Psi$ at the end of the execution of the algorithm by assigning distinct vertices of $V$ to every variable of $\Psi$ such that
$(s(u),s(v)) \in E$ 
if and only if $s(x_{\{u,v\}})=0$. We also get a solution to the originally given set of clauses (before contractions of variables) by setting contracted variables to the same value. 

The three claims show the correctness of the algorithm. It is easy to see that the algorithm can be
implemented in polynomial (in fact, in quadratic) time in the input size. 
\end{proof}


\subsection{Tractability of types max and min}\label{subsect:maxMin}

We are left with proving tractability of the CSP for reducts $\Gamma$ as in Case~(f) of Theorem~\ref{thm:higherArity}, i.e., for reducts which have a canonical binary injective polymorphism of type $\maxi$ or $\mini$. We first observe that we can assume that this polymorphism is either balanced, or of type $\maxi$ and $E$-dominated, or of type $\mini$ and $N$-dominated.

\begin{proposition}\label{prop:maxMinStandardizing}
    Let $\Gamma$ be a reduct of $G$. If $\Gamma$ has a canonical binary injective polymorphism of type $\maxi$, then it also has a canonical binary injective polymorphism of type $\maxi$ which is balanced or $E$-dominated. If it has a canonical binary injective polymorphism of type $\mini$, then it also has a canonical binary injective polymorphism of type $\mini$ which is balanced or $N$-dominated.
\end{proposition}
\begin{proof}
    We prove the statement for type $\maxi$ (the situation for $\mini$ is dual). Let $p$ be the polymorphism of type $\maxi$.
    Then $h(x,y):=p(x,p(x,y))$ is not $N$-dominated in the first argument: we only have to show that if $u=(u_1,u_2)$, $v=(v_1,v_2)$ in $V^2$
are so that $\EEQ(u,v)$ holds, then $E(h(u),h(v))$ holds. 
To see this, note that $\ENEQ((u_1,p(u)),(v_1,p(v))$, 
and thus the application of $p$ to
$(u_1,p(u))$ and $(v_1,p(v))$, respectively, yields two elements of $V$
connected by an edge, since $p$ is of type $\maxi$.

Since $h(x,y)$ is not $N$-dominated in the first argument, it is either $E$-dominated or balanced in the first argument. Suppose it is
$E$-dominated in the first argument. Let $u,v \in V^2$ such that
$\NEQEQ(u,v)$. 
Then $\ENEQ(h(x,y)h(y,x))$, and since $p$ is of type $\maxi$, we
have $E(p(h(x,y),h(y,x)))$, so $h$ is $E$-dominated in the first argument.
The argument for the second argument follows since 
$p(h(x,y),h(y,x))$ is
symmetric in its arguments. 
The situation where $h$ is balanced in the
first argument implies that $p(h(x,y),h(y,x))$ 
is balanced in both
arguments, by a similar argument. 
   Therefore, $p(h(x,y),h(y,x))$ is either balanced or $E$-dominated, and still of type $\maxi$.
\end{proof}

We will need the following result which was shown in~\cite[Proposition~14]{Maximal}.
For a relational structure $\Gamma$, we denote
by $\hat \Gamma$ the expansion of $\Gamma$ that also contains
the complement for each relation in $\Gamma$. We call a homomorphism between two structures $\Gamma$ and $\Delta$ \emph{strong} if it is
also a homomorphism between $\hat \Gamma$ and $\hat \Delta$.

\begin{proposition}\label{prop:alg}
    Let $\Gamma$ be an $\omega$-categorical homogeneous structure such that $\Csp(\hat \Gamma)$ is tractable,
    and let $\Delta$ be a reduct of $\Gamma$.
    If $\Delta$ has a polymorphism which is a strong homomorphism from $\Gamma^2$ to $\Gamma$, then
    $\Csp(\Delta)$ is tractable as well.
\end{proposition}

In the following, a strong homomorphism from a power
of $\Gamma$ to $\Gamma$ will be called \emph{strong polymorphism}. We apply Proposition~\ref{prop:alg} to our setting as follows.

\begin{proposition}\label{prop:horn}
    Let $\Gamma$ be a reduct of $G$ with a finite signature, and which is preserved by a binary canonical injection which is of type $\maxi$ and balanced or $E$-dominated, or of type $\mini$ and balanced or $N$-dominated. Then $\Csp(\Gamma)$ can be solved in polynomial time.
\end{proposition}

\begin{proof}
    We have the following.

    \begin{itemize}
        \item A canonical binary injection which is of type $\mini$ and $N$-dominated is a strong polymorphism of $(V;E,=)$.
        \item A canonical binary injection which is of type $\maxi$ and $E$-dominated is a strong polymorphism of $(V;N,=)$.
        \item A canonical binary injection which is of type $\maxi$ and balanced is a strong polymorphism of $(V;\neg E,=)$.
        \item A canonical binary injection which is of type $\mini$ and balanced is a strong polymorphism of $(V;\neg N,=)$.
    \end{itemize}

    The tractability result follows from Proposition~\ref{prop:alg}, because $$\Csp(V;E,\neg E,N,\neg N,=,\neq)$$ can be solved in polynomial time.
    One way to see this is to verify that all
    relations are preserved by a balanced polymorphism of type majority,
    and to use the algorithm presented in Section~\ref{subsect:edgeMajorityBalanced}.
\end{proof}

This completes the proof of the dichotomy statement of Theorem~\ref{thm:main}! 

%% file: classification.tex
\section{Classification}
\label{sect:classification}
We have proven so far that all reducts of the random graph with finitely many relations define a CSP which is either tractable or NP-complete. This section is devoted to a more explicit description of the border between tractable and hard reducts.

\begin{definition}
    Let $B$ be a behavior for functions from $G^2$ to $G$. A ternary injection $f \colon V^3\To V$ is \emph{hyperplanely of type $B$} if the binary functions $(x,y)\mapsto f(x,y,c)$, $(x,z)\mapsto f(x,c,z)$, and $(y,z)\mapsto f(c,y,z)$ have behavior $B$ for all $c\in V$.
\end{definition}

We have already met a special case of this concept in Definition~\ref{defn:balanced} of Section~\ref{subsect:edgeMinorityBalanced}: a ternary function is balanced if and only if it is hyperplanely balanced and of type $p_1$.
Let us now define some more behaviors of binary functions which will appear 
hyperplanely in ternary functions in our classification.

\begin{definition}
    A binary injection $f \colon V^2\To V$ is 
    \begin{itemize}
        \item \emph{$E$-constant} if the image of $f$ is a clique;
        \item \emph{$N$-constant} if the image of $f$ is an independent set;
        \item of type \emph{$\xnor$} if for all $u,v\in V^2$ with $\NEQNEQ(u,v)$ the relation $E(f(u),f(v))$ holds if and only if $\EE(u,v)$ or $\NN(u,v)$ holds;
        \item of type \emph{$\xor$} if for all $u,v\in V^2$ with $\NEQNEQ(u,v)$ the relation $E(f(u),f(v))$ holds if and only if neither $\EE(u,v)$ nor $\NN(u,v)$ hold.
    \end{itemize}
\end{definition}

Observe that if two canonical functions $f, g \colon V^n \To V$ satisfy the same type conditions, then they generate the same clone. This follows easily from the homogeneity of $G$ and by local closure.

Let $I_6$ be the 6-ary relation defined by
\begin{align*} \{(x_1,x_2,y_1,y_2,z_1,z_2) \in V^6\; | \; &
 (x_1=x_2 \wedge y_1 \neq y_2 \wedge z_1 \neq z_2) \\
  & \vee  \;
 (x_1 \neq x_2 \wedge y_1 = y_2 \wedge z_1 \neq z_2) \\
 & \vee  \; (x_1 \neq x_2 \wedge y_1 \neq y_2 \wedge z_1 = z_2) \} \; .
 \end{align*}
It is easy to see that the polymorphisms of $I_6$ are precisely the essentially unary operations which after deletion of all dummy variables are injective.

Similarly, we define relations $E_6$ and $N_6$ by altering the above definition and replacing all occurrences of $\neq$ by $E$ and $N$, respectively. One can show that polymorphisms of $E_6$ are essentially unary and preserve $E$; on the other hand, as opposed to the situation for $I_6$, they need not be injective.

\begin{theorem}\label{thm:minimalTractableClones}
Let $\Gamma$ be a reduct of $G$. Then 
either one of the relations $E_6$, $N_6$, $I_6$, $H_1$, $H_1'$,  $H_2$, or $H_2'$ 
has a primitive positive definition in $\Gamma$,
or 
$\Gamma$ has a canonical polymorphism of one of the following 17 types. 
 \begin{enumerate}
        \item \label{mt:const} A constant operation.
        \item \label{mt:maxbalanced} A balanced binary injection of type $\maxi$.
        \item \label{mt:minbalanced} A balanced binary injection of type $\mini$.
        \item \label{mt:maxEdominated} An $E$-dominated binary injection of type $\maxi$.
        \item \label{mt:minNdominated} An $N$-dominated binary injection of type $\mini$.
        \item \label{mt:majHpProjBalanced} A ternary injection of type majority which is hyperplanely balanced and of type projection.
        \item \label{mt:majHpNfanatic} A ternary injection of type majority which is hyperplanely $E$-constant.
        \item \label{mt:majHpNphobic} A ternary injection of type majority which is hyperplanely $N$-constant.
        \item \label{mt:majHpMaxEdominated} A ternary injection of type majority which is hyperplanely of type $\maxi$ and $E$-dominated.
        \item \label{mt:majHpMinNdominated} A ternary injection of type majority which is hyperplanely of type $\mini$ and $N$-dominated.
        \item \label{mt:minHpProjBalanced} A ternary injection of type minority which is hyperplanely balanced and of type projection.
        \item \label{mt:minHpProjEdominated} A ternary injection of type minority which is hyperplanely of type projection and $E$-dominated.
        \item \label{mt:minHpProjNdominated} A ternary injection of type minority which is hyperplanely of type projection and $N$-dominated.
        \item \label{mt:minHpMatchfanaticEdominated} A ternary injection of type minority which is hyperplanely balanced of type $\xnor$.
        \item \label{mt:minHpMatchphobicNdominated} A ternary injection of type minority which is hyperplanely balanced of type $\xor$.
        \item \label{mt:Nfanatic} A binary injection which is $E$-constant.
        \item \label{mt:Nphobic} A binary injection which is $N$-constant.
    \end{enumerate}
\end{theorem}

\begin{proof}
Assume that none of the relations $E_6$, $N_6$, $I_6$, $H_1$, $H_1'$,  $H_2$, or $H_2'$ 
has a primitive positive definition in $\Gamma$. Then $\Gamma$ has polymorphisms violating these relations.

Consider the case where all polymorphisms of $\Gamma$ are essential unary. Then let $f$ be a unary polymorphism violating $I_6$; clearly, $f$ cannot be injective, so say without loss of generality that it sends two adjacent vertices to the same vertex. Now let $g$ be a unary polymorphism violating $N_6$. Then $g$ sends some non-edge to an edge, or to a single vertex. By virtue of $f$ and $g$, we then get that in either case $\Gamma$ also has a unary polymorphism $h$ which sends a non-edge to a single vertex.
Now a standard iterative argument using local closure shows that $\Gamma$ has a constant polymorphism, and we are done.

So assume henceforth that $\Gamma$ has an essential polymorphism. By Lemma~5.3.10 in~\cite{Bodirsky-HDR},
$\Gamma$ also has a binary essential polymorphism $f$. 

We apply Proposition~\ref{prop:endos}. There is nothing to show when the first case of that proposition holds, i.e., when $\Gamma$ has a constant endomorphism. 

Assume the second case holds, i.e.,  $\Gamma$ has the endomorphism $e_E$ or $e_N$; without loss of generality, we consider the case where $e_E$ preserves $\Gamma$. Then consider the structure $\Delta$ induced in $\Gamma$ on the image $e_E[V]$. 
This structure $\Delta$ is invariant under all permutations of its domain,
and hence is first-order definable in $(e_E[V];=)$. 
It follows from the results in~\cite{ecsps} that it either has a 
constant polymorphism, or a binary injection, or all polymorphisms of $\Delta$ are essentially unary. 
The structure $\Delta$ cannot have a constant endomorphism as otherwise also $\Gamma$ has a constant polymorphism by composing the constant of $\Delta$ with $e_E$. We now show that $\Delta$ has an essential operation. 
Suppose that $f(a,a)=f(a,b)$ for all $a,b \in V$ with $E(a,b)$. 
We claim that $f(u,u)=f(u,v)$ for \emph{every} $u,v \in V$. To see this,
let $w \in V$ be such that $E(u,w)$ and $E(v,w)$. 
Then $f(u,u)=f(u,w)=f(u,v)$, as required. 
It follows that $f$ does not depend
on its first variable, a contradiction. Hence, 
there exist $a,b \in V$ such that $E(a,b)$ and $f(a,a) \neq f(a,b)$. 
Similarly, there exist $c,d \in V$ such that $E(c,d)$ and $f(c,c) \neq f(d,c)$. 
Let $T$ be an infinite clique adjacent to $a,b,c,d$. 
Then $f$ is either essential on $T \cup \{a,b\}$ or on $T \cup \{c,d\}$, both cliques. Suppose without loss of generality
that $f$ is essential on $T \cup \{a,b\}$. Since
all operations with the same behavior as $e_E$ generate each other,
we can also assume that the image of $e_E$ equals $T \cup \{a,b\}$.
Then the 
restriction of the mapping $(x_1,x_2) \mapsto e_E(f(x_1,x_2))$ to $e_E[V]$
is an essential polymorphism of $\Delta$. 
Hence, the above-mentioned result from~\cite{ecsps} implies that
$\Delta$ has a binary injective polymorphism $h'$. 
Then $h(x,y):=h'(e_E(x),e_E(y))$ is a polymorphism of $\Gamma$. 
But $h$ is a binary injection which is $E$-constant, 
and so $\Gamma$ has a polymorphism from Item~16 of our list. 
The argument when $\Gamma$ is preserved by $e_N$ is similar,
with Item~17 instead of Item~16.

It remains to discuss the last four cases of Proposition~\ref{prop:endos}. Consider the very last case, i.e., where the endomorphisms of $\Gamma$ are generated by $\Aut(G)$. Then Theorem~\ref{thm:higherArity} applies, and recall that we assume that $H_1$ has no primitive positive definition in $\Gamma$, excluding the first case of that theorem. 
If $\Gamma$ has a binary canonical injective polymorphism of type $\maxi$ or $\mini$, then by Proposition~\ref{prop:maxMinStandardizing} 
one of items~2 to~5 applies. 
Otherwise, $\Gamma$ has a ternary injective polymorphism $t$ of type  minority or majority, and one of the binary canonical injective polymorphisms of type projection listed in Theorem~\ref{thm:higherArity} -- denote it by $p$. Set $$w(x,y,z):=t(p(p(x,y),p(y,z)),p(p(y,z),p(z,x)),p(p(z,x),p(x,y)))\; .
$$ Then the function $w$ has one of the behaviors that describe functions from Items~6 to~15 -- which of the behaviors depends on the precise behavior of $p$, and is shown in Figure~\ref{fig:classification-correspondence}. We leave the verification to the reader. 

When the endomorphisms of $\Gamma$ are generated by the function $- \colon V\To V$, then we may refer to Theorem~\ref{thm:minus}, which brings us back to the preceding case. Similarly, when the endomorphisms of $\Gamma$ are generated by $\sw$ or by $\{-,\sw\}$, then we may refer to Theorems~\ref{thm:sw} and~\ref{thm:minus-sw}, respectively, concluding the proof.
\end{proof}

\begin{figure*}
\begin{center}
{\small
 \begin{tabular}{|l|l|l|}
    \hline
    Binary injection type $p_1$ & Type majority & Type minority\\
    \hline
    Balanced & Hp.\  balanced, type $p_1$ & Hp.\  balanced, type $p_1$\\
    $E$-dominated & Hp.\  $E$-constant & Hp.\  type $p_1$, $E$-dominated\\
    $N$-dominated & Hp.\  $N$-constant & Hp.\  type $p_1$, $N$-dominated\\
    Balanced in 1st, $E$-dom. in 2nd arg.& Hp.\  type $\maxi$, $E$-dom. & Hp.\  type $\xnor$, balanced. \\
    Balanced in 1st, $N$-dom. in 2nd arg.& Hp.\  type $\mini$, $N$-dom. & Hp.\  type $\xor$, balanced. \\
    \hline
 \end{tabular}
 }
\end{center}
\caption{Minimal tractable canonical functions of type majority / minority and their corresponding canonical binary injections of type projection.} \label{fig:classification-correspondence}
\end{figure*}

The following is an operational tractability criterion for reducts of $G$. 

\begin{corollary}\label{cor:operational-main}
Let $\Gamma$ be a reduct of $G$ with finite relational signature. Then:
\begin{itemize}
\item  either $\Gamma$ has a canonical polymorphism of one of the 17 types listed in Theorem~\ref{thm:minimalTractableClones},\\ and $\Csp(\Gamma)$ is tractable, or
\item one of the relations $E_6$, $N_6$, $I_6$, $H_1$, $H_1'$, $H_2$, $H_2'$ has a primitive positive definition in $\Gamma$, and $\Csp(\Gamma)$ is NP-complete.
\end{itemize}
\end{corollary}
\begin{proof}
First suppose that one of the relations $E_6$, $N_6$, $I_6$, $H_1$, $H_1'$, $H_2$, $H_2'$ has a primitive positive definition in $\Gamma$. 
In the case of $H_1$, NP-hardness of $\Csp(\Gamma)$ follows from Proposition~\ref{prop:h-hard}, in the case of $H_1'$ from Proposition~\ref{prop:h1prime}, in the case of $H_2$ from Proposition~\ref{prop:h2}, and in the case of $H_2'$ from Proposition~\ref{prop:h2prime}. In the case of $I_6$, NP-hardness of $\Csp(\Gamma)$ follows
from~\cite{ecsps}. NP-hardness for the relations $E_6$, $N_6$ can be shown similarly as for $I_6$ by reduction from positive 1-in-3-SAT. Another way to see it is to show that all polymorphisms of those relations are essential unary but none is constant, and then apply the recent results from~\cite{Topo-Birk}.

Otherwise, by Theorem~\ref{thm:minimalTractableClones}
the reduct $\Gamma$ has a polymorphism of one of 17 described types, and
 we have to prove that $\Csp(\Gamma)$ is in P.
 If Item~1 applies, that is if $\Gamma$ is preserved by a constant polymorphism, then 
 $\Csp(\Gamma)$ is trivially tractable as already stated in Proposition~\ref{prop:endos}. In the case of items~2 to~5, $\Csp(\Gamma)$ is tractable by Proposition~\ref{prop:horn}. If $\Gamma$ is preserved by a function of type majority or minority (Item~6 to 15) then $\Csp(\Gamma)$ is tractable by Propositions~\ref{prop:tract:bc},~\ref{prop:tract:d} and~\ref{prop:tract:e}. In those cases, certain binary canonical injections of type projection are required -- these are obtained by identifying the first two variables of the function of type majority / minority, and possibly exchanging the two arguments -- Figure~\ref{fig:classification-correspondence} shows which function of type majority / minority yields which type of binary injection. We leave the verification to the reader. 
 
 Finally, suppose that $\Gamma$ is preserved by an operation $f$ which is an $E$-constant binary injection from Item~16; the case of Item~17 is similar. Then $g(x):=f(x,x)$ is a homomorphism from $\Gamma$ to the structure $\Delta$ induced by the image $g[V]$ in $\Gamma$. 
 This structure $\Delta$ is invariant under 
 all permutations of its domain, and hence is first-order definable 
 in $(g[V];=)$; such structures definable by equality only have been called \emph{equality constraint languages} in~\cite{ecsps}, and their computational complexity has been classified. The structure $\Delta$ has a binary injection among its polymorphisms, namely, the restriction of $f$ to $\Delta$. It then follows from the results in~\cite{ecsps} that $\Csp(\Delta)$ is tractable. Hence, by Proposition~\ref{prop:homoEqui} $\Csp(\Gamma)$ tractable as well, since $\Gamma$ and $\Delta$ are homomorphically equivalent. 
\end{proof}

Clearly, if we add relations to a reduct $\Gamma$, then the CSP of the structure thus obtained is computationally at least as complex as the CSP of $\Gamma$. On the other hand, by Lemma~\ref{lem:pp-reduce}, adding relations with a primitive positive definition to a reduct does not increase the computational complexity of the corresponding CSP more than polynomially. Therefore, it makes sense to call a reduct \emph{primitive positive closed} if it contains all relations that are primitive positive definable from it, and work with such reducts. Observe that primitive positive closed reducts will have infinitely many relations, and hence do not define a CSP; however, as we have already discussed in Section~\ref{sect:toolsua}, it is convenient to consider a primitive positive closed reduct $\Gamma$ tractable if every reduct which has finitely many relations, all taken from $\Gamma$, has a tractable CSP.

 The primitive positive closed reducts of $G$ form a complete lattice, in which the meet of an arbitrary set $S$ of reducts is their \emph{intersection}, i.e., the reduct which has precisely those relations that are relations of all reducts in $S$.  Call a primitive positive closed reduct \emph{maximal tractable} if it is tractable and any extension of it by relations that are first-order definable in $G$ is not tractable anymore. Under the assumption that P does not equal NP, we will now list the maximal tractable reducts of $G$; there are 17 of them. It will also follow from our proof that a reduct of $G$ is tractable if and only if its relations are contained in the relations of one of the reducts of our list.

Recall the notion of a \emph{clone} from Section~\ref{sect:toolsua}. It follows from Theorem~\ref{conf:thm:inv-pol} and Proposition~\ref{prop:locclos} that the lattice of primitive positive closed reducts of $G$ and the lattice of locally closed clones containing $\Aut(G)$ are antiisomorphic via the mappings $\Gamma\mapsto\Pol(\Gamma)$ (for reducts $\Gamma$) and ${\mathcal C}\mapsto \Inv({\mathcal C})$ (for clones ${\mathcal C}$). We refer to the introduction of \cite{BodChenPinsker} for a detailed exposition of this well-known connection. Therefore, the maximal tractable reducts correspond to \emph{minimal tractable} clones, which are precisely the clones of the form $\Pol(\Gamma)$ for a maximal tractable reduct. We can use Corollary~\ref{cor:operational-main} 
to determine the minimal tractable clones; the maximal tractable reducts then are those with relations $\Inv({\mathcal C})$ for a minimal tractable clone ${\mathcal C}$.

\begin{corollary}
Assume $\text{P} \neq \text{NP}$. There are 17 minimal tractable clones that contain $\Aut(G)$; equivalently, there are 17 maximal tractable reducts of $G$. 
\end{corollary}
\begin{proof}
By Corollary~\ref{cor:operational-main} and the previous discussion, 
every minimal tractable clone that contains $\Aut(G)$ must contain
an operation from one of the 17 types of operations listed in Theorem~\ref{thm:minimalTractableClones}. Also recall that 
each operation of one of those 17 types generates a clone that contains
every other operation with the same type. 
It therefore suffices to verify that all of these 17 clones are incomparable 
(i.e., no clone of the list contains another clone of the list), 
and hence that the clones in our list are indeed minimal. 

This task is automatically verifiable:
all functions in a clone generated by a set of canonical functions
from a finite power of $G$ to $G$ are canonical -- this can be shown by a straightforward induction over terms, since
type conditions propagate through composition. Given a finite set $\cal F$
of canonical functions in form of their behaviors, for fixed $n\geq 1$
we can calculate all behaviors of the $n$-ary functions generated by
$\cal F$ by composing the behaviors in all possible ways until we do not
obtain any new behaviors. 
By this method, an algorithm can check
that indeed, the behaviors of the ternary functions of each of the clones in our list are distinct.
\end{proof}

\begin{figure*}
\begin{center}
\includegraphics[scale=0.7]{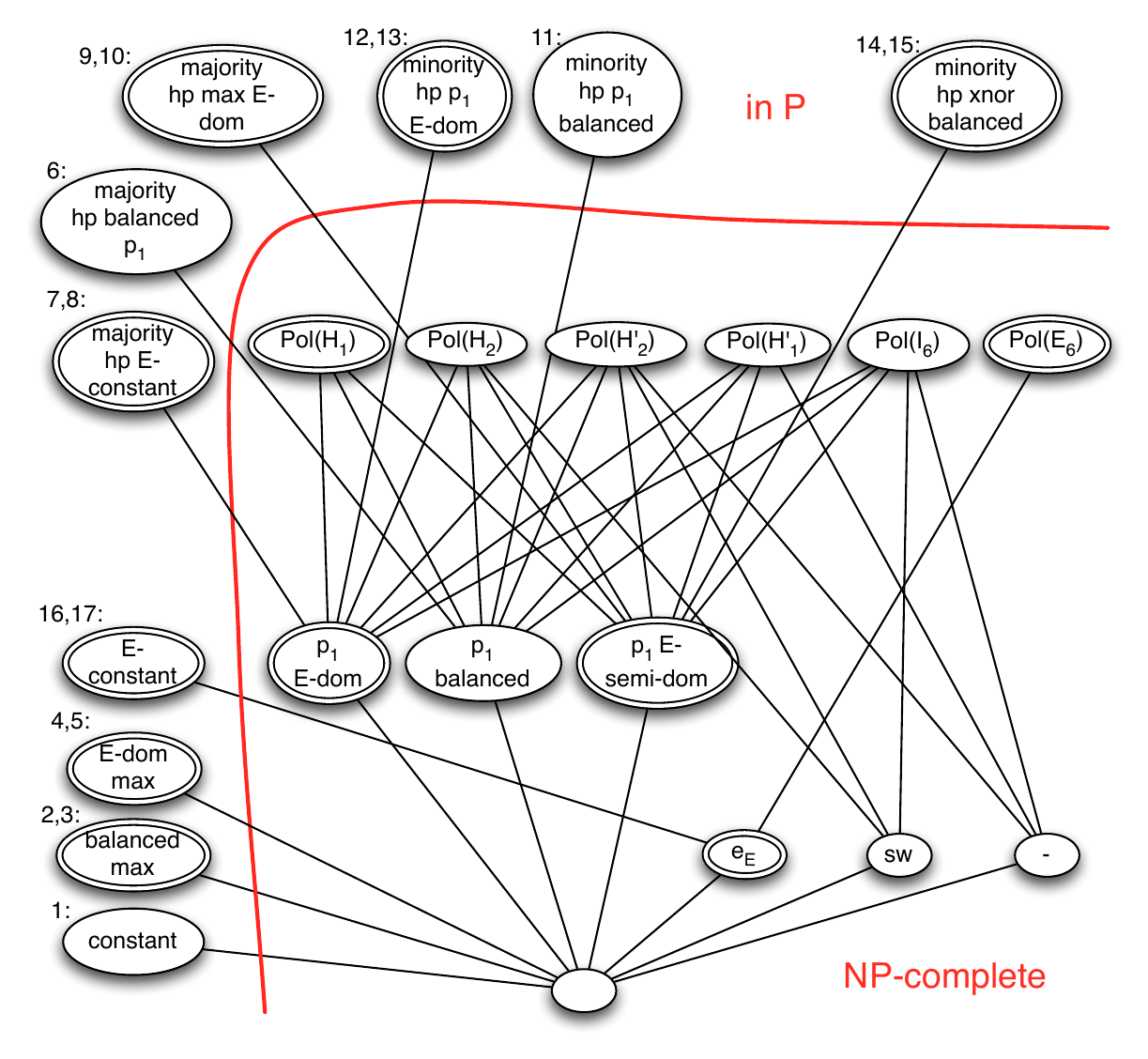}
\end{center}
\caption{The border: Minimal tractable and maximal hard clones containing $\Aut(G)$.}\label{fig:main}
\end{figure*}

Figure~\ref{fig:main} shows the border between 
the clones of reducts with hard, and those with tractable CSP. 
The picture contains all minimal tractable clones as well as all maximal hard clones (with their obvious definition), plus some other clones that are of interest in this context. Lines between the circles that symbolize clones indicate containment (however, we do not mean to imply that there are no other clones between them which are not shown in the picture). Clones are symbolized with a double border when they have a dual clone (generated by the dual function in the sense of Definition~\ref{defn:dual}, whose behavior is obtained by exchanging $E$ with $N$, $\maxi$ with $\mini$, and $\xnor$ with $\xor$). Of two dual clones, only one representative (the one which has $E$ and $\maxi$ in its definition) is included in the picture. The numbers of the minimal tractable clones refer to the numbers in Theorem~\ref{thm:minimalTractableClones}. ``$E$-semidominated'' refers to ``balanced in the first and $E$-dominated in the second argument''. 

We conclude by giving the argument for the decidability claim of Theorem~\ref{thm:main}.

\begin{proposition}\label{prop:decidability}
	There is an algorithm which given a finite set $\Psi$ of graph formulas decides whether or not the problem Graph-SAT($\Psi$) is tractable.
\end{proposition}
\begin{proof}
	By Corollary~\ref{cor:operational-main}, the algorithm only has to check whether one of the canonical functions in Theorem~\ref{thm:minimalTractableClones} preserves all formulas $\psi$ in $\Psi$. To do so it applies the canonical operation to orbit representatives from tuples satisfying $\psi$ in all possible ways, and checks whether the result satisfies $\psi$, too.
\end{proof}

We remark that it also follows from the more recent and more general result in~\cite{BodPinTsa} that it is decidable whether or not one of the relations in Corollary~\ref{cor:operational-main}	has a primitive positive definition from a given finite language reduct $\Gamma$ of $G$ (of which the relations are given as graph formulas). This again yields Proposition~\ref{prop:decidability}.

Observe that the algorithm in the proof of Proposition~\ref{prop:decidability} even decides tractability of Graph-SAT($\Psi$) in polynomial time if the formulas $\psi$ in $\Psi$ are given as follows: if $R$ is the, say, $k$-ary relation defined by $\psi$ in $G$, then for every orbit of $k$-tuples in $G$ that is contained in $R$ the representation of $\psi$ has a $k$-tuple representing this orbit (with the information which relations $E$, $N$, and $=$ hold on the tuple). Now since the operations the algorithm has to consider are at most ternary, the number of possibilities for applying a canonical function to orbit representatives is at most cubic in the number of orbits satisfying $\psi$, which equals the representation size of $\psi$ under this assumption.